\documentclass[review,authoryear,3p]{elsarticle}

\usepackage{amsmath,amssymb, lscape, bm, ascmac}
\usepackage{float}
\usepackage{caption}
\usepackage{subcaption}
\usepackage{appendix}
\usepackage{color}
\definecolor{red}{gray}{0}

\usepackage[colorlinks=true]{hyperref}
\journal{Journal of Icarus}

\biboptions{authoryear}

\begin{document}

\begin{frontmatter}
\newpageafter{abstract}

\title{Dependence of the initial internal structure\\ of chondrule rim\\ on dust size distribution}

\author[mymainaddress]{Hiroaki Kaneko\corref{mycorrespondingauthor}}
\cortext[mycorrespondingauthor]{Corresponding author}
\ead{kaneko.h.aq@m.titech.ac.jp}

\author[mysecondaryaddress]{Sota Arakawa}

\author[mymainaddress]{Taishi Nakamoto}

\address[mymainaddress]{Department of Earth and Planetary Sciences, Tokyo Institute of Technology, 2-12-1 Ookayama, Meguro-ku, Tokyo 152-8550 Japan}
\address[mysecondaryaddress]{Division of Science, National Astronomical Observatory of Japan, 2-21-1 Osawa, Mitaka-shi, Tokyo 181-8588 Japan}

\begin{abstract}
Coarse objects in chondrites such as chondrules and CAIs are mostly coated with fine-grained rims (FGRs). FGRs can be formed on the surface of free floating chondrules in a turbulent nebula, where dust aggregation also occurs. A former study has reported that the morphology of the dust populations accreting onto chondrules affects the initial structures of FGRs. It was revealed that, if monomer grains accrete onto chondrules, the smaller grains tend to accumulate near the surface of chondrules, and FGRs exhibit grain size coarsening from the bottom to the top. However, the study did not consider the effect of temporal growth of dust aggregates on FGRs formation. In this study, we calculate the aggregation of polydisperse monomer grains and their accretion onto chondrules. The following two different stages of dust aggregation can be identified: the monomer-aggregation stage and the BCCA-like stage. In the monomer-aggregation stage, monomer grains are incorporated into aggregates when the average aggregate size reaches the size of the monomer. In the BCCA-like stage, aggregates evolve fractally in a fashion similar to that of single size monomer grains. Based on the results of the previous study, we obtain the requisite conditions for chondrules to acquire monomer-accreting FGRs with grain size coarsening observed in some chondrites. In the case of similar size distribution as that of Inter Stellar Medium (ISM), the maximum grain size of $>1 \ \mathrm{\mu m}$ is widely ($\alpha<10^{-3}$) required for monomer accretion, while if turbulent intensity in a nebula is extremely weak ($\alpha<10^{-5}$), a maximum grain size $\sim 10 \ \mathrm{\mu m}$ is required. The monomer size distributions having larger mass fraction in the large grains compared to ISM might be necessary for the effective grain size coarsening.
\end{abstract}


\end{frontmatter}

%\linenumbers

\section{Introduction}
Chondrules and calcium-aluminum-rich refractory inclusions (CAIs) are the major components of primitive meteorites, i.e. the chondrites. These components are often coated with fine-grained rims (FGRs) with grain size ranging from $\mathrm{nm}$ to $\mathrm{\mu m}$ and FGRs are visibly distinct from the interstitial matrix.

The origin of FGRs has been debated and still unsolved. Several models on the origin of FGR have been suggested so far. The nebular origin models \citep{Metzler.etal1992} suggest that FGRs were developed around the freely floating chondrules in the solar nebular due to their collisions with fine dust grains \citep{Morfill.etal1998, Cuzzi2004}. On the other hand, the parent body origin models suggest that dust grains were first attached to chondrules and FGRs were formed via compaction of the attached dust grains in the regolith \citep{Sears.etal1993, Takayama.Tomeoka2012, Trigo-Rodriguez.etal2006}, or they were formed via alteration of chondrules in the parent bodies \citep{Sears.etal1993, Takayama.Tomeoka2012}. 

The origin of FGRs via alteration of chondrules may be unlikely because the chemical composition of FGRs is independent of the type of core chondrules, and several authors consider the nebular origin models to be the most likely \citep[e.g.][]{Brearley1993}. One of the most important observations that supports the nebular origin model is the positive correlation between chondrule radii and the volumes of FGRs \citep[e.g.][]{Hanna.Ketcham2018}. Such a positive correlation can be achieved in a turbulent nebular, where solid particles are dynamically decoupled from the surrounding gas depending on their size \color{red} \citep{Morfill.etal1998, Cuzzi2004, Ormel.etal2008, Carballido2011, Xiang.etal2019}\color{black}. Larger chondrules have larger relative velocity to the gas and fine dust grains, which are well coupled with the gas. \color{red} Therefore, larger chondrules floating in the nebular sweep more dust grains compared to smaller chondrules \color{black} and, thereby, they develop larger volume of FGRs. It is likely that both FGRs and interstitial matrix were compressed; however, strain analysis on Allende (CV3) chondrites indicate that FGR and interstitial matrix experienced different compaction events \citep{Bland.etal2011}. \color{red}\citet{Bland.etal2011} suggested that the initial porosity of FGRs was approximately 70--80\% before compression, which is in agreement with the theoretical and experimental predictions of the nebular origin of FGRs \citep{Ormel.etal2008, Beitz.etal2013, Xiang.etal2019}\color{black}. 
 
\citet{Xiang.etal2019} examined the formation processes of FGRs on chondrules by $N$-body and Monte Carlo simulations, and they found that the resultant internal structures of FGRs depend on the turbulent intensity and the chondrule radius. Moreover, the morphology of dust particles (i.e., monomer grains or aggregates) accreting onto chondrules also affects the structures of FGRs. When monomer grains (individual grains of $\mathrm{nm}$ to $\mathrm{\mu m}$ size) accrete onto chondrules (monomer-accretion case), the smaller grains preferentially accumulate near surface of chondrules and the larger grains tend to pile up at the upper layer of FGRs. A radial gradient in the size of monomer grains is found from the bottom to the top of FGRs. On the other hand, when dust aggregates (coalescence of monomer grains) accrete onto chondrules (aggregate-accretion case), no radial structure of FGRs is found, and the size of monomer grains is uniformly distributed throughout rims.

\color{red}In the nebula, monomer grains quickly collide and stick with each other to form aggregates \citep[e.g.][]{Matsumoto.etal2019, Matsumoto.etal2021}. Dust particles evolve fractally, and the filling factor of aggregates decreases as they grow. \citet{Xiang.etal2019} treated a monomer accretion case that ignored the ongoing and inevitable, time-dependent transformation of monomers into aggregates. \color{black}Dust aggregation controls the phase of monomer accretion and as a result the structures of FGRs. Moreover, \citet{Xiang.etal2019} fixed the size distribution of monomer grains that was similar to that of the Inter Stellar Medium (ISM). However, the grain size population must have evolved in the nebula \citep[e.g.][]{Ashworth1977, Toriumi1989}. For example, during chondrule-forming flash heating events, small dust particles might have evaporated completely and recondensed as monomer grains through homogeneous nucleation, which could have changed the grain size of the population from ISM and also potentially generated nm size ($<100 \ \mathrm{nm}$) grains \citep{Miura.etal2010, Arakawa.Nakamoto2016}. The purpose of this study is to investigate the effect of dust aggregation before their accretion onto chondrules on the structure of FGRs. We set the initial population of monomer grains with various power law size distributions and calculate their collisional growth considering the porosity change. We aim to understand the conditions of the disk environment and grain size distribution for the monomer-accretion case and the aggregate-accretion case. 

The remainder of this paper is organized as follows. In section 2, we introduce our model, which includes dust aggregation and porosity change. In section 3, we present the results of our numerical calculations. In section 4, we discuss the caveats of our model and the expected structures of FGRs. In section 5, we conclude the implications of the results of this study. \color{red}In the following sections, "grain" and "monomer grain" refer to monomer, "aggregate" refers to aggregate, and "dust" and "dust particle" refer to both monomer and aggregate.\color{black}

\section{Model}
\subsection{Gas disk}
In this study, we consider dust aggregation and rim formation on chondrules in a local box. We neglect the radial transport of solid materials and confine our simulations to the disk midplane where chondrule formation might have occurred.  We use the minimum mass solar nebular model \citep{Hayashi1981}; but for the surface density gradient, power law index of $-1$ is assumed instead of $-1.5$ as indicated by astronomical observations \citep[e.g.][]{Andrews2020}. The gas surface density and the midplane temperature at the distance $r$ from the central star are given as follows:
\begin{equation}
\Sigma_{g} = 1700\left(\frac{r}{\mathrm{au}}\right)^{-1}\mathrm{g/cm^{2}} =  680\left(\frac{r}{2.5 \ \mathrm{au}}\right)^{-1}\mathrm{g/cm^{2}} \ \ ,
\end{equation}
\begin{equation}
T =  280\left(\frac{r}{\mathrm{au}}\right)^{-0.5}\mathrm{K}  =  177\left(\frac{r}{2.5 \ \mathrm{au}}\right)^{-0.5}\mathrm{K} \ \ .
\end{equation}
When a vertically isothermal temperature profile is assumed, we obtain the gas mass density at the midplane as follows:   
\begin{equation}
\rho_{g} = 1.72 \times 10^{-10}\left(\frac{r}{2.5 \ \mathrm{au}}\right)^{-9/4}\mathrm{g/cm^{3}} \ \ .
\end{equation}
We note that our results can be easily scaled with different gas disk profiles. \color{red}In this work, we fix the location at $r=2.5 \ \mathrm{au}$ and do not use $r$-dependence of gas surface density and temperature anymore.\color{black}

\subsection{Dust motion}
In a protoplanetary disk, dust grains stick together when they collide with each other. For the source of relative velocity $\Delta v$ between two dust particles or between the dust particles and (rimmed) chondrules, we only consider the Brownian motion $\Delta v_{\mathrm{BM}}$, which is dominant during the early stage of aggregation of  $\mathrm{\mu m}$ size grains, and the random motion caused by gas turbulence $\Delta v_{t}$, which is dominant at the succeeding stage of aggregation of $100 \ \mathrm{\mu m}$ -  $\mathrm{mm}$ size particles. Relative velocity due to the Brownian motion is given as follows:
\begin{equation}
\Delta v_{\mathrm{BM}} = \sqrt{\frac{8k_{\mathrm{b}}T (m_{1}+m_{2}) }{\pi m_{1}m_{2}} },
\label{brownian motion}
\end{equation}
where $k_{\mathrm{b}}$ is Boltzmann constant. Relative velocity due to the turbulent motion can be obtained from the following closed form
\color{red}
\begin{equation}
\Delta v_{t} = \sqrt{\alpha}c_{s}\begin{cases} \mathrm{Re}^{1/4}(\mathrm{St}_{1}-\mathrm{St}_{2})   &(t_{s,1}<t_{\eta}) \\ \\
\left[ 2y_{a}-(1+\delta)+\frac{2}{1+\delta}\left(\frac{1}{1+y_{a}}+\frac{\delta^{3}}{y_{a}+\delta}\right) \right]^{1/2}\sqrt{\mathrm{St}_{1}}   &(5t_{\eta}\lesssim t_{s,1}\leq t_{L}) \\ \\
\left(\frac{1}{1+\mathrm{St}_{1}}+\frac{1}{1+\mathrm{St}_{2}}\right)^{1/2}   &(t_{L}<t_{s,1})
\end{cases}
\label{turbulent motion}
\end{equation}\\
\color{black}
derived by \citet{Ormel.Cuzzi2007}. The relative velocity $\Delta v$ is given as $\Delta v=\sqrt{\Delta v_{\mathrm{BM}}^{2}+\Delta v_{\mathrm{t}}^{2}}$. In the above formulae, subscripts 1 and 2 denote the colliding particles; $m_{1}$ and $m_{2}$, $t_{s,1}$ and $t_{s,2}$ ($t_{s,1}>t_{s,2}$), and $\mathrm{St}_{1}$ and $\mathrm{St}_{2}$ denote masses, stopping time against the surrounding gas, and Stokes number of colliding particles, respectively. Stokes number is the ratio of the stopping time to the turnover time of the largest turbulent eddy, $t_{L}$ (Keplerian time $\Omega^{-1}$ is assumed). Turbulent Reynolds number $\mathrm{Re}$ is the ratio of the turbulent viscosity $\nu_{t}$ to the molecular viscosity $\nu_{m}$, $t_{\eta}=\mathrm{Re}^{-0.5}t_{L}$ is Kolmogorov timescale, $y_{a}\approx 1.6(t_{s,1}\ll t_{L})$ is a numerical parameter, and \color{red} $\delta=t_{s,2}/t_{s,1}<1$\color{black}. For the turbulent viscosity, we use $\alpha$ prescription \citep{Shakura.Sunyaev1973}
\begin{equation}
\nu_{t}=\alpha c_{s}^{2} \Omega^{-1},
\end{equation}
where $\alpha$ is the dimensionless parameter that describes the intensity of turbulence. Unless otherwise noted, $\alpha=10^{-4}$ is assumed. Isothermal sound speed is $c_{s}=\sqrt{k_{\mathrm{b}}T/m_{g}}$ and the mean molecular mass is $m_{g}=3.9 \times 10^{-24} \ \mathrm{g/cm^{3}}$. At the disk midplane, $\mathrm{Re}$ can be expressed as follows:
\begin{equation}
\mathrm{Re}=\frac{\alpha\sigma_{\mathrm{mol}}\Sigma_{g}}{2m_{g}},
\end{equation}
where $\sigma_{\mathrm{mol}}=2.0 \times 10^{-15} \ \mathrm{cm^{2}}$ is the molecular cross-section \citep{Adachi.etal1976}.\\

In this study, particle sizes (both dust and chondrules) are always smaller than the gas molecular mean free path ($11 \ \mathrm{cm}$ at $2.5 \ \mathrm{au}$ in our gas disk model); hence the Epstein drag regime can be applied. In this case, Stokes number of particles with size $a$ and filling factor $\phi$ at the midplane can be obtained as follows:
\begin{equation}
\mathrm{St}=\frac{\pi a\rho_{\mathrm{mat}}\phi}{2\Sigma_{g}},
\end{equation}
where $\rho_{\mathrm{mat}}$ is the material density of the dust grains for which a value of $3.0 \ \mathrm{g/cm^{3}}$ is assumed. We obtain $a$ and $\phi$ from the volume $V$ and mass $m$ of the particles as $V=(4\pi/3)a^{3}$ and $\phi=(m/V)/\rho_{\mathrm{mat}}$ by apporoximating them as spheres. Here, we do not distinguish the characteristic radius of aerodynamic behavior from the geometric radius, which is valid when the rims are not highly porous \color{red}(dust aggregation mainly proceeds through the Brownian motion which does not depend on Stokes numbers of the dust particles). \color{black} We note that for the turbulent relative velocity of \citet{Ormel.Cuzzi2007}, we use the first regime in Eq. \ref{turbulent motion} for simplicity, as shown later.

\subsection{Time evolution} \label{Time evolution}
In most of our calculations, the initial mass density of dust and chondrules at the midplane, $\rho_{d,0}$ and $\rho_{c,0}$, respectively, where subscripts c and d denote chondrule and dust, are both set to be $0.005\rho_{g}$ (then total solid to gas mass ratio at the midplane is $0.01$). Zero denotes the initial value. We fix $\rho_{c,0}$ and change $\rho_{d,0}$. Subsequently, we solve Smoluchowski equation which is given as follows:
\begin{align}
\frac{\partial N(m,\phi)}{\partial t}=&\frac{1}{2}\iiiint K(m^{'},\phi^{'},m^{''},\phi^{''})N(m^{'},\phi^{'})N(m^{''},\phi^{''}) \notag \\
& \ \ \ \ \ \ \ \ \ \ \ \ \ \ \ g(m,\phi,m^{'},\phi^{'},m^{''},\phi^{''}) dm^{'}d\phi^{'}dm^{''}d\phi^{''} \notag \\
&-\iint K(m,\phi ,m^{'},\phi^{'})N(m,\phi)N(m^{'},\phi^{'}) dm^{'}d\phi^{'} \notag \\
&-K_{c}N(m,\phi)n_{c}.
\label{Smoluchowski}
\end{align}
This equation describes the temporal evolution of $N(m,\phi)$, the number density of the dust particles per unit mass $m$ and unit filling factor $\phi$. On the right-hand side of Eq. \ref{Smoluchowski}, the first term corresponds to an increase of $N(m,\phi)$ due to coagulation of two dust particles with $g=1$ for total mass $m$ and the resulting filling factor $\phi$, but otherwise $g=0$. The second term corresponds to a decrease due to collisions of the particles, whose mass and filling factor are $m$ and $\phi$, with others, and the last term implies a decrease due to their accretion onto chondrules to form rims, where $n_{c}$ is the number density of chondrules. The collisional kernels $K$ and $K_{c}$ are defined as $K=\pi (a^{'}+a^{''})^{2} \times \Delta v$ and $K_{c}=\pi (a+a_{c})^{2} \times \Delta v$, respectively. We assume that the collisional cross-section between dust particles and chondrules is determined by the radius of (rimmed) chondrule $a_{c}$, and neglect the contribution of dust particles. Furthermore, we consider that the relative velocity between dust particles and chondrules is dominated by the turbulent relative motion of chondrules against surrounding gas for which $\mathrm{St}_{2}$ = 0 is substituted in the first formula of Eq. \ref{turbulent motion}; thus we use $K_{c} \simeq \pi a_{c}^{2} \times \Delta v_{t}$.

\subsection{Porosity change} \label{Porosity change}
We set the initial size distributions of monomer grains as power law number density distributions, $n_{0}(a) \propto a^{\beta}$ with $\beta=-4.5, -3.5$, and $-2.5$. \color{red}These three values are representative of three cases; $\beta=-4.5$ corresponds to the case where the smallest grains dominate the grain populations, $\beta=-2.5$ corresponds to the case where the largest grains dominate the grain populations, and $\beta=-3.5$ corresponds to the intermediate case (see subsection \ref{other beta value} for further explanations). \color{black} First, dust particles collide with each other mainly due to the Brownian motion. Subsequently, the turbulent motion induced by the surrounding gas starts to play its role and they quickly accrete onto chondrules \citep{Ormel.etal2008} because the collisional timescale between dust particles is roughly the same as that of accretion onto chondrules (see their section 2.4). As the collisional energy of the Brownian motion is low enough to rotate monomer grains inside the aggregates, the resultant collisions between dust particles might lead to hit-and-stick growth, which implies no deformation occurs. In this case, the newly formed aggregates have voids that reduce their internal density. To describe this internal evolution, we use the empirical formula for aggregation of the single size monomer grains that \citet{Okuzumi.etal2009} obtained from $N$-body simulations. Although we consider the growth of monomer grains with various size distributions, we note that the volume of voids that are generated from the sticking of two aggregates is independent of the size distribution of constituent monomers in the hit-and-stick regime. The volume of voids is given by the following equation:
\begin{equation}
V_{\mathrm{void}}=\mathrm{min}\left[0.99-1.03\ln{\left( \frac{2}{V_{1}/V_{2}+1} \right)}, 6.94 \right]  \times  V_{2} \ ,
\end{equation}\\
where $V_{1}$ and $V_{2}$ ($V_{1}>V_{2}$) are the volume of the colliding particles. The volume of the resultant aggregates after collisions $V_{1+2}$ is given by
\begin{equation}
V_{1+2}=V_{1}+V_{2}+V_{\mathrm{void}}.
\end{equation} 

\subsection{Numerical setting}
\begin{figure}[H]
  \centering
  \includegraphics[width=15cm,pagebox=cropbox,clip]{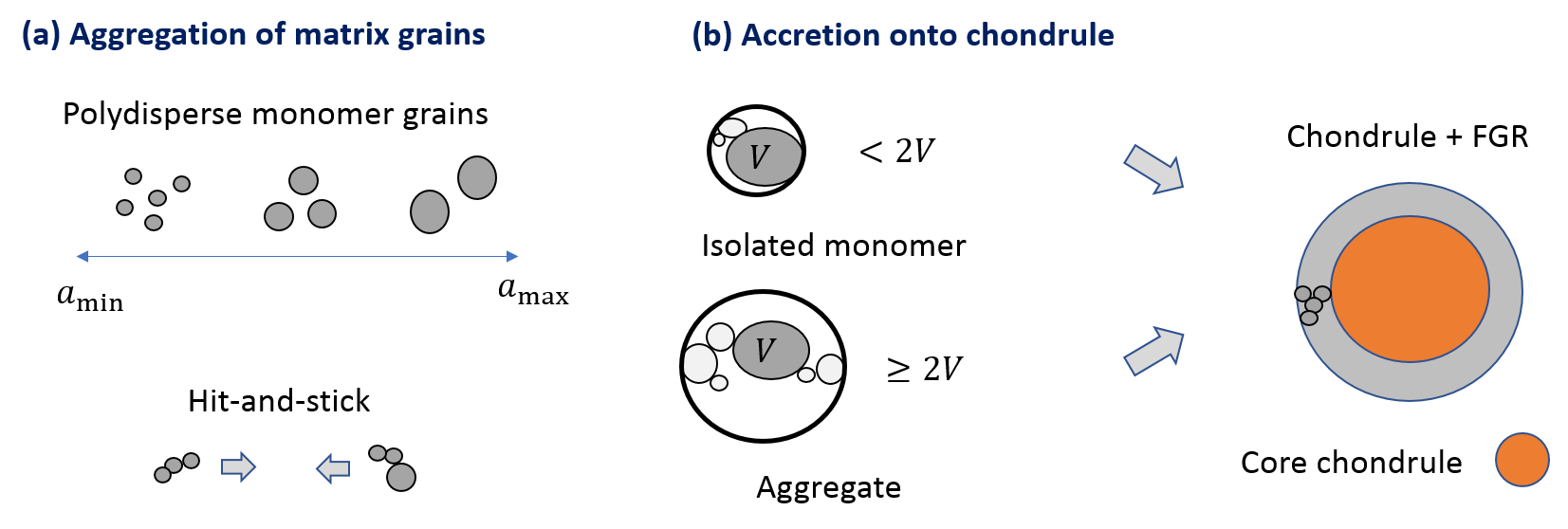}
  \caption{Outline of dust aggregation and their accretion onto chondrules. (a) First, polydisperse monomer grains (gray circle) and their aggregates collide with each other leading to hit-and-stick growth. (b) Then, the isolated monomer grains or aggregates accrete onto chondrules (orange circle) to form FGRs. If a dust particle has a volume less than twice the largest grain inside it, we define this largest grain as an isolated monomer grain.}
  \label{outline}
\end{figure}

\color{red} We divide the mass distribution logarithmically into bins such that mass $m[k+1]$ implies $10^{1/10}$ times mass $m[k]$, with $m_{\mathrm{min}}$, $10^{1/10}m_{\mathrm{min}}$, $10^{2/10}m_{\mathrm{min}} \cdots$, where $m_{\mathrm{min}}$ is the mass of the smallest grain. We also divide the filling factor logarithmically into $1$, $10^{-1/20}$, $10^{-2/20}$ to $10^{-20/20}(=0.1)$ to resolve the difference between monomer grains and aggregates (see below), and $10^{-11/10}(=0.1\times10^{-1/10})$, $10^{-12/10} \cdots$. To judge whether the monomer-accreting or aggregate-accreting rim formation occurs, we also track the maximum size of monomer grains inside the individual dust particles of mass $m$ and filling factor $\phi$ to distinguish between monomer grains and aggregates by the rule below. If a dust particle has a volume less than twice the largest grain inside it, we define this largest grain as an isolated monomer grain, which is not incorporated into an aggregate yet (Figure \ref{outline}). In contrast, aggregation between two isolated monomer grains with the same size leads to the loss of these isolated grains because the resultant aggregate has a volume larger than twice the volume of these grains. We integrate the Smoluchowski equation explicitly with a default time increment of $1 \ \mathrm{yr}$, but time increment is determined at every timestep to avoid a decrease in mass by more than $10 \%$ at a single timestep for every bin. In our model, most of the dust particles accrete onto chondrules at the timescale of $\sim 100 \ \mathrm{yr}$. We calculate the growth of dust particles for $1000 \ \mathrm{yr}$. We note that the ratios of $10^{1/10}$ between the adjacent bins are somewhat coarse \citep[e.g.][]{Lee2000} and we must restrict a decrease in mass at a single timestep to a lower fraction to converge the numerical solutions. However, more accurate calculations are numerically expensive. We confirmed that our results differ just slightly from the more accurate calculations, and there are no qualitative differences. The influence of such a small difference to our conclusions is mush smaller than that of the uncertainties in the gas surface density and turbulent intensity of the protoplanetary disks. In \ref{resolution test}, we show an example of our test calculations with the higher resolutions for single size monomer grains, which can be compared with the previous work by \citet{Okuzumi.etal2009}. \color{black}

\section{Results}
\color{red}At the beginning of this section, we overview the procedures we take in this and following sections. As mentioned in the previous section, small dust particles grow due to the Brownian motion, and as soon as the turbulent motion takes over, they accrete onto chondrules at a faster rate than their collisions with each other, so further growth of aggregates can be neglected. In subsection \ref{Aggregation of dust particles without accretion onto chondrules}, we solve dust aggregation by considering only the Brownian motion and neglecting dust aggregation by the turbulent motion and their accretion onto chondrules to see the behavior of dust aggregation by the Brownian motion. We fit the results with the power law functions with two different stages based on the nature of dust aggregation. In the next subsection \ref{Including accretion of dust particles onto chondrules}, we introduce dust aggregation by the turbulent motion and accretion of dust particles onto chondrules. Using the power law fitting formulae in subsection \ref{Aggregation of dust particles without accretion onto chondrules} and analytical treatments that combine dust aggregation and sweep up by chondrules (see \ref{appendixB}), we obtain the analytically modified fitting formulae for this more realistic cases with dust accretion onto chondrules (these formulae are more complicated due to the additional effects of accretion onto chondrules and this is the reason why we first consider only dust aggregation by the Brownian motion in subsection \ref{Aggregation of dust particles without accretion onto chondrules}). The elapsed time can be replaced by the mass fraction of dust rims that already accrete onto chondrules against core chondrules. The timescale needed to form FGRs that are observed in the meteorites \citep[e.g.][]{Hanna.Ketcham2018} can be expected using the actual rim mass fraction, and we can calculate the dust growth within this timescale.

We consider the case of $\beta=-3.5$, which corresponds to the size distribution of ISM \citep{Mathis.etal1977} that was also used by \citet{Xiang.etal2019}, in subsections \ref{Aggregation of dust particles without accretion onto chondrules} and \ref{Including accretion of dust particles onto chondrules}, and then move onto the cases of $\beta=-4.5$ and $\beta=-2.5$ in subsection \ref{other beta value}. We obtain the analytically modified fitting formulae for $\beta=-4.5$ and $\beta=-2.5$ including the effect of dust accretion onto chondrules. These analytically modified formulae for each $\beta$ value are applied to obtain the requirements for monomer-accreting and aggregate-accreting rim formation to occur in subsection \ref{requirement}.\color{black}

\subsection{Aggregation of dust particles without accretion onto chondrules} \label{Aggregation of dust particles without accretion onto chondrules}
\color{red}In this and next subsections, we consider the case of $\beta=-3.5$. \color{black} We define the average size of aggregates $a_{\mathrm{agg}}$ and the average filling factor $\phi_{\mathrm{agg}}$ as follows:
\begin{align} 
\frac{4\pi}{3}a_{\mathrm{agg}}^{3} &:= \int V(a)n_{\mathrm{agg}}(a)da/\int n_{\mathrm{agg}}(a)da, \\
\rho_{\mathrm{mat}}\phi_{\mathrm{agg}} &:= \iint mN_{\mathrm{agg}}(m,\phi)dmd\phi/\int V(a)n_{\mathrm{agg}}(a)da.
\end{align}
The number density distribution of the aggregates and the isolated monomers grains with respect to their size $a$ are given by $n_{\mathrm{agg}}(a)$ and $n_{\mathrm{mono}}(a)$, respectively. Two types of number density, $N(m,\phi)$ and $n(a)$, have different dimensions but satisfy $\iint N(m,\phi)dmd\phi=\int n(a)da$. At the beginning, $n_{\mathrm{mono}}(a)=n_{0}(a)$ and  $n_{\mathrm{agg}}(a)=0$ (see section \ref{Porosity change}). Then, the number density of monomer grains decreases and that of the aggregates increases. Note that the aggregates with the same size $a$ can have various filling factors.

Figure \ref{isolated monomer grains} shows the fraction of isolated grains for each size of monomer grains,  $n_{\mathrm{mono}}(a)/n_{0}(a)$, in the case of $(a_{\mathrm{min}},a_{\mathrm{max}})=(0.01 \ \mathrm{\mu m},1 \ \mathrm{\mu m})$. Throughout this paper, $a_{\mathrm{min}}$ and $a_{\mathrm{max}}$ refer to the maximum and minimum size of the initial population of monomer grains. In this figure, different colors represent monomer grains with different sizes. The horizontal axis shows the average size of aggregates, $a_{\mathrm{agg}}$, which increases as time progresses. The number density of monomer grains, $n_{\mathrm{mono}}(a)$, is found to sharply decrease as $a_{\mathrm{agg}}$ reaches $a$. Smaller grains are more rapidly incorporated into the aggregates compared to the larger grains. 

\color{red}This trend \color{black} also can be observed in the volume distribution. Figure \ref{volume distribution a 3.5} shows the total volume per each size bin ($V(a)n(a)a$) divided by the initial volume of the smallest grains ($V(a_{\mathrm{min}})n_{0}(a_{\mathrm{min}})a_{\mathrm{min}}$), when the average aggregate size is $0.2 \ \mathrm{\mu m}$. At this time, most of the monomer grains whose size is more than $0.5 \ \mathrm{\mu m}$ retain their initial volume distribution and remain isolated (note that the slight increase in volume is due to the adhesion of much smaller grains). After the average size of aggregates reaches the maximum size of the monomer grains (Figures \ref{volume distribution c 3.5} and \ref{volume distribution d 3.5}), the shape of the distribution does not change while keeping its range relatively narrow. 

The timescale of the growth of dust aggregates is important when we consider their accretion onto chondrules to form rims. We normalize the average aggregate size, $a_{\mathrm{agg}}$, by the minimum size of monomer grains, $a_{\mathrm{min}}$, and the elapsed time by the collisional timescale between the smallest monomer grains, $t_{\mathrm{min}}$, where
\color{red}
\begin{align}
t_{\mathrm{min}}& =  \frac{1}{4\pi a_{\mathrm{min}}^{2} \times n_{0}(a_{\mathrm{min}})a_{\mathrm{min}} \times \Delta v_{\mathrm{BM}}} \notag \\
& =\begin{cases} \frac{1}{4\pi a_{\mathrm{min}}^{2} \times \frac{(4+\beta)\rho_{d}a_{\mathrm{min}}^{4+\beta}}{m_{\mathrm{gr}}(a_{\mathrm{min}}) \left\{ a_{\mathrm{max}}^{4+\beta}-a_{\mathrm{min}}^{4+\beta} \right\} } \times \sqrt{\frac{16k_{\mathrm{b}}T}{\pi m_{\mathrm{gr}}(a_{\mathrm{min}})}}}  & (\beta \neq -4)\\
\frac{1}{ 4\pi a_{\mathrm{min}}^{2} \times \frac{ \rho_{d} }{ m_{\mathrm{gr}}(a_{\mathrm{min}}) \ln(a_{\mathrm{max}}/a_{\mathrm{min}}) } \times \sqrt{\frac{16k_{\mathrm{b}}T}{\pi m_{\mathrm{gr}}(a_{\mathrm{min}})}} }  & (\beta=-4).
\end{cases}
\label{def_t_min}
\end{align}\\
\color{black}
The mass of a monomer grain, $m_{\mathrm{gr}}(a)$, is defined as $m_{\mathrm{gr}}(a)=(4\pi/3)a^{3}\rho_{\mathrm{mat}}$. Using these normalizations, we can scale our results with those of other disk environments.

Figure \ref{evolution of aggregate size 3.5} shows the normalized time evolution of the average size of aggregates with the fitting formulae (the black dotted lines in this figure), which is given as follows:
\color{red}
\begin{align}
a_{\mathrm{agg}}/a_{\mathrm{min}}=& \begin{cases}1.10\left(\frac{t}{t_{\mathrm{min}}}\right)^{0.794}  & (a_{\mathrm{agg}}/a_{\mathrm{min}}<Y)  \\
1.13Y^{-0.322}\left(\frac{t}{t_{\mathrm{min}}}\right)^{1.05} & (a_{\mathrm{agg}}/a_{\mathrm{min}}>Y),
\end{cases}
\label{fitting1}
\end{align}\\
where
\begin{equation}
Y=2.96\left(a_{\mathrm{max}}/a_{\mathrm{min}}\right)+131 \label{Y}.
\end{equation}
See \ref{appendixA} for the detail of fitting. Eqs. \ref{fitting1} and \ref{Y} fit well except $a_{\mathrm{max}}/a_{\mathrm{min}} \ll 50.0$ (due to too simple linear fitting, see Figure \ref{YX}). There are two stages due to two different natures of dust aggregation (see below). \color{black} To test the validity of the normalization, we compare the results from the same $a_{\mathrm{max}}/a_{\mathrm{min}}$ ratio, $ (a_{\mathrm{min}}, a_{\mathrm{max}})=(0.02 \ \mathrm{\mu m}, 1 \ \mathrm{\mu m})$ (red) and $(0.05 \ \mathrm{\mu m}, 2.5 \ \mathrm{\mu m})$ (yellow). Both results are found to agree well with each other.

Figure \ref{evolution of filling factor 3.5} shows the normalized temporal evolution of the average filling factor of aggregates, where the black dotted lines indicate the fitting formulae given as follows:
\begin{align}
\phi_{\mathrm{agg}}=& \begin{cases}0.903\left(\frac{a_{\mathrm{agg}}}{a_{\mathrm{min}}}\right)^{-0.420}  &(a_{\mathrm{agg}}/a_{\mathrm{min}}<X)\\
0.903X^{0.650}\left(\frac{a_{\mathrm{agg}}}{a_{\mathrm{min}}}\right)^{-1.07} &(a_{\mathrm{agg}}/a_{\mathrm{min}}>X),
\end{cases}
\label{fitting2}
\end{align}\\
where
\begin{equation}
X=0.284\left(a_{\mathrm{max}}/a_{\mathrm{min}}\right)+11.9 \label{X},
\end{equation}
which fit well except $a_{\mathrm{max}}/a_{\mathrm{min}} \ll 50.0$. Two size distributions, $ (a_{\mathrm{min}},a_{\mathrm{max}})=(0.05 \ \mathrm{\mu m},2.5 \ \mathrm{\mu m})$ and $(0.02 \ \mathrm{\mu m},1 \ \mathrm{\mu m})$, both agree well with each other in this figure as well. 

In Figures \ref{evolution of aggregate size 3.5} and \ref{evolution of filling factor 3.5}, two stages of aggregation are observed and their branch point is approximately the time at which the average size of aggregates becomes equal to the maximum size of monomer grains. \color{red} In Figures \ref{volume distribution 3.5}, during the first stage (Figures \ref{volume distribution a 3.5} and \ref{volume distribution b 3.5}), there are remaining large monomer grains, and monomers and aggregates coevolve to the larger aggregates. On the other hand, during the second stage (Figures \ref{volume distribution c 3.5} and \ref{volume distribution d 3.5}), there are almost no monomer grains, and aggregates with relatively narrow size spread evolve to the larger aggregates. \color{black} After the branch point, aggregates begin BCCA-like growth, which refers to the aggregation of the particles having the same volume (BCCA refers to Ballistic Cluster-Cluster Aggregation). Similar to BCCA, the filling factor is approximately inversely proportional to the size of aggregates, which corresponds to the fractal dimension of $\sim2$ \citep[e.g.][]{Okuzumi.etal2009}. Furthermore, in the case of aggregation due to the Brownian motion, its growth timescale increases nearly proportionally to the size of aggregates. These behaviors can be interpreted as aggregates of aggregates in BCCA fashion. We refer to this second aggregation stage as BCCA-like stage hereafter. Before the BCCA-like stage, isolated monomer grains with different sizes are incorporated into aggregates when the average aggregate size reaches the size of monomers (Figure \ref{isolated monomer grains}). We refer to this first aggregation stage as the monomer-aggregation stage, which has the fractal dimension of $2.5$ to $2.6$. 

\subsection{Including accretion of dust particles onto chondrules}  \label{Including accretion of dust particles onto chondrules}
Analytic discussions on the accretion of dust particles onto chondrules were well established by \citet{Morfill.etal1998} and \citet{Cuzzi2004}, and we briefly review the things that are relevant to this study in the beginning of this subsection. 
\color{red}We demonstrate here that the total accreted mass of dust onto chondrules can be used as a clock of dust aggregation instead of $t$ without tracking the actual rim internal structures by some assumptions and approximations mentioned in the last of subsection \ref{Time evolution} and reconsidered in subsection \ref{Setting of the model}. \color{black}

In our model, the mass density of rimmed chondrules evolves as follows:
\begin{equation}
\frac{\mathrm{d}\rho_{c}}{\mathrm{d}t}=\rho_{d} \pi a_{c}^{2} n_{c} \Delta v_{t}(\mathrm{St}_{c}), \\
\label{accretion onto chondrule}
\end{equation}
where $n_{c}$ is the number density of chondrules, and $\mathrm{St}_{c}(t)$ is Stokes number of rimmed chondrules, which is given as follows:
\color{red}
\begin{equation}
\mathrm{St}_{c}(t)=\frac{\pi a_{c}(t)\rho_{\mathrm{mat}}\phi_{\mathrm{c}}(t)}{2\Sigma_{g}}. \\ 
\label{Stc(t)}
\end{equation}
The radius of rimmed chondrules is $a_{c}(t)$, which is the sum of the radius of core chondrules and the thickness of the rims, and $\phi_{\mathrm{c}}(t)$ is the filling factor of the rimmed chondrules averaging the contributions from rims and core chondrules through,
\begin{equation}
\phi_{\mathrm{c}}(t)=\frac{m_{\mathrm{core}}+m_{\mathrm{rim}}(t)}{m_{\mathrm{core}}+m_{\mathrm{rim}}(t)/ \phi_{\mathrm{rim}}(t)}, \label{phi-c}\\ 
\end{equation}
where $m_{\mathrm{core}}$ and $m_{\mathrm{rim}}$ are the mass of core chondrules and rims, respectively, $\phi_{\mathrm{rim}}$ is the filling factor of rims, and the internal density of core chondrules of $\rho_{\mathrm{mat}}$ is assumed. The properties of rims, $m_{\mathrm{rim}}$ and  $\phi_{\mathrm{rim}}(t)$, are time-dependent and thus $\mathrm{St}_{c}(t)$ is as well. The filling factor of rims, $\phi_{\mathrm{rim}}$, is strongly related to $\phi_{\mathrm{agg}}$ but not equivalent to it. Compaction of rims, which is discussed in subsection \ref{Setting of the model}, or voids generated when dust particles accrete onto chondrules affect the internal structures of rims \citep[e.g.][]{Ormel.etal2008, Xiang.etal2019}. The porous rims around chondrules make their coupling with the gas stronger and lead to lower Stokes number, which is indicated in reduced $\phi_{c}$ in Eq. \ref{Stc(t)}. We note that the rims also increase the mass of rimmed chondrules and the objects with larger mass have larger Stokes number due to their larger inertia. However, if $\phi_{\mathrm{rim}} < (2-m_{\mathrm{rim}}/m_{\mathrm{core}})/3$, the accumulation of the rims around chondrules reduces Stokes number. The turbulent relative velocity between chondrules and dust particles, $\Delta v_{t}(\mathrm{St}_{c})=\sqrt{\alpha}\mathrm{Re}^{1/4}\mathrm{St}_{c}c_{s}$, is obtained from Eq. \ref{turbulent motion} with $\mathrm{St}_{1}=\mathrm{St}_{c}(t)$ and $\mathrm{St}_{2}=0$. The left-hand side of Eq. \ref{accretion onto chondrule} describes the increase in mass due to the accretion of dust particles. Eq. \ref{accretion onto chondrule} assumes perfect sticking of rim materials, and the mass of dust particles that accrete onto chondrules are regarded as the mass of chondrules. We return to this point in subsection \ref{Setting of the model}. Eq. \ref{accretion onto chondrule} can be transformed into
\begin{align}
\frac{\mathrm{d}\rho_{c}}{\mathrm{d}t}= &\rho_{d} \pi a_{c}^{2} n_{c} \sqrt{\alpha}\mathrm{Re}^{1/4}\mathrm{St}_{c}c_{s} \notag \\ 
= &\rho_{d} \pi a_{c}^{3}\rho_{\mathrm{mat}}\phi_{\mathrm{c}} n_{c} \left(\frac{\pi \sqrt{\alpha}\mathrm{Re}^{1/4}c_{s}}{2\Sigma_{g}}\right) \notag \\ 
= &\rho_{d} m_{c}n_{c} \left(\frac{3\pi \sqrt{\alpha}\mathrm{Re}^{1/4}c_{s}}{8\Sigma_{g}}\right), 
\label{accretion onto chondrule1.5}
\end{align}
where $m_{c}=m_{\mathrm{core}}+m_{\mathrm{rim}}$ is the mass of rimmed chondrules, and we obtain,
\begin{equation}
\frac{\mathrm{d}\rho_{c}}{\mathrm{d}t}= \left(\frac{3\pi \sqrt{\alpha} \mathrm{Re}^{1/4} c_{s}}{8\Sigma_{g}}\right)\rho_{c}(\rho_{\mathrm{tot}}-\rho_{c}),
\label{accretion onto chondrule2}
\end{equation}
with
\begin{equation}
\rho_{\mathrm{tot}}=\rho_{d}+\rho_{c}=\mathrm{const}. 
\end{equation}
Eq. \ref{accretion onto chondrule1.5} indicates that sweep up rate of rimmed chondrules is proportional to the mass of them and independent of the detail structures of rims. \color{black} Solving Eq. \ref{accretion onto chondrule2}, we can obtain 
\begin{align}
\rho_{c}=&\frac{ \exp \left\{ A_{\mathrm{min}}\left(\frac{t}{ t_{\mathrm{min}} }\right) \right\} \rho_{\mathrm{tot}} }{ X_{cd}+\exp \left\{ A_{\mathrm{min}}\left(\frac{t}{ t_{\mathrm{min}} }\right) \right\} }, \\
\rho_{d}=&\frac{ X_{cd}\rho_{\mathrm{tot}} }{X_{cd}+\exp \left\{ A_{\mathrm{min}}\left(\frac{t}{ t_{\mathrm{min}} }\right) \right\}},
\label{accretion}
\end{align}
with
\begin{align}
A_{\mathrm{min}}=& \left(\frac{3\pi \sqrt{\alpha} \mathrm{Re}^{1/4} \rho_{\mathrm{tot}}c_{s}}{8\Sigma_{g}}\right) \times t_{\mathrm{min}}, 
\label{A} \\
X_{cd}=&\frac{\rho_{d,0}}{\rho_{c,0}},
\end{align}
where the elapsed time is normalized by $t_{\mathrm{min}}$. \color{red} We checked that when $t=0$, $\rho_{c}=\rho_{c,0}$ and $\rho_{d}=\rho_{d,0}$ are satisfied. We also confirmed that when $t \to \infty$, $\rho_{c} \to \rho_{\mathrm{tot}}$ and $\rho_{d} \to 0$ are satisfied. \color{black} Dimensionless parameter $A_{\mathrm{min}}$ expresses the ratio of the timescale of the aggregate growth (precisely, the collisional timescale between two smallest monomer grains) to that of accretion onto chondrules. When $\beta=-3.5$, $A_{\mathrm{min}}$ and $t_{\mathrm{min}}$ are scaled as follows:
\begin{align}
A_{\mathrm{min}}=5.63 \times 10^{-4} &\left(\frac{\alpha}{10^{-4}}\right)^{3/4}\left(\frac{\Sigma_{g}}{680 \ \mathrm{g/cm^{2}}}\right)^{-3/4} \notag \\
&\left(\frac{1+X_{cd}}{X_{cd}}\right)\left(\frac{a_{\mathrm{min}}}{0.1 \ \mathrm{\mu m}}\right)^{2.5}\left(\sqrt{\frac{a_{\mathrm{max}}}{a_{\mathrm{min}}}}-1\right), 
\label{Amin} \\
t_{\mathrm{min}}=2.35 \times 10^{-1} & \ \mathrm{yr} \left(\frac{r}{2.5 \ \mathrm{au}}\right)^{3/2} \left(\frac{\Sigma_{g}}{680 \ \mathrm{g/cm^{2}}}\right)^{-1} \notag \\
&\left(\frac{\rho_{d,0}/\rho_{g}}{5.0\times 10^{-3}}\right)^{-1} \left(\frac{a_{\mathrm{min}}}{0.1 \ \mathrm{\mu m}}\right)^{2.5}\left(\sqrt{\frac{a_{\mathrm{max}}}{a_{\mathrm{min}}}}-1\right).
\end{align}\\
\color{red} We note that we do not apply $r$-dependence of gas surface density and temperature from our gas disk model. The dependence on $r$ above comes from only stellar vertical gravity (or Keplerian angular velocity). \color{black} The mass ratio of the rims (dust particles that accrete onto chondrules) to chondrules core\color{red}, $m_{\mathrm{rim}}/m_{\mathrm{core}}$, \color{black} is given as
\begin{equation}
\epsilon_{\mathrm{rim}}=\frac{ \exp \left\{ A_{\mathrm{min}}\left(\frac{t}{t_{\mathrm{min}}}\right) \right\}-1 }{ 1+\frac{1}{X_{cd}}\exp \left\{ A_{\mathrm{min}}\left(\frac{t}{t_{\mathrm{min}}}\right) \right\} }.
\end{equation}
The time $t_{\mathrm{rim}}(\epsilon_{\mathrm{rim}})$ at which chondrules acquire the rims, whose mass ratio to chondrules core is $\epsilon_{\mathrm{rim}}$ can be expressed as
\begin{align}
\frac{t_{\mathrm{rim}}(\epsilon_{\mathrm{rim}})}{t_{\mathrm{min}}} &= \frac{1}{A_{\mathrm{min}}}\ln \left[\frac{ 1+\epsilon_{\mathrm{rim}} }{ 1-\frac{\epsilon_{\mathrm{rim}}}{X_{cd}} }\right] \notag \\
&= \left(\frac{4.17\times 10^{2} \ \mathrm{yr}}{t_{\mathrm{min}}}\right) \left(\frac{\alpha}{10^{-4}}\right)^{-3/4} \left(\frac{r}{2.5 \ \mathrm{au}}\right)^{3/2} \left(\frac{\Sigma_{g}}{680 \ \mathrm{g/cm^{2}}}\right)^{-1/4} \notag \\
& \ \ \ \ \ \ \ \ \ \ \ \ \ \ \ \ \ \ \ \ \ \ \ \ \ \left(\frac{\rho_{d,0}/\rho_{g}}{5.0\times 10^{-3}}\right)^{-1} \left(\frac{X_{cd}}{1+X_{cd}}\right)\ln \left[\frac{ 1+\epsilon_{\mathrm{rim}} }{ 1-\frac{\epsilon_{\mathrm{rim}}}{X_{cd}} }\right].
\label{trim}
\end{align}
\color{red} The mass fraction of rims to chondrule core, $\epsilon_{\mathrm{rim}}$, and time, $t$, are in one-to-one correspondence through $t=t_{\mathrm{rim}}(\epsilon_{\mathrm{rim}})$. In the limit of $\epsilon_{\mathrm{rim}} \ll 1$, the rim formation timescale can be approximated as $t_{\mathrm{rim}}(\epsilon_{\mathrm{rim}}) \simeq m_{\mathrm{core}}\epsilon_{\mathrm{rim}}/\rho_{d,0}\pi a_{c,0}^{2}\Delta v_{t}(\mathrm{St}_{c,0})$, where $a_{c,0}$ and $\mathrm{St}_{c,0}$ are the radius and Stokes number of core chondrules, respectively.
\color{black} 

Figure \ref{Including} shows the numerical results, which include dust aggregation by the turbulent motion and the accretion of dust particles onto chondrules. We fix the initial chondrule density, $\rho_{c,0}=0.005\rho_{g}$, and change the initial dust density $\rho_{d,0}=0.5\rho_{c,0}$, $\rho_{c,0}$, and $10\rho_{c,0}$ ($\rho_{d,0}/\rho_{g}=0.0025,\ 0.005,\ 0.05$, respectively). \color{red} In Figure \ref{Including size}, the size evolutions of aggregates terminate at the certain sizes, but do not in Figure \ref{evolution of aggregate size 3.5}. This difference between Figures \ref{evolution of aggregate size 3.5} and \ref{Including size} is instructive for understanding the timescale arguments. At first, dust particles collide with each other within much shorter timescale than their accretion onto chondrules. However, as they grow, collisional timescale between them increases and eventually exceeds accretion timescale. After that, decrease in the number density of dust particles due to the accretion onto chondrules prevents their collisional growth so the size evolutions of aggregates terminate in Figure \ref{Including size}. From Eq. \ref{fitting1} and \ref{accretion}, we can obtain the following formulae that describe the size evolution of aggregates before their accretion onto chondrules (see \ref{appendixB} for the details of the derivation). 
\begin{align}
a_{\mathrm{agg}}/a_{\mathrm{min}}=& \begin{cases}1.10\left[ \ \left(\frac{1+X_{cd}}{X_{cd}}\right) \left(\frac{t}{t_{\mathrm{min}}}\right) \right.\\
\ \ \ \ \ \ \left. - \left(\frac{1+X_{cd}}{X_{cd}A_{\mathrm{min}}}\right) \ln \left[ \frac{ X_{cd}+\exp \left\{ A_{\mathrm{min}}\left(\frac{t}{t_{\mathrm{min}}}\right) \right\} }{X_{cd}+1}  \right] \ \right]^{0.794}  &(a_{\mathrm{agg}}/a_{\mathrm{min}}<Y)
\\ \\
1.13Y^{-0.322}\left[ \ \left(\frac{1+X_{cd}}{X_{cd}}\right) \left(\frac{t}{t_{\mathrm{min}}}\right) \right.\\
\ \ \ \ \ \ \left. - \left(\frac{1+X_{cd}}{X_{cd}A_{\mathrm{min}}}\right) \ln \left[ \frac{ X_{cd}+\exp \left\{ A_{\mathrm{min}}\left(\frac{t}{t_{\mathrm{min}}}\right) \right\} }{X_{cd}+1}  \right] \ \right]^{1.05}  &(a_{\mathrm{agg}}/a_{\mathrm{min}}>Y).
\end{cases}
\label{fitting3}
\end{align}\\
In Figure \ref{Including size}, the black dotted lines indicate Eq. \ref{fitting3}. Figure \ref{Including filling factor} shows the evolution of the average filling factor of aggregates and the black dotted lines indicate Eq. \ref{fitting2}. It is found that, although we include the turbulent motion as the source of collisions between dust particles, Eqs. \ref{fitting3} and \ref{fitting2} derived from dust aggregation only by the Brownian motion in subsection \ref{Aggregation of dust particles without accretion onto chondrules} fit well with the numerical results. This supports the fact that the accretion of dust particles onto chondrules is faster than their further growth by the turbulent motion. For comparison, we show the result without the turbulent relative velocity between dust particles but including the accretion onto chondrules for the parameter set of $(a_{\mathrm{min}},a_{\mathrm{max}})=(0.05 \ \mathrm{\mu m},2.5 \ \mathrm{\mu m})$ and $X_{cd}=10$ (the brown dashed line in Figure \ref{Including}). This parameter set is the most significant one because the largest $X_{cd}$ and the growth without turbulent motion terminate at the slightly smaller size than the yellow line. We can also observe in Figure \ref{Including} that an increase in the dust density leads to the further growth of aggregates to the larger final size. Within the halving time of dust density due to sweep up by chondrules (note that this halving time depends on the initial dust to chondrule mass ratio because rimmed chondrules have enhanced sweep up rate proportionally to their mass), dust particles with the higher density can grow larger.

However, as shown in Figure \ref{e-rim}, the mass of the rims ($\epsilon_{\mathrm{rim}}$) and the evolution of aggregates ($a_{\mathrm{agg}}$ and $\phi_{\mathrm{agg}}$) have no dependence on the initial dust to chondrules mass ratio. This is because a higher dust density leads to not only a faster growth of aggregates due to more frequent collisions but also a faster rim formation. Substituting $t_{\mathrm{rim}}$ (Eq. \ref{trim}) into Eq. \ref{fitting3}, we can derive the following equations for $a_{\mathrm{agg}}$:  
\begin{align}
a_{\mathrm{agg}}/a_{\mathrm{min}}=& \begin{cases}1.10\left[ \left(\frac{1+X_{cd}}{X_{cd}A_{\mathrm{min}}}\right) \ln \left( 1+\epsilon_{\mathrm{rim}}  \right) \ \right]^{0.794}  &(a_{\mathrm{agg}}/a_{\mathrm{min}}<Y)
 \\ \\
1.13Y^{-0.322}\left[\left(\frac{1+X_{cd}}{X_{cd}A_{\mathrm{min}}}\right) \ln \left( 1+\epsilon_{\mathrm{rim}}  \right) \ \right]^{1.05}  &(a_{\mathrm{agg}}/a_{\mathrm{min}}>Y),
\end{cases}
\label{fitting4}
\end{align}\\
which are plotted as the black dotted lines in Figure \ref{e-rim size}, and $(1+X_{cd})/(X_{cd}A_{\mathrm{min}})$ are independent of $\rho_{d,0}$ and $\rho_{c,0}$ (of cource $\epsilon_{\mathrm{rim}}$ must be lower than $X_{cd}$). Eq. \ref{fitting4} describes the average aggregate size when chondrules complete the formation of rims with the mass fraction $\epsilon_{\mathrm{rim}}$ to core chondrules, and it will be used in the section \ref{requirement} to derive the conditions for the monomer-accretion and aggregate-accretion cases. We can obtain $\phi_{\mathrm{agg}}$ as a function of $\epsilon_{\mathrm{rim}}$ by substituting Eq. \ref{fitting4} into Eq. \ref{fitting2}, which is plotted by the black dotted lines in Figure \ref{e-rim filling factor}. \color{black}

\subsection{Other $\beta$ values of the power law size distributions} 
\label{other beta value}
\color{red} In this subsection, we evaluate the results with $\beta=-4.5$ and $-2.5$, and the dependence on $\beta$. What we do here are almost the same as what we did in the previous subsections \ref{Aggregation of dust particles without accretion onto chondrules} and \ref{Including accretion of dust particles onto chondrules}. First, we consider the dust aggregation by the Brownian motion and establish the fitting formulae for it, then, include the effects of the turbulent motion and sweeping up by chondrules and derive the analytically modified fitting formulae. However, there is an important difference between $\beta=-3.5$ case and $\beta=-4.5$ and $-2.5$ cases. We define the new reference time $t_{\mathrm{single}}$ (Eq. \ref{def_t_single}) instead of $t_{\mathrm{min}}$ (Eq. \ref{def_t_min}). We explain on this in the followings. \color{black}

The most important issue is which end of the size distribution dominates the mass and collisional cross-section. The mass (or volume) and collisional cross-section for each size bin of the grains population is proportional to $(4\pi/3)a^{3}\rho_{\mathrm{mat}}n_{0}(a)a\propto a^{\beta+4}$ and $\pi a^{2}n_{0}(a)a\propto a^{\beta+3}$, respectively. When $\beta=-3.5$, the largest grains dominate the mass, while the smallest grains dominate the collisional cross-section. However, when $\beta=-4.5$ or $-2.5$, either the smallest grains or the largest grains dominate both mass and collisional cross-section. In this case, we can approximate the grain population as almost single size grains if we focus on the average growth of aggregates.
Futhermore, when $\beta=-2.5$, the scaling law cannot be obtained by the normalization with $a_{\mathrm{min}}$ and $t_{\mathrm{min}}$ as shown in Figure \ref{evolution of aggregates 2.5a}. \color{red} We can observe that the tracks of the lines differ among various size distributions, and they do not collapse into one except during the BCCA-like stage ($a_{\mathrm{agg}}>a_{\mathrm{max}}$) in the left panel. We note that this exception exists only when $\beta=-2.5$ and is not the general feature of both $-3.0 \lesssim \beta < -2.5$ and $\beta \geqq -2.5$. The reason why the lines collapse when $\beta=-2.5$ is shown in \ref{beta2.5 reason}. \color{black} Instead of $t_{\mathrm{min}}$, the elapsed time is normalized by $t_{\mathrm{single}}(a)$ in the following procedures, where
\begin{align}
t_{\mathrm{single}}(a) =& \frac{1}{4\pi a^{2} \times \frac{\rho_{d}}{m_{\mathrm{gr}}(a)} \times \sqrt{\frac{16k_{b}T}{\pi m_{\mathrm{gr}}(a)}}} \notag \\
=& 3.71 \times 10 \ \mathrm{yr} \left(\frac{r}{2.5 \ \mathrm{au}}\right)^{3/2} \left(\frac{\Sigma_{g}}{680 \ \mathrm{g/cm^{2}}}\right)^{-1} \left(\frac{\rho_{d,0}/\rho_{g}}{5.0\times 10^{-3}}\right)^{-1} \left(\frac{a}{1.0 \ \mathrm{\mu m}}\right)^{2.5}.
\label{def_t_single}
\end{align}
\color{red} Here, $t_{\mathrm{single}}(a)$ is the timescale at which the single size monomer grains with the size of $a$ collide with each other. We compare the results with that obtained from the monodisperse monomer grains (in other words, with no size range limit). \color{black}

In the case of $\beta=-4.5$, the smallest grains dominate both mass and collisional cross-section, and at the monomer-aggregation stage, they collide and stick with mostly the same smallest grains. Figure \ref{evolution of aggregates 4.5} shows the evolution of aggregates for $\beta=-4.5$ that are normalized by $a_{\mathrm{min}}$ and $t_{\mathrm{single}}(a_{\mathrm{min}})$. 
\color{red} Compared with the case of $\beta=-3.5$ (Figure \ref{evolution of aggregates 3.5}), the slopes of the lines during the monomer-aggregation stage ($a_{\mathrm{agg}}/a_{\mathrm{min}}\lesssim a_{\mathrm{max}}/a_{\mathrm{min}}$) and during the BCCA-like stage ($a_{\mathrm{agg}}/a_{\mathrm{min}}\gtrsim a_{\mathrm{max}}/a_{\mathrm{min}}$) are almost the same and exhibit the slope of BCCA growth. Moreover, even after entering the BCCA-like stage, the evolutionary paths of both average filling factor and aggregate size overlap when $a_{\mathrm{max}}/a_{\mathrm{min}}>20.0$. The reason for this is that, when $\beta=-4.5$, the evolution of aggregates is controlled by the aggregates composed of the smallest monomer grains during the monomer-aggregation stage due to their dominant presence in the grain population. However, we refer to the growth during $a_{\mathrm{agg}} \lesssim a_{\mathrm{max}}$ as the monomer-aggregation stage because there are remaining larger monomer grains. The monodisperse grains exhibit a faster growth, and the grains with narrower size ranges, especially $a_{\mathrm{max}}/a_{\mathrm{min}}<20.0$, grow with closer rates. The fitting formulae of the evolutionary paths of the average aggregate filling factor ($\phi_{\mathrm{agg}}$) and size ($a_{\mathrm{agg}}$) for monodisperse monomer grains are given as
\begin{equation}
a_{\mathrm{agg}}/a_{\mathrm{single}}=0.386\left(t/t_{\mathrm{single}}(a_{\mathrm{single}})\right)^{1.06}
\label{fit-a-single}
\end{equation}
and
\begin{equation}
\phi_{\mathrm{agg}}=0.661\left(a_{\mathrm{agg}}/a_{\mathrm{single}}\right)^{-1.04},
\label{fit-phi-single}
\end{equation}
where $a_{\mathrm{single}}$ is the radius of the single size monomer grains. \color{black} The growth trucks are converged by the following equations when $a_{\mathrm{max}}/a_{\mathrm{min}}>20.0$:
\begin{equation}
a_{\mathrm{agg}}/a_{\mathrm{min}}=0.0739\left(t/t_{\mathrm{single}}(a_{\mathrm{min}})\right)^{1.06}
\label{fit-a-4.5}
\end{equation}
and
\begin{equation}
\phi_{\mathrm{agg}}=1.96\left(a_{\mathrm{agg}}/a_{\mathrm{min}}\right)^{-1.05}.
\label{fit-phi-4.5}
\end{equation}
\color{red} From here on, we mainly focus on cases of $a_{\mathrm{max}}/a_{\mathrm{min}}>20.0$. Eq. \ref{fit-phi-4.5} shows that aggregates in wide size range cases are more compact than those in monodisperse case (see eq. \ref{fit-phi-single}) at the same normalized size due to the contribution of the larger monomer grains. On the other hand, eq. \ref{fit-phi-4.5} shows that aggregates in wide size range cases with $\beta=-4.5$ are more porous than those with $\beta=-3.5$ (because $0.903X^{0.650} > 1.96$, see eq. \ref{fitting2}) at the same normalized size, since there are less large monomer grains in the grain population in $\beta=-4.5$ cases than $\beta=-3.5$ cases. \color{black} 
The volume distributions of dust particles for $\beta=-4.5$ are shown in Figure \ref{volume distribution 4.5}. Similar to the case of $\beta=-3.5$ (Figure \ref{volume distribution 3.5}), the larger grains are isolated longer compared to the smaller grains (Figure \ref{volume distribution a 4.5}). \color{red} Compared to Figure \ref{volume distribution 3.5}, the aggregates have much larger total volume than monomer grains during the monomer-aggregation stage in Figure \ref{volume distribution 4.5} due to the dominant presence of the smallest grains when $\beta=-4.5$. 

If we consider the accretion of dust particles onto chondrules, we have (see \ref{appendixB} for derivation), 
\begin{equation}
a_{\mathrm{agg}}/a_{\mathrm{min}} = 0.0739\left[ \left(\frac{1+X_{cd}}{X_{cd}A_{\mathrm{single}}(a_{\mathrm{min}})}\right) \ln \left( 1+\epsilon_{\mathrm{rim}}  \right) \ \right]^{1.06}, 
\label{fitting4.5acc}
\end{equation}
where
\begin{align}
A_{\mathrm{single}}(a_{\mathrm{min}}) &= \left(\frac{3\pi \sqrt{\alpha} \mathrm{Re}^{1/4} \rho_{\mathrm{tot}}c_{s}}{8\Sigma_{g}}\right) \times t_{\mathrm{single}}(a_{\mathrm{min}}) \notag \\
&=2.81 \times 10^{-4} \left(\frac{\alpha}{10^{-4}}\right)^{3/4}\left(\frac{\Sigma_{g}}{680 \ \mathrm{g/cm^{2}}}\right)^{-3/4} \left(\frac{1+X_{cd}}{X_{cd}}\right)\left(\frac{a_{\mathrm{min}}}{0.1 \ \mathrm{\mu m}}\right)^{2.5}. 
\end{align} 
These analytically modified fitting formulae are plotted with the numerical results in Figure \ref{e-rim size 4.5} to check the validity of our analytical treatments. We can obtain the evolution of the average filling factor by substituting Eq. \ref{fitting4.5acc} into Eq. \ref{fit-phi-4.5}, which is shown in Figure \ref{e-rim filling factor 4.5} with the numerical results. We use Eq. \ref{fitting4.5acc} to derive the requirement for monomer-accretion case in subsection \ref{requirement}. \color{black} \\ 

On the other hand, in the case of $\beta=-2.5$, the largest grains dominate both mass and collisional cross-section. 
Figure \ref{evolution of aggregates 2.5} shows the evolution of aggregates for $\beta=-2.5$ normalized by $a_{\mathrm{max}}$ and $t_{\mathrm{single}}(a_{\mathrm{max}})$. \color{red} The monodiperse grains show a slower growth, and the grains with the narrower size ranges, especially $a_{\mathrm{max}}/a_{\mathrm{min}}<20.0$, grow at closer rates. \color{black} The fitting formulae of the evolutionary paths of $\phi_{\mathrm{agg}}$ and $a_{\mathrm{agg}}$ for grains with wider size ranges converge to
\begin{equation}
a_{\mathrm{agg}}/a_{\mathrm{max}}=1.56\left(t/t_{\mathrm{single}}(a_{\mathrm{max}})\right)^{1.04}
\label{fit-a-2.5}
\end{equation}
and
\begin{equation}
\phi_{\mathrm{agg}}=0.281\left(a_{\mathrm{agg}}/a_{\mathrm{max}}\right)^{-1.04}.
\label{fit-phi-2.5}
\end{equation}
The largest grains grow faster than the monodiperse grains, which were assisted by more porous aggregation with smaller grains. \color{red} Eq. \ref{fit-phi-2.5} shows that aggregates in wide size range cases with $\beta=-2.5$ are more porous than those in monodisperse case (see eq. \ref{fit-phi-single}) at the same normalized size due to the contribution of the smaller monomer grains. We note that Eqs. \ref{fit-a-2.5} and \ref{fit-phi-2.5} denote the BCCA-like stage ($a_{\mathrm{agg}} \gtrsim a_{\mathrm{max}}$) and we must be careful in using these equations (see subsection \ref{requirement}). Figure \ref{volume distribution 2.5} shows the volume distribution of dust particles for the case of $\beta=-2.5$. Similar to the cases of $\beta=-3.5$ (Figure \ref{volume distribution 3.5}) and $\beta=-4.5$ (Figure \ref{volume distribution 4.5}), monomer grains that are larger than the average size of aggregates remain isolated during the monomer-aggregation stage (Figures \ref{volume distribution a 2.5} and \ref{volume distribution b 2.5}). In comparison with the case of $\beta=-3.5$ and $-4.5$, the total volume of aggregates is less than that of isolated larger monomer grains during this stage. Moreover, more survivability of the grains smaller than the average aggregate size can be observed in Figures \ref{volume distribution a 2.5} and \ref{volume distribution b 2.5} than $\beta=-3.5$. This is because both small and large grains are incorporated into aggregates through the collision and sticking with the largest grains having the largest cross-section, and the total of cross-section is lower when $\beta=-2.5$ so the capturing of the smaller grains into aggregates is slower. 

Then, we include the accretion of dust particles onto chondrules. The analytically modified fitting formulae are
\begin{equation}
a_{\mathrm{agg}}/a_{\mathrm{max}} = 1.56\left[ \left(\frac{1+X_{cd}}{X_{cd}A_{\mathrm{single}}(a_{\mathrm{max}})}\right) \ln \left( 1+\epsilon_{\mathrm{rim}}  \right) \ \right]^{1.04}, 
\label{fitting2.5acc}
\end{equation}
where
\begin{align}
A_{\mathrm{single}}(a_{\mathrm{max}}) &= \left(\frac{3\pi \sqrt{\alpha} \mathrm{Re}^{1/4} \rho_{\mathrm{tot}}c_{s}}{8\Sigma_{g}}\right) \times t_{\mathrm{single}}(a_{\mathrm{max}}) \notag \\
&=8.90 \times 10^{-2} \left(\frac{\alpha}{10^{-4}}\right)^{3/4}\left(\frac{\Sigma_{g}}{680 \ \mathrm{g/cm^{2}}}\right)^{-3/4} \left(\frac{1+X_{cd}}{X_{cd}}\right)\left(\frac{a_{\mathrm{max}}}{1.0 \ \mathrm{\mu m}}\right)^{2.5}. 
\end{align}
These equations are plotted with the numerical results in Figure \ref{e-rim size 2.5} to check the validity of our analytical treatments. The green line with the grain size range of $(a_{\mathrm{min}},a_{\mathrm{max}})=(0.002 \ \mathrm{\mu m},2 \ \mathrm{\mu m})$ does not fit Eq. \ref{fitting2.5acc} although it seems to asymptote gradually to the black dotted lines. This is because these monomer grains are too large (or massive) to collide and make aggregates by the Brownian motion before their accretion onto chondrules. In this case, the large monomer grains survive and accrete onto chondrules as monomer grains (see subsection \ref{requirement} for more detail discussion). Moreover, Eq. \ref{fitting2.5acc} does not fit with the numerical results during $a_{\mathrm{agg}} \lesssim a_{\mathrm{max}}$ because previous Eq. \ref{fit-a-2.5} denotes only BCCA-like stage during $a_{\mathrm{agg}} \gtrsim a_{\mathrm{max}}$. These equations can be substituted into Eq. \ref{fit-phi-2.5} to obtain $\phi_{\mathrm{agg}}$ as a function of $\epsilon_{\mathrm{rim}}$ which is shown in Figure \ref{e-rim filling factor 2.5} with the numerical results. In the next section (especially in subsection \ref{requirement}), we will use the equations of $a_{\mathrm{agg}}$ as a function of $\epsilon_{\mathrm{rim}}$ (Eqs. \ref{fitting4}, \ref{fitting4.5acc}, and \ref{fitting2.5acc}), which include the effect of accretion onto chondrules, to obtain the requirement for the monomer-accreting and aggregate-accreting rim formations.
\color{black}

\section{Discussion}

\subsection{Setting of the model} \label{Setting of the model}
\subsubsection{Accretion of dust particles onto chondrules}
In this study, the first regime of Eq. \ref{turbulent motion} \citep{Ormel.Cuzzi2007} was used for the turbulent motion. Stokes number of chondrules and Kolmogorov timescale normalized by the turnover time of the largest eddy (Keplerian timescale $\Omega^{-1}$ was assumed) are given as follows:
\begin{align}
\mathrm{St}_{c}=&2.08 \times 10^{-4} \phi_{c} \left(\frac{a_{c}}{0.3 \ \mathrm{mm}}\right) \left(\frac{\Sigma_{g}}{680 \ \mathrm{g/cm^{2}}}\right)^{-1}, \label{St-c}\\
\mathrm{St}_{\eta}=&\mathrm{Re}^{-1/2}=2.39 \times 10^{-4}\left(\frac{\alpha}{10^{-4}}\right)^{-1/2} \left(\frac{\Sigma_{g}}{680 \ \mathrm{g/cm^{2}}}\right)^{-1/2}. \label{St-eta}
\end{align}
We use $0.3 \ \mathrm{mm}$ as the typical radius of chondrules. In a weakly turbulent ($\alpha \leq10^{4}$) and gas rich nebula, $\mathrm{St}_{c}<\mathrm{St}_{\eta}$ and our treatment is valid. The fact that Stokes number of chondrules coated with fluffy FGRs is lower than that of bare core chondrules also validates our treatment. However, \citet{Hanna.Ketcham2018} found that a positive correlation between core chondrule radii and the volumes of FGRs, which matched the power law function with an index of 2.5. This correlation corresponds to the second regime of Eq. \ref{turbulent motion} ($\mathrm{St}_{c}>\mathrm{St}_{\eta}$), indicates a stronger turbulence or a lower gas density. This is because when the second regime of Eq. \ref{turbulent motion} is applied, the volume that individual chondrules sweep up in unit time is $\pi a_{c}^{2}\Delta v_{t}\propto a_{c}^{2}\sqrt{\mathrm{St}_{c}}$, which is proportional to $a_{c}^{2.5}$. 

We estimate the difference between the first and second regime of the turbulent relative velocity, which are referred to as
\begin{align} 
\Delta v_{t,1} =& \sqrt{\alpha}\mathrm{Re}^{1/4}\mathrm{St}_{c}c_{s} \notag \\
=& 10.6 \phi_{c} \left(\frac{a_{c}}{0.3 \ \mathrm{mm}}\right) \left(\frac{\alpha}{10^{-4}}\right)^{3/4} \left(\frac{\Sigma_{g}}{680 \ \mathrm{g/cm^{2}}}\right)^{-3/4} \left(\frac{T}{177 \ \mathrm{K}}\right)^{1/2}  \ \mathrm{cm/s},
\label{del_vt1}
\end{align} 
and
\begin{align} 
\Delta v_{t,2} =& 1.72\sqrt{\alpha\mathrm{St}_{c}}c_{s} \notag \\
=& 19.6 \phi_{c}^{1/2} \left(\frac{a_{c}}{0.3 \ \mathrm{mm}}\right)^{1/2} \left(\frac{\alpha}{10^{-4}}\right)^{1/2} \left(\frac{\Sigma_{g}}{680 \ \mathrm{g/cm^{2}}}\right)^{-1/2} \left(\frac{T}{177 \ \mathrm{K}}\right)^{1/2} \ \mathrm{cm/s},
\label{del_vt2}
\end{align} 
where $\delta=0$ in Eq. \ref{turbulent motion}. Using Eq. \ref{St-c} and \ref{St-eta}, the difference between these velocities can be calculated as
\begin{equation} 
\frac{\Delta v_{t,1}}{\Delta v_{t,2}} = \frac{1}{1.85} \phi_{c}^{1/2} \left(\frac{a_{c}}{0.3 \ \mathrm{mm}}\right)^{1/2} \left(\frac{\alpha}{10^{-4}}\right)^{1/4} \left(\frac{\Sigma_{g}}{680 \ \mathrm{g/cm^{2}}}\right)^{-1/4}.
\label{vt1/vt2}
\end{equation}
Even if FGRs are formed in an environment having stronger turbulence or lower gas density, the difference between $\Delta v_{t,1}$ and $\Delta v_{t,2}$ should be small (within a factor of $3$ for wide parameter ranges) because of weak dependence among the parameters in Eq. \ref{vt1/vt2}. To deal with the second regime ($\Delta v_{t,2}$), we need to solve the rim structures and chondrule size distribution, which complicates the calculation although the arguments remain mostly unchanged. \color{red} Both $\Delta v_{t,1}<\Delta v_{t,2}$ and $\Delta v_{t,1}>\Delta v_{t,2}$ are possible due to the jump between the two regimes of turbulent relative velocity \citep[see for example, FIG 1 of][]{Ormel.etal2008}, but when $\mathrm{St}_{c} \gg \mathrm{St}_{\eta}$, $\Delta v_{t,1}$ is larger than $\Delta v_{t,2}$ so our calculations overestimate the efficiency of the rim formation and dust particle would grow larger than our estimates before accretion. \color{black} We also note that if FGRs are formed in a nebula with very strong turbulence or very low gas density, then the monomer-accretion is expected to occur as discussed in section \ref{requirement}.

Another assumption on the accretion of dust particles onto chondrules was that we neglected the contribution of dust particles to the collisional cross-section and the relative velocity between chondrules and dust particles. When the radius of dust particles is smaller than the radius of chondrules, Stokes number of dust particles is much lower than that of chondrules due to their porous structures, therefore, the turbulent velocity of dust particles becomes negligible. Furthermore, our main focus is to evaluate if FGRs are mostly formed via monomer accretion or aggregate accretion. It was observed that the growth stage shifts from the monomer-aggregation stage to the BCCA-like stage when the average aggregate size becomes comparable to the maximum size of monomer grains, $a_{\mathrm{max}}$. Since $a_{\mathrm{max}}$ is much smaller than $a_{c}$, our assumption of the collisional cross-section and the relative velocity is valid in the scope of this study. \color{red} We note that, however, neglecting the contribution of dust particles is not always valid especially during BCCA-like stage. If monomer grains are small enough (e.g. $a \lesssim 0.1 \ \mathrm{\mu m}$ for monodisperse monomer grains when $\alpha=10^{-4}$ and $\Sigma_{g}=680 \ \mathrm{g/cm^2}$), the final aggregate size can reach $\gtrsim 100 \ \mathrm{\mu m}$ before accretion, but in such cases, rims must be formed through the accretion of very porous aggregates. \color{black}

\subsubsection{Hit-and-stick growth}
We assumed the hit-and-stick growth of dust aggregates. Here, we check the validity of this assumption. The collisional energy, $0.5m_{1}m_{2}\left(\Delta v_{\mathrm{BM}} \right)^{2}/(m_{1}+m_{2})$, of the Brownian motion is expressed as follows: 
\color{red}
\begin{equation}
E_{\mathrm{coll,BM}}=\frac{4k_{\mathrm{b}}T}{\pi} = 3.11\times 10^{-14} \left(\frac{T}{177 \ \mathrm{K}}\right) \ \mathrm{erg}.
\end{equation}
\color{black}
If this collional energy exceeds the critical rolling energy, $E_{\mathrm{roll}}$ \citep{Dominik.Tielens1995}, then monomer grains inside the colliding aggregates rotate and the aggregates start to deform. The critical rolling energy, $E_{\mathrm{roll}}$ is given as
\color{red}
\begin{equation}
E_{\mathrm{roll}} = 6\pi^{2}\gamma a_{0}\xi = 2.96\times10^{-11}\left(\frac{\gamma}{25 \ \mathrm{erg/cm^{2}}}\right)\left(\frac{a_{\mathrm{gr}}}{0.01 \ \mathrm{\mu m}}\right)\left(\frac{\xi}{0.2 \ \mathrm{nm}}\right)\mathrm{erg},
\end{equation}
\color{black}
where $\gamma=25 \ \mathrm{erg/cm^{2}}$ is the surface energy of silicate grains, $a_{\mathrm{gr}}=0.01 \ \mathrm{\mu m}$ is the typical radius of the smallest monomer grains in our model, and $\xi$ is the critical displacement with $0.2 \ \mathrm{nm}$ being its lower limit \citep{Dominik.Tielens1997}. As $E_{\mathrm{coll,BM}} \ll E_{\mathrm{roll}}$, our assumption that collisions due to the Brownian motion lead to hit-and-stick growth of the aggregates is valid.

\color{red}
We check the turbulent motion as well for completeness. After aggregates become large enough for the turbulent motion to be a significant source of the relative velocity (i.e., the turbulent motion exceeds the Brownian motion), they accrete onto chondrules faster than their further growth. Decrease in the number density of aggregates terminates the growth of aggregates. So, the turbulent motion cannot be much larger than the Brownian motion because Stokes number of aggregates do not increase much more. Then, collisional energy of turbulent motion can be only comparable to or less than the Brownian motion.
\color{black}

\color{red}
\subsubsection{Erosion, fragmentation, sputtering and compaction of FGRs}
We assumed that dust particles swept by chondrules perfectly accumulate as rims ignoring erosion and fragmentation of rims. We did not consider the collisions between rimmed chondrules which can lead to sputtering \citep{Gunkelmann.etal2017} or compaction \citep{Ormel.etal2008} of rims. If $\alpha \sim 10^{-2}$, the turbulent relative velocity of rimmed chondrules can reach $> 1 \ \mathrm{m/s}$ (Eqs. \ref{del_vt1} and \ref{del_vt2}), so events mentioned above might occur. 
The filling factor of rims, $\phi_{\mathrm{rim}}$, is an important parameter which affects the Stokes number of rimmed chondrules considerably (Eq. \ref{phi-c}). If porous aggregates with $\phi_{\mathrm{agg}} \lesssim 0.01$ accrete onto chondrules as rims, the turbulent relative velocity of rimmed chondrules decreases, which can prevent the above events. However, much lower $\phi_{\mathrm{agg}}$ does not necessarily mean much lower $\phi_{\mathrm{rim}}$ because porous aggregates can be compressed when they accrete onto chondrules. Further more, compound aggregates \citep{Arakawa2017, Matsumoto.etal2019} composed of rimmed chondrules and matrix aggregates might suffer ejection of chondrules through collisions \citep{Arakawa2017, Umstatter.Urbassek2021a}. This mechanism can be one of the causes of unrimmed chondrules \citep[e.g][]{Simon.etal2018}. Compaction of rims and matrix might change the fate of compound aggregates. Although our methodologies are valid even if compaction of rims occurs because sweep up rate of chondrules is proportional to the mass of rimmed chondrules as far as the first regime of the turbulent motion (Eq. \ref{turbulent motion}) can be applied, we must look at carefully what kinds of evolution of chondrules, rims, and matrix happen in a nebula in the future work. 
\color{black}

\color{red}
\subsection{Expected structures of FGRs}
\label{requirement}
\begin{figure}[H]
  \centering
  \includegraphics[width=12cm,pagebox=cropbox,clip]{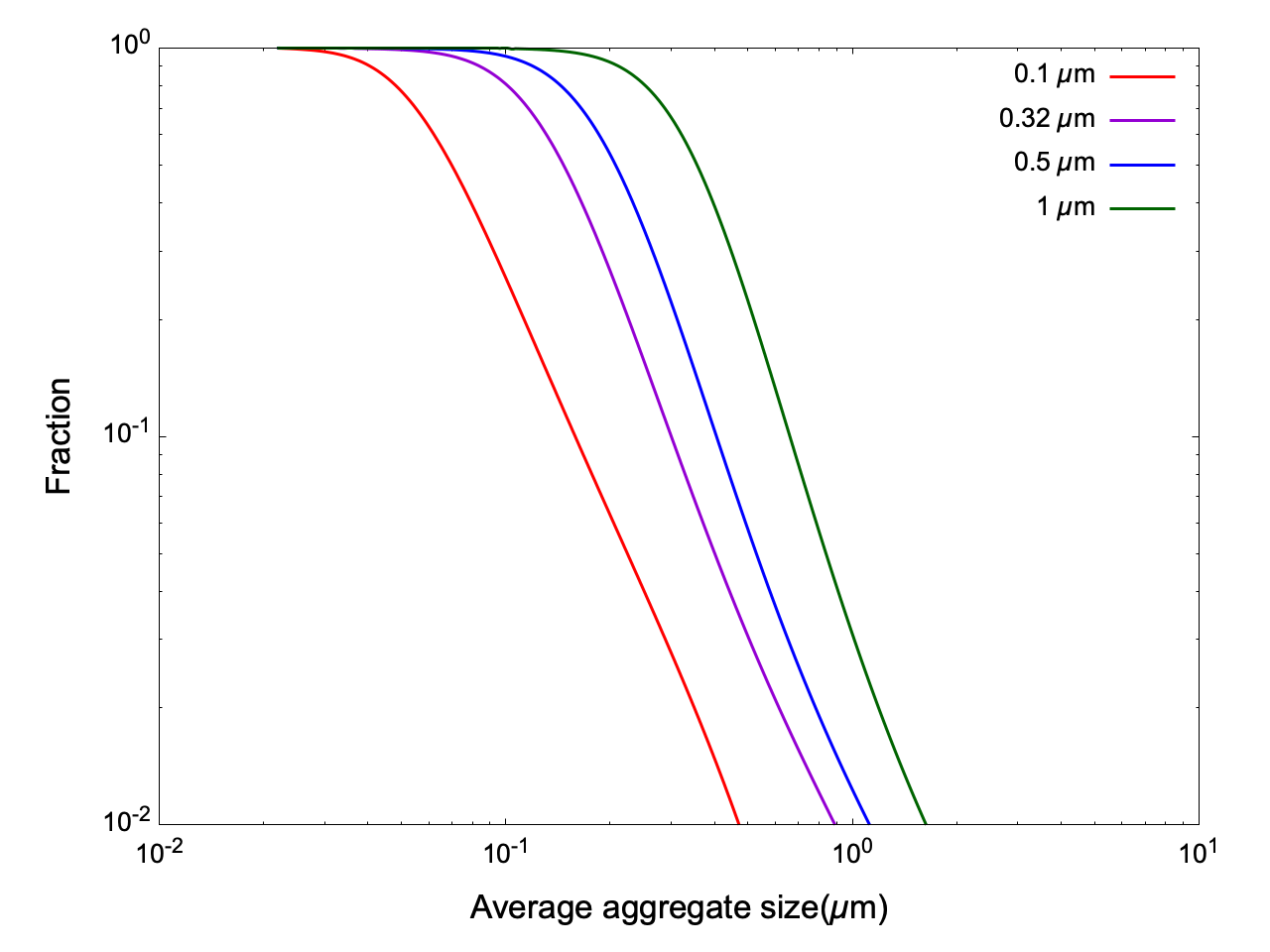}
  \caption{The fraction of isolated monomer grains as a function of time. The horizontal axis shows the average size of aggregates, $a_{\mathrm{agg}}$, which increases with time. The different colors of the curves represent monomer grains with different sizes; $0.1 \ \mathrm{\mu m}$ (red), $0.32 \ \mathrm{\mu m}$ (violet), $0.5 \ \mathrm{\mu m}$ (blue), and $1 \ \mathrm{\mu m}$ (green).}
  \label{isolated monomer grains}
\end{figure}

\begin{figure}[H]
\begin{tabular}{cc}
\begin{minipage}[t]{0.5\hsize}
  \centering
  \includegraphics[width=8.5cm,pagebox=cropbox,clip]{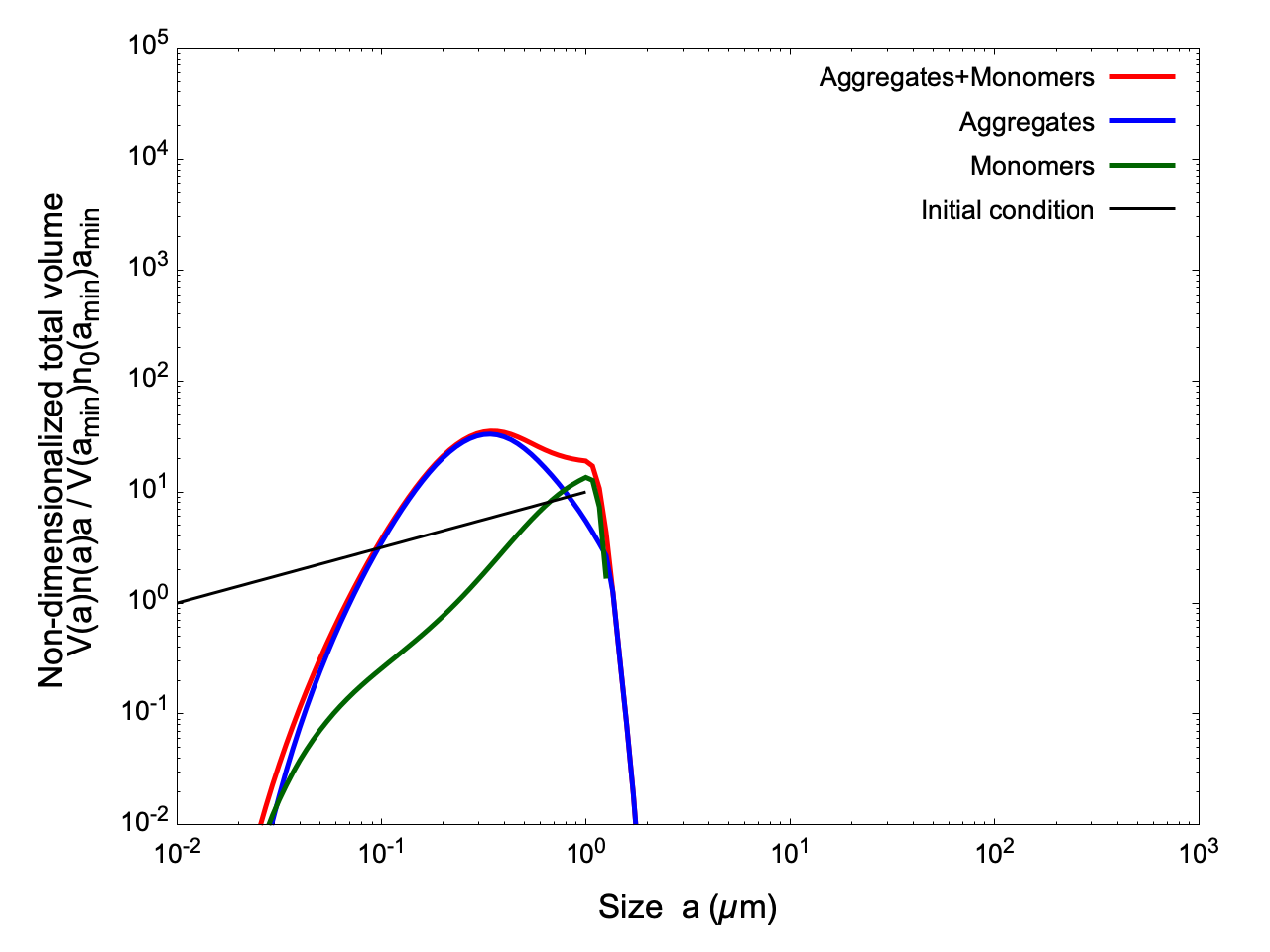}
  \subcaption{Average aggregate size is $0.2 \ \mathrm{\mu m}$}
  \label{volume distribution a 3.5}
\end{minipage}
\begin{minipage}[t]{0.5\hsize}
  \centering
  \includegraphics[width=8.5cm,pagebox=cropbox,clip]{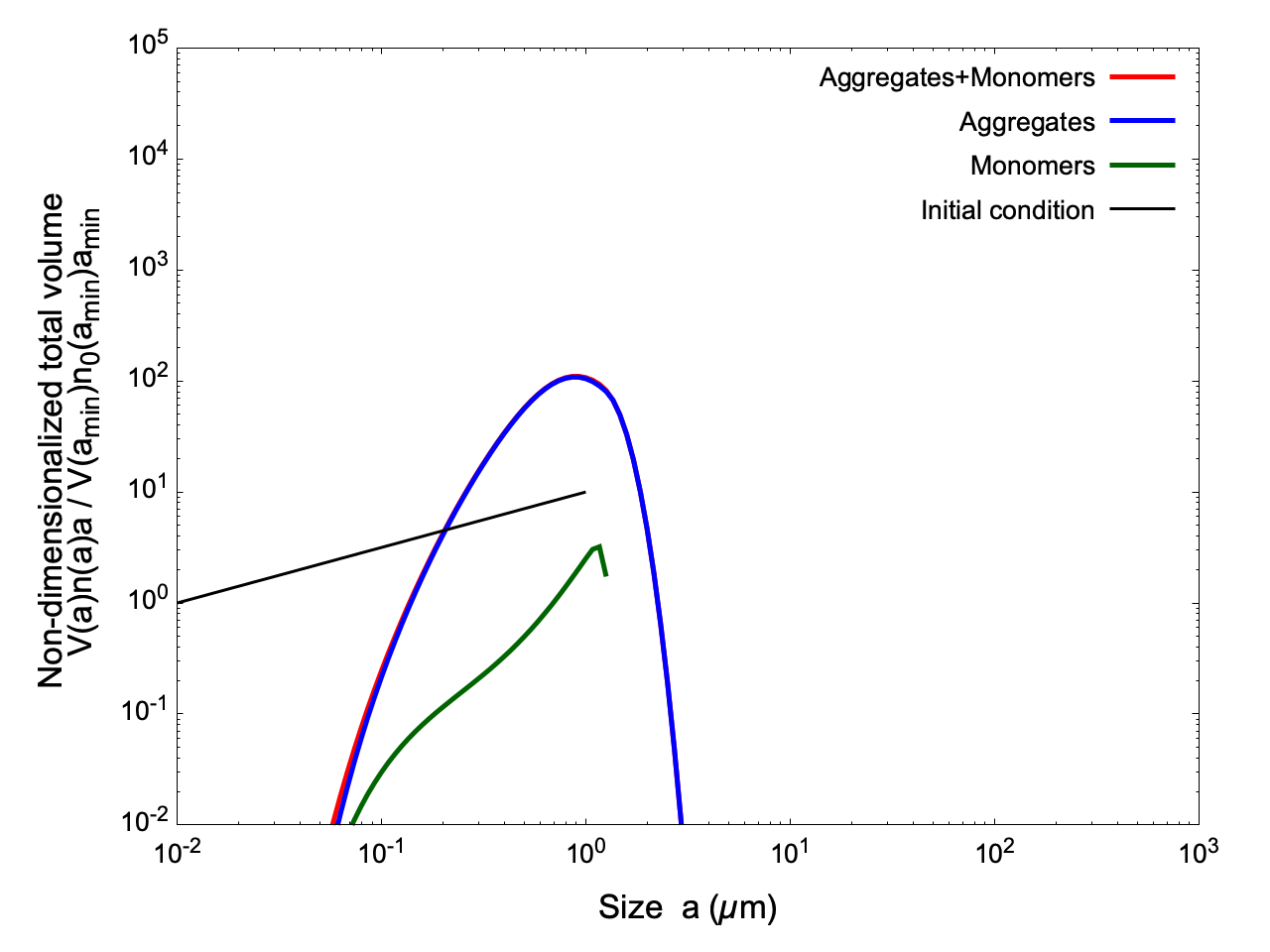}
  \subcaption{Average aggregate size is $0.5 \ \mathrm{\mu m}$}
  \label{volume distribution b 3.5}
\end{minipage}\\ \\
\begin{minipage}[t]{0.5\hsize}
  \centering
  \includegraphics[width=8.5cm,pagebox=cropbox,clip]{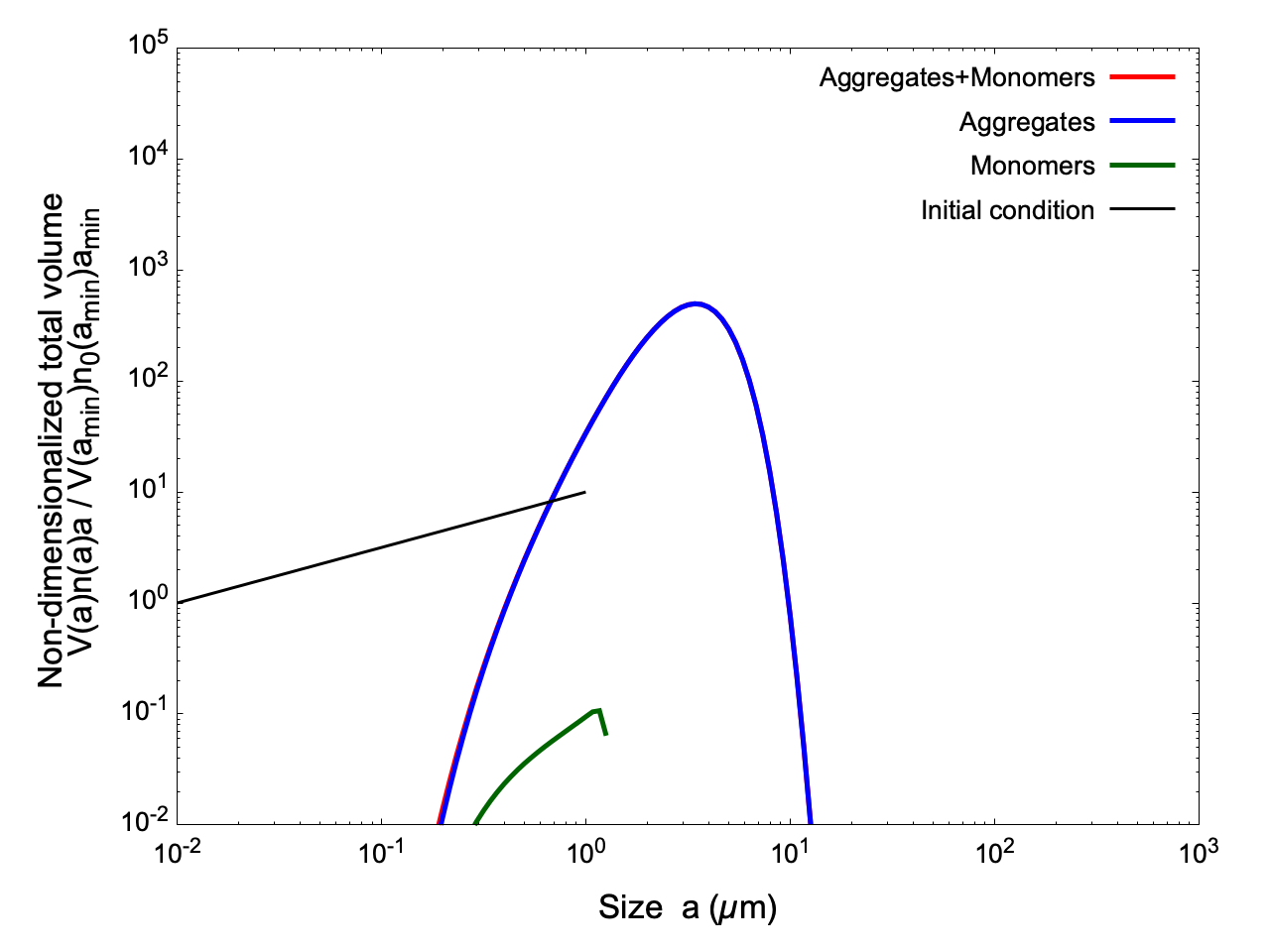}
  \subcaption{Average aggregate size is $2 \ \mathrm{\mu m}$}
  \label{volume distribution c 3.5}
\end{minipage}
\begin{minipage}[t]{0.5\hsize}
  \centering
  \includegraphics[width=8.5cm,pagebox=cropbox,clip]{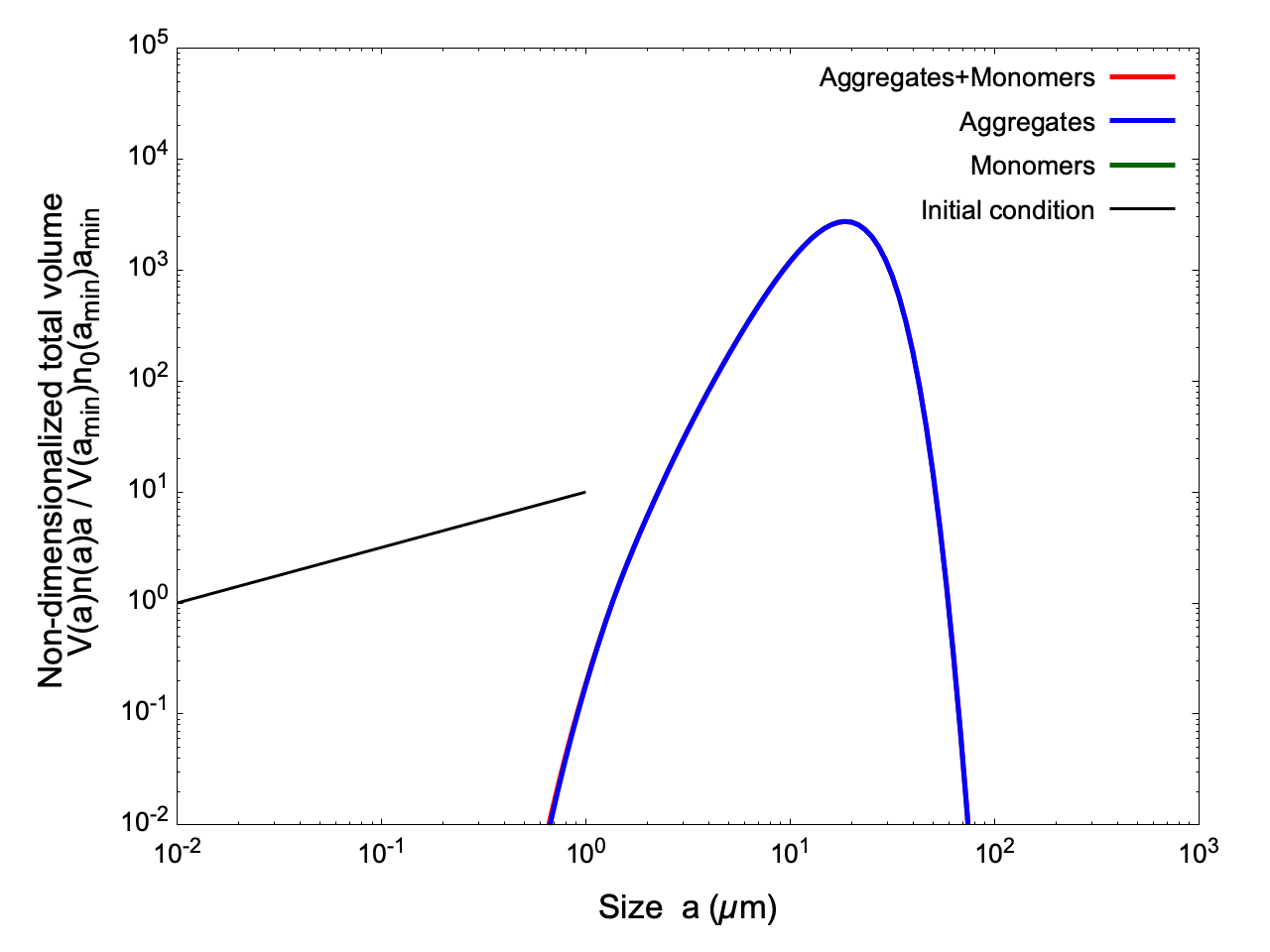}
  \subcaption{Average aggregate size is $10 \ \mathrm{\mu m}$}
  \label{volume distribution d 3.5} 
  \end{minipage}
\end{tabular}
\caption{Evolution of the volume distribution for the case of $\beta=-3.5$ and $(a_{\mathrm{min}},a_{\mathrm{max}})=(0.01 \ \mathrm{\mu m},1 \ \mathrm{\mu m})$. The volume distribution of aggregates (blue), monomer grains (green), and all dust particles; that is, aggregates and monomer grains (red) are shown for each time. The black line indicates the initial distribution of monomer grains. The total volume of dust particles within each size bin, $V(a)n(a)a$, is divided by the initial volume of the smallest monomer grains, $V(a_{\mathrm{min}})n_{0}(a_{\mathrm{min}})a_{\mathrm{min}}$.}
\label{volume distribution 3.5}
\end{figure}

In this subsection, we obtain the conditions for monomer-accretion and aggregate-accretion cases to occur. The initial structures of FGRs differ between these two cases \citep{Xiang.etal2019}.  First, we consider the case of $\beta=-3.5$. Using Eq. \ref{fitting4} and $\epsilon_{\mathrm{rim}}$ of the observed value in meteorites, we can obtain $a_{\mathrm{agg}}(\epsilon_{\mathrm{rim}})$, the average aggregate size at the time when chondrules acquire the mass of FGRs as observed in meteorites. If $a_{\mathrm{max}} \ll a_{\mathrm{agg}}(\epsilon_{\mathrm{rim}})$, i.e., the average aggregate size already exceeds the maximum size of monomer grains when chondrules complete rim formation, and in other words, growth timescale of aggregates to the size of $a_{\mathrm{max}}$ is much shorter than the timescale of rim formation, all the monomer grains turn into aggregates before their accretion as seen in Figures \ref{isolated monomer grains} and \ref{volume distribution 3.5} (aggregate-accretion case). Grains of various sizes are mixed inside aggregates so FGRs composed of these aggregates display uniform and porous structures \citep{Xiang.etal2019}. On the other hand, if $a_{\mathrm{max}} \gg a_{\mathrm{agg}}(\epsilon_{\mathrm{rim}})$, i.e., the larger monomer grains are still isolated (Figures \ref{isolated monomer grains} and \ref{volume distribution 3.5}), FGRs show layering, size-coarsening, and compact structures \citep[monomer-accretion case,][]{Xiang.etal2019}. 

We consider the condition for the monomer-accretion and aggregate-accretion cases when $\beta=-3.5$. From the above consideration and Eq. \ref{fitting4} in the regime of $a_{\mathrm{agg}}/a_{\mathrm{min}}<Y$, the condition for the monomer-accreting case, $a_{\mathrm{max}} \gg a_{\mathrm{agg}}(\epsilon_{\mathrm{rim}})$, can be transformed into 
\begin{align}
a_{\mathrm{max}}/a_{\mathrm{min}} \gg &1.10\left[ \left(\frac{1+X_{cd}}{X_{cd}A_{\mathrm{min}}}\right) \ln \left( 1+\epsilon_{\mathrm{rim}}  \right) \ \right]^{0.794},  \\
(a_{\mathrm{max}}/a_{\mathrm{min}})^{1/0.794} \gg &\frac{1.10^{1/0.794}}{5.63 \times 10^{-4}} \left(\frac{\alpha}{10^{-4}}\right)^{-3/4}\left(\frac{\Sigma_{g}}{680 \ \mathrm{g/cm^{2}}}\right)^{3/4} \notag \\
& \ \ \ \ \ \ \ \ \ \ \ \ \ \ \ \ \left(\frac{a_{\mathrm{min}}}{0.1 \ \mathrm{\mu m}}\right)^{-2.5}\left(\sqrt{\frac{a_{\mathrm{max}}}{a_{\mathrm{min}}}}-1\right)^{-1} \ln \left( 1+\epsilon_{\mathrm{rim}}  \right). 
\end{align}
Then, separating information on the monomer size distribution from that on the others (gas surface density, turbulent intensity, and rim mass ratio), we finally obtain
\begin{equation}
f_{-3.5}(a_{\mathrm{max}}, a_{\mathrm{min}}) \gg Q_{-3.5}(\alpha, \Sigma_{g}, \epsilon_{\mathrm{rim}})
\label{monomer-accretion}
\end{equation}
where 
\begin{align}
f_{-3.5}(a_{\mathrm{max}}, a_{\mathrm{min}}) =& \left(\frac{a_{\mathrm{min}}}{0.1 \ \mathrm{\mu m}}\right)^{2.5}\left(\frac{a_{\mathrm{max}}}{a_{\mathrm{min}}}\right)^{1.26}\left(\sqrt{\frac{a_{\mathrm{max}}}{a_{\mathrm{min}}}}-1\right), \label{f_3.5_eq} \\
Q_{-3.5}(\alpha, \Sigma_{g}, \epsilon_{\mathrm{rim}}) =& 2.00 \times 10^{3}\left(\frac{\alpha}{10^{-4}}\right)^{-3/4} \left(\frac{\Sigma_{g}}{680 \ \mathrm{g/cm^{2}}}\right)^{3/4} \ln \left( 1+\epsilon_{\mathrm{rim}}  \right). \label{Q_3.5_eq}
\end{align}
If Eq. \ref{monomer-accretion} is satisfied, there are isolated large grains when chondrules complete rim formation. Among these grains, the larger grains make the voids in rims, and the smaller grains or perhaps smaller aggregates pass through the voids. Layered internal structures of rims with grain size coarsening are achieved. On the other hand, the condition for the aggregate-accretion case is as follows:
\begin{equation}
f_{-3.5}(a_{\mathrm{max}}, a_{\mathrm{min}}) \ll Q_{-3.5}(\alpha, \Sigma_{g}, \epsilon_{\mathrm{rim}}).
\label{aggregate-accretion}
\end{equation}
If Eq. \ref{aggregate-accretion} is satisfied, porous aggregates with the average filling factor obtained from Eq. \ref{fitting2} accrete onto chondrules. According to the simulation results of \citet{Xiang.etal2019}, uniform distributions of monomer grains in the rims are achieved. We show the values of $f_{-3.5}(a_{\mathrm{max}}, a_{\mathrm{min}})$ in Figure \ref{f-3.5} for fixed $a_{\mathrm{min}}$ and compare them with $Q_{-3.5}(\alpha, \Sigma_{g}, \epsilon_{\mathrm{rim}})$. We assume $\epsilon_{\mathrm{rim}}=0.4$ and $\Sigma_{g}=680 \ \mathrm{g/cm^{2}}$. The rim mass fraction $\epsilon_{\mathrm{rim}}=0.4$ corresponds to the thickness of $0.12$ times the core chondrules radius for compact rims. \citet{Hanna.Ketcham2018} found the slope of $0.11$ for the linear fitting on the core chondrules radius and rim thickness in Murchison meteorite although they also reported that smaller chondrules have relatively thicker rims. If $\alpha$ is larger (faster accretion due to faster turbulent velocity) or $\Sigma_{g}$ is lower (faster turbulent velocity due to weaker coupling), the requisite value of $f$ for the monomer-accretion case becomes lower (the decrease of $\Sigma_{g}$ has the same effect as the increase of $\alpha$ so we fixed $\Sigma_{g}$ in Figure \ref{f-3.5}).

In wide parameter ranges, it might be required that the maximum size of monomer grains should be larger than $1 \ \mathrm{\mu m}$ for the monomer-accretion case. If the turbulent intensity is very weak ($\alpha=10^{-5}$), $a_{\mathrm{max}} \sim 10 \ \mathrm{\mu m}$ is required. We note that there is no or weak dependence of such requirements on the total solid density and the dust/chondrule mass density ratio (we note that the simplified model of dust accretion onto chondrules in a turbulent disk was used). However, these parameters control the timescales of the collisonal growth and the formation of FGRs (Eq. \ref{trim}). If these timescales are short, then, mixing of materials between inside and outside the chondrule forming regions is limited. If chondrules were formed in the regions with significant concentration of solid particles \citep[e.g.][]{Alexander.etal2008}, $t_{\mathrm{rim}}$ in Eq. \ref{trim} would have been considerably shorter than the Kepler timescale. It was reported that FGRs and the matrix have similar chemical compositions, hence, they were originated from the same reservoir \citep[e.g.][]{Brearley1993}. However, FGRs and the matrix exhibit different presolar grain abundance and alteration degree \citep[e.g.][]{Leitner.etal2016, Haenecour.etal2018, Zanetta.etal2021}. FGRs may have been formed soon after or during chondrule forming events \citep[e.g.][]{Liffman2019}. If it is true, we have to consider the formation of FGRs in the context of chondrule formation.  

For other $\beta$ values, we can derive the conditions for the monomer-accretion and aggregate-accretion cases in the form similar to Eqs. \ref{monomer-accretion} and \ref{aggregate-accretion}. When $\beta=-4.5$, using Eq. \ref{fitting4.5acc}, the requirement for the monomer-accretion case is
\begin{align}
a_{\mathrm{max}}/a_{\mathrm{min}} \gg & 0.0739\left[ \left(\frac{1+X_{cd}}{X_{cd}A_{\mathrm{single}}(a_{\mathrm{min}})}\right) \ln \left( 1+\epsilon_{\mathrm{rim}}  \right) \ \right]^{1.06}, \notag \\
(a_{\mathrm{max}}/a_{\mathrm{min}})^{1/1.06} \gg & \frac{0.0739^{1/1.06}}{2.81 \times 10^{-4}} \left(\frac{\alpha}{10^{-4}}\right)^{-3/4}\left(\frac{\Sigma_{g}}{680 \ \mathrm{g/cm^{2}}}\right)^{3/4} \left(\frac{a_{\mathrm{min}}}{0.1 \ \mathrm{\mu m}}\right)^{-2.5} \ln \left( 1+\epsilon_{\mathrm{rim}}  \right).
\end{align}
We can obtain the final form of the requirement for the monomer-accretion case as
\begin{equation}
f_{-4.5}(a_{\mathrm{max}}, a_{\mathrm{min}}) \gg Q_{-4.5}(\alpha, \Sigma_{g}, \epsilon_{\mathrm{rim}}),
\label{monomer-accretion4.5}
\end{equation}
where
\begin{align}
f_{-4.5}(a_{\mathrm{max}}, a_{\mathrm{min}})=& \left(\frac{a_{\mathrm{min}}}{0.1 \ \mathrm{\mu m}}\right)^{2.5} \left(\frac{a_{\mathrm{max}}}{a_{\mathrm{min}}}\right)^{0.943}, \label{f_4.5_eq} \\
Q_{-4.5}(\alpha, \Sigma_{g}, \epsilon_{\mathrm{rim}})=& 3.05 \times 10^{2} \left(\frac{\alpha}{10^{-4}}\right)^{-3/4} \left(\frac{\Sigma_{g}}{680 \ \mathrm{g/cm^{2}}}\right)^{3/4} \ln \left( 1+\epsilon_{\mathrm{rim}}  \right). \label{Q_4.5_eq}
\end{align}
We note again that we consider the cases of $a_{\mathrm{max}}/a_{\mathrm{min}} > 20$. 

When $\beta=-2.5$, we have to be careful in using this argument because we do not have the well-normalized solution during the monomer-aggregation stage. Eq. \ref{fitting2.5acc} only covers the BCCA-like stage. However, what we obtain here are rough estimates (in principle, timescale comparison), and Eq. \ref{fitting2.5acc} works for such a purpose. The timescale or alternatively $\epsilon_{\mathrm{rim}}(t)$ at which $a_{\mathrm{max}}/a_{\mathrm{max}} \sim a_{\mathrm{agg}}(t)/a_{\mathrm{max}}$ from Eq. \ref{fitting2.5acc} is approximately equivalent to the timescale of the collisions between the largest grains and also the end of the monomer-aggregation stage. Then, the requirement for the monomer-accretion case is
\begin{align}
a_{\mathrm{max}}/a_{\mathrm{max}} \gg & 1.56\left[ \left(\frac{1+X_{cd}}{X_{cd}A_{\mathrm{single}}(a_{\mathrm{max}})}\right) \ln \left( 1+\epsilon_{\mathrm{rim}}  \right) \ \right]^{1.04}, \notag \\
(a_{\mathrm{max}}/a_{\mathrm{max}})^{1/1.04} \gg & \frac{1.56^{1/1.04}}{8.90 \times 10^{-2}} \left(\frac{\alpha}{10^{-4}}\right)^{-3/4}\left(\frac{\Sigma_{g}}{680 \ \mathrm{g/cm^{2}}}\right)^{3/4} \notag \\
& \ \ \ \ \ \ \ \ \ \ \ \ \ \ \ \ \ \ \ \ \ \ \ \ \ \ \ \ \ \ \ \ \ \left(\frac{a_{\mathrm{max}}}{1.0 \ \mathrm{\mu m}}\right)^{-2.5} \ln \left( 1+\epsilon_{\mathrm{rim}}  \right).
\end{align}
The final form of the requirement for the monomer-accretion case can be obtained as 
\begin{equation}
f_{-2.5}(a_{\mathrm{max}}, a_{\mathrm{min}}) \gg Q_{-2.5}(\alpha, \Sigma_{g}, \epsilon_{\mathrm{rim}}),
\label{monomer-accretion2.5}
\end{equation}
where
\begin{align}
f_{-2.5}(a_{\mathrm{max}}, a_{\mathrm{min}})=& \left(\frac{a_{\mathrm{max}}}{1.0 \ \mathrm{\mu m}}\right)^{2.5}, \label{f_2.5_eq} \\
Q_{-2.5}(\alpha, \Sigma_{g}, \epsilon_{\mathrm{rim}})=& 1.72\times10^{1} \left(\frac{\alpha}{10^{-4}}\right)^{-3/4}\left(\frac{\Sigma_{g}}{680 \ \mathrm{g/cm^{2}}}\right)^{3/4} \ln \left( 1+\epsilon_{\mathrm{rim}}  \right). \label{Q_2.5_eq}
\end{align}
There is no dependency on $a_{\mathrm{min}}$ in $f_{-2.5}$ because grain populations and dust aggregation are dominated and controlled by the largest grains, and we compared the collisional timescale between two largest grains and the timescale of rim formation.

We show $f_{-4.5}(a_{\mathrm{max}}, a_{\mathrm{min}})$ and $f_{-2.5}(a_{\mathrm{max}}, a_{\mathrm{min}})$ in Figure \ref{f-4.5,2.5}. Even if $\beta=-2.5$, the condition of $a_{\mathrm{max}}>1 \ \mathrm{\mu m}$ is required for the monomer-accretion case to occur. We plot the threshold values of $a_{\mathrm{max}}$, when $a_{\mathrm{min}}=0.1 \ \mathrm{\mu m}$ in Figure \ref{amax-beta}. The maximum monomer size, $a_{\mathrm{max}}$, must be larger than these thresholds for the monomer-accretion case.

We looked at wide size range cases ($a_{\mathrm{max}}/a_{\mathrm{min}}>20$) above. Here, we roughly examine narrower (or general) size range cases using the single size case. This may be helpful for the reconsideration of the previous discussions for the wide size range cases. If we assume $\epsilon_{\mathrm{rim}}=0.4$, the formation timescale of rims from Eq. \ref{trim} is $177 \ \mathrm{yr}$ when $X_{cd}=1$ and other parameters are the same as those in the denominators in Eq. \ref{trim}. On the other hand, the collisional time between single size monomer grains is $6.55 \ \mathrm{yr}, 37.1 \ \mathrm{yr}$, and $210 \ \mathrm{yr}$ for the size of $a=0.5 \ \mathrm{\mu m}, 1 \ \mathrm{\mu m}$, and $2 \ \mathrm{\mu m}$, respectively, from Eq. \ref{def_t_single}. Monomer grains smaller than $1 \ \mathrm{\mu m}$ quickly coagulate into larger aggregates before accretion onto chondrules in the cases of narrow size ranges as well. When $\beta=-2.5$, although the largest grains dominates the grains population, their growth can be assisted by the porous aggregation of the smaller grains, so the threshold value of the maximum size of monomer grains larger than $1 \ \mathrm{\mu m}$ is required in Figure \ref{amax-beta}. Moreover, grain size distributions of $\beta=-3.5$ and $-4.5$ have more mass fraction in the smaller grains so larger size of $a_{\mathrm{max}}$ is required.

In the model of \citet{Xiang.etal2019}, $\beta=-3.5$ was considered, and the larger grains had larger total volume compared to the smaller grains. These larger grains make the voids the smaller grains can pass through. However, if we consider simultaneous dust aggregation, the small aggregates already have several orders of magnitude larger volume compared to the large isolated monomer grains even during the monomer-aggregation stage (Figures \ref{volume distribution 3.5} and \ref{volume distribution 4.5}), and might control the overall structures of the rims when $\beta \leq -3.5$. This might diminish the effect of the larger grains, so we infer that $-3.5<\beta$ is necessary for grain size coarsening. We show our prediction in Figure \ref{prediction}.

If FGRs are mainly composed of $\mathrm{nm}$-size grains, the aggregate-accreting rim formation occurs unless the turbulent intensity is very high ($\alpha>10^{-3}$) or gas surface density is very low (due to late stage of the nebular evolution or large heliocentric distance). We reconsider what observations indicate for grain size coarsening and layered structures of rims. \citet{Metzler.etal1992} studied fourteen CM chondrites and confirmed that multi-layered structures of rims are common in this chondrite group. Most of these layered rims show the chemical difference between inner and outer layers that Fe/Mg ratio increases from inner to outer layers (but, we note that there are rare exceptions of rims which show the opposite trends in chemical compositions). \citet{Brearley.etal1999a} also reported that inner Mg-rich layers in Murchison meteorite are finer grained than outer Fe-rich layers. Of course, it is possible that the grain sizes of dust materials were linked to their chemistry, and the monomer-accreting rim formation processes \citep{Xiang.etal2019} separated the grains with different chemical compositions into the inner and outer layers. But another possibility is that just the chemical composition of dust materials in the neighbor of chondrules changed with time. \citet{Zega.Buseck2003a} suggested episodic accretion of dust onto chondrules. We have to seek the both chemically and mechanically consistent rim formation histories, presumably from chondrule formation events if dust to gas ratio is much higher than the typical solar abundance, in the future work.   
 
During the monomer-aggregation stage, the average filling factor of aggregates is found to be larger than $0.1$ for various size ranges if $\beta \gtrsim -3.5$ (Figures \ref{evolution of filling factor 3.5} and \ref{filling factor 2.5b}). FGRs formed in this stage might have a filling factor larger than $\sim 0.1$. However, if the dust growth enters the BCCA-like stage, the filling factor of aggregates decreases rapidly. When $\beta \lesssim -4.5$, the smallest grains already enter BCCA growth during the monomer-aggregation stage so the filling factor of aggregates decreases rapidly even during the monomer-aggregation stage (Figure \ref{filling factor 4.5b}). The filling factor of accreting FGRs affects the rim formation processes and the subsequent growth of chondrules, rims and matrix as discussed in subsection \ref{Setting of the model}. Observational data of the grains size distributions in unaltered FGRs along with their maximum and minimum size will provide important insights on the environment of FGRs and chondrule formation and the following growth of chondrules and fine dust grains.
\color{black}

\section{Conclusion}
In this study, we have investigated the aggregation of polydisperse monomer grains \color{red}by the Brownian motion\color{black}. To solve porosity change, we used the empirical formula given by \citet{Okuzumi.etal2009} based on $N$-body simulations, assuming that the volume of voids arising from the hit-and-stick growth is independent of the size of constituent monomer grains inside the aggregates. Two stages exist in the evolution of aggregates, the monomer-aggregation stage and the BCCA-like stage. The first monomer-aggregation stage lasts until the average aggregate size reaches the maximum size of the monomer grains. In the subsequent BCCA-like stage, the average size of aggregates increases proportionally to the elapsed time, and the average filling factor of aggregates decreases inversely proportionally to the average aggregate size.
We obtained the fitting formulae (Eqs. \ref{fitting1}, \ref{fitting2}), which allow us to calculate the evolution of aggregates taking the accretion of dust particles onto chondrules into account for the grain size distribution of $\beta=-3.5$, which corresponds to the size distribution of ISM \citep{Mathis.etal1977}. These formulae can be scaled to other nebular environments through $A_{\mathrm{min}}$ (defined in Eq. \ref{A}).
We also conducted numerical simulations with $\beta=-4.5$ and $-2.5$. When $\beta=-4.5$, the differences between the monomer-aggregation stage and the BCCA-like stage are less clear compared to when $\beta=-3.5$. The growth of aggregates can be approximated by the minimum size of the monomer grains, but larger grains still affect the evolution. When $\beta=-2.5$, the aggregate growth can be scaled by the maximum size instead of the minimum size.
According to \citet{Xiang.etal2019}, the initial internal structures of FGRs are different in the monomer-accretion case and the aggregate-accretion case. \color{red}The conditions where the monomer-accretion case occurs were derived from Eqs. \ref{fitting4}, \ref{fitting4.5acc}, and \ref{fitting2.5acc}, and expressed as Eqs. \ref{monomer-accretion}, \ref{monomer-accretion4.5}, and \ref{monomer-accretion2.5}, which can be used to infer the environment of FGRs and chondrules formation\color{black}. In most of scenarios, the maximum size of the power law based grain size distributions of larger than $1 \ \mathrm{\mu m}$ is required for the monomer-accretion case to occur. As for the slope of the power law grain size distributions, $\beta>-3.5$ might be necessary for effective grain size coarsening. Detail observational data of the grain size distributions in unaltered FGRs with their maximum and minimum sizes might provide valuable information to infer the formation environment of FGRs and chondrules, along with the coevolution of chondrules and dust grains.

\section{Acknowledgments}
\color{red}We thank H. Miura and S. Okuzumi for useful comments. S.A. is supported by JSPS KAKENHI Grant No. JP20J00598 and T.N. is supported by JSPS KAKENHI Grant No. JP18K03721.\color{black}

\clearpage

\section*{Figures}

\clearpage

\clearpage

\clearpage

\begin{figure}[H]
\begin{tabular}{cc}
\begin{minipage}[t]{0.5\hsize}
  \centering
  \includegraphics[width=8.5cm,pagebox=cropbox,clip]{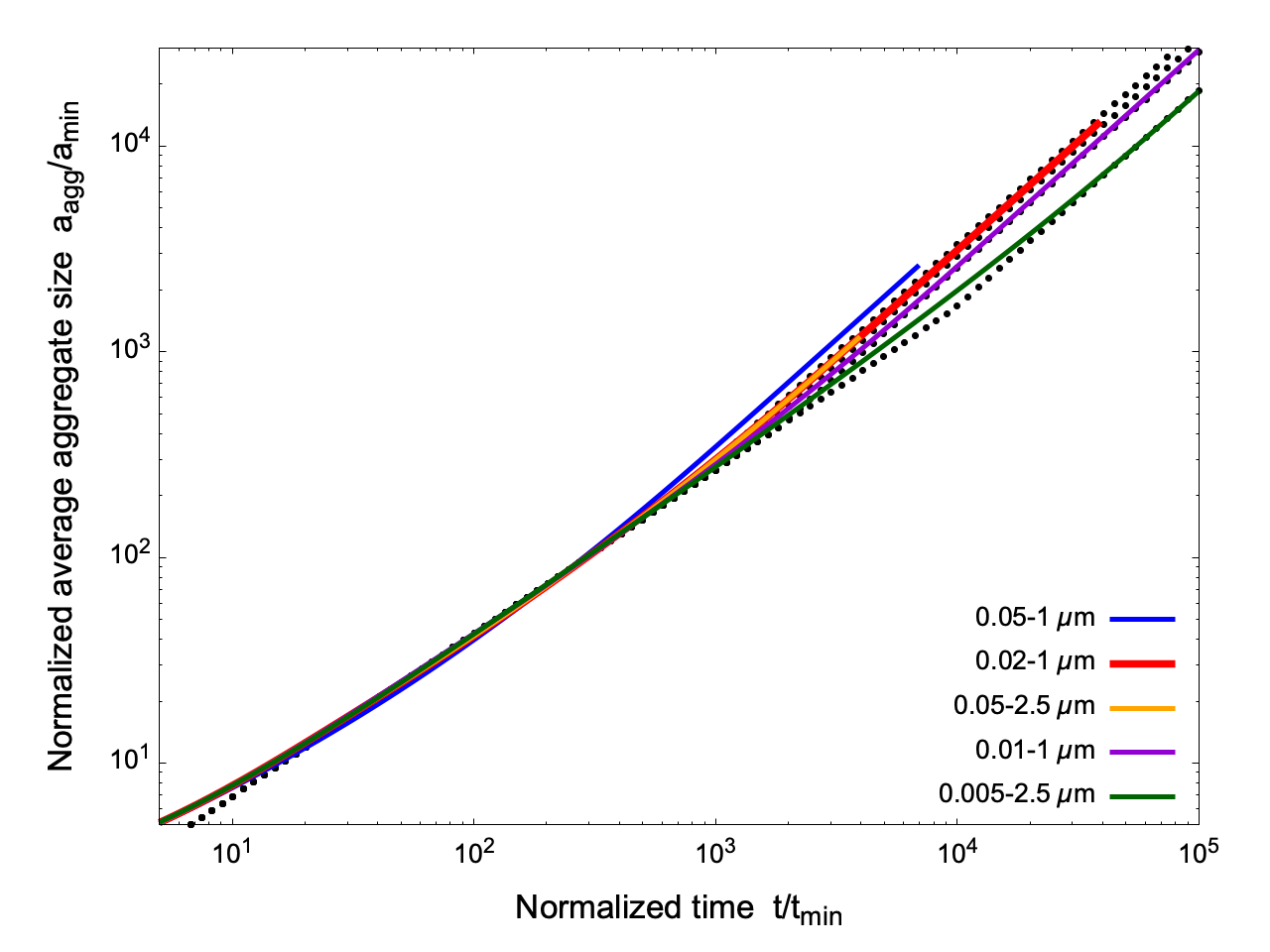}
  \subcaption{Average size of aggregates $a_{\mathrm{agg}}$}
  \label{evolution of aggregate size 3.5}
\end{minipage}
\begin{minipage}[t]{0.5\hsize}
  \centering
  \includegraphics[width=8.5cm,pagebox=cropbox,clip]{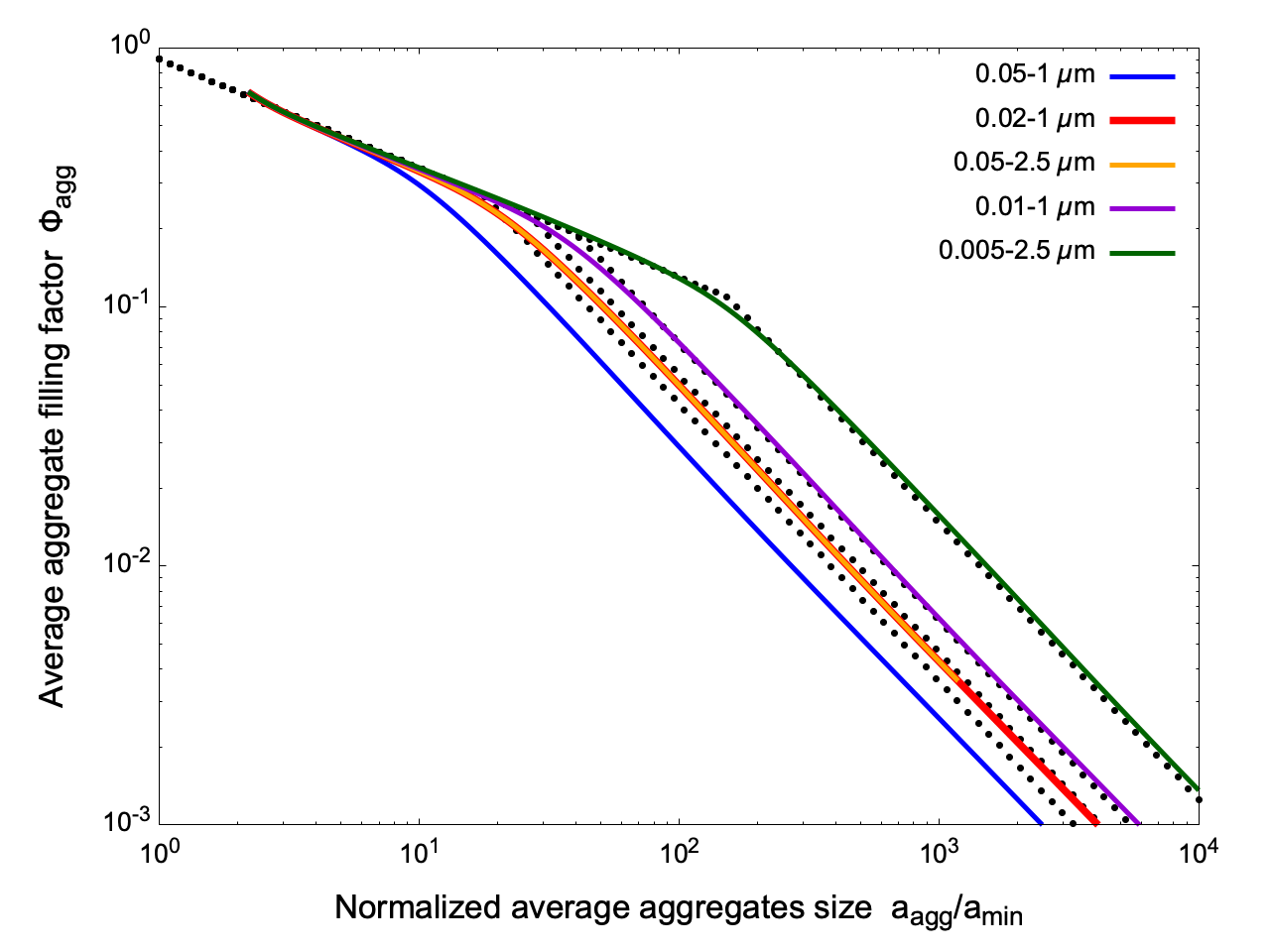}
  \subcaption{Average filling factor of aggregates $\phi_{\mathrm{agg}}$}
  \label{evolution of filling factor 3.5}
\end{minipage}
\end{tabular}
\caption{Evolution of aggregates for the cases of $\beta=-3.5$. (a) Evolution of the average size of the aggregates. Different colors represent different size distributions of the initial monomer grains; $(a_{\mathrm{min}},a_{\mathrm{max}})=(0.05 \ \mathrm{\mu m}, 1 \ \mathrm{\mu m})$ (blue), $(0.02 \ \mathrm{\mu m}, 1 \ \mathrm{\mu m})$ (red), $(0.05 \ \mathrm{\mu m}, 2.5 \ \mathrm{\mu m})$ (yellow), $(0.01 \ \mathrm{\mu m}, 1 \ \mathrm{\mu m})$ (violet), and $(0.005 \ \mathrm{\mu m}, 2.5 \ \mathrm{\mu m})$ (green). The horizontal axis shows the normalized time, which is defined as the elapsed time divided by $t_{\mathrm{min}}$; $1.44\times10^{-1}, \ 2.55\times10^{-2}, \ 2.52\times10^{-1}, \ 6.67\times10^{-3}, \ 2.80\times10^{-3} \mathrm{yr}$ for $(a_{\mathrm{min}},a_{\mathrm{max}})=(0.05 \ \mathrm{\mu m}, 1 \ \mathrm{\mu m})$, $(0.02 \ \mathrm{\mu m}, 1 \ \mathrm{\mu m})$, $(0.05 \ \mathrm{\mu m}, 2.5 \ \mathrm{\mu m})$, $(0.01 \ \mathrm{\mu m}, 1 \ \mathrm{\mu m})$, and $(0.005 \ \mathrm{\mu m}, 2.5 \ \mathrm{\mu m})$, respectively. The vertical axis shows the normalized size which is defined as the average aggregate size ($a_{\mathrm{agg}}$) devided by the minimum grain size ($a_{\mathrm{min}}$). The black dotted lines indicate the fitting formulae (Eq. \ref{fitting1}). (b)  Evolution of the average fiiling factor of aggregates. The black dotted lines show the fitting formulae (Eq. \ref{fitting2}). }
\label{evolution of aggregates 3.5}
\end{figure}

\clearpage

\begin{figure}[H]
\begin{tabular}{cc}
\begin{minipage}[t]{0.5\hsize}
  \centering
  \includegraphics[width=8.5cm,pagebox=cropbox,clip]{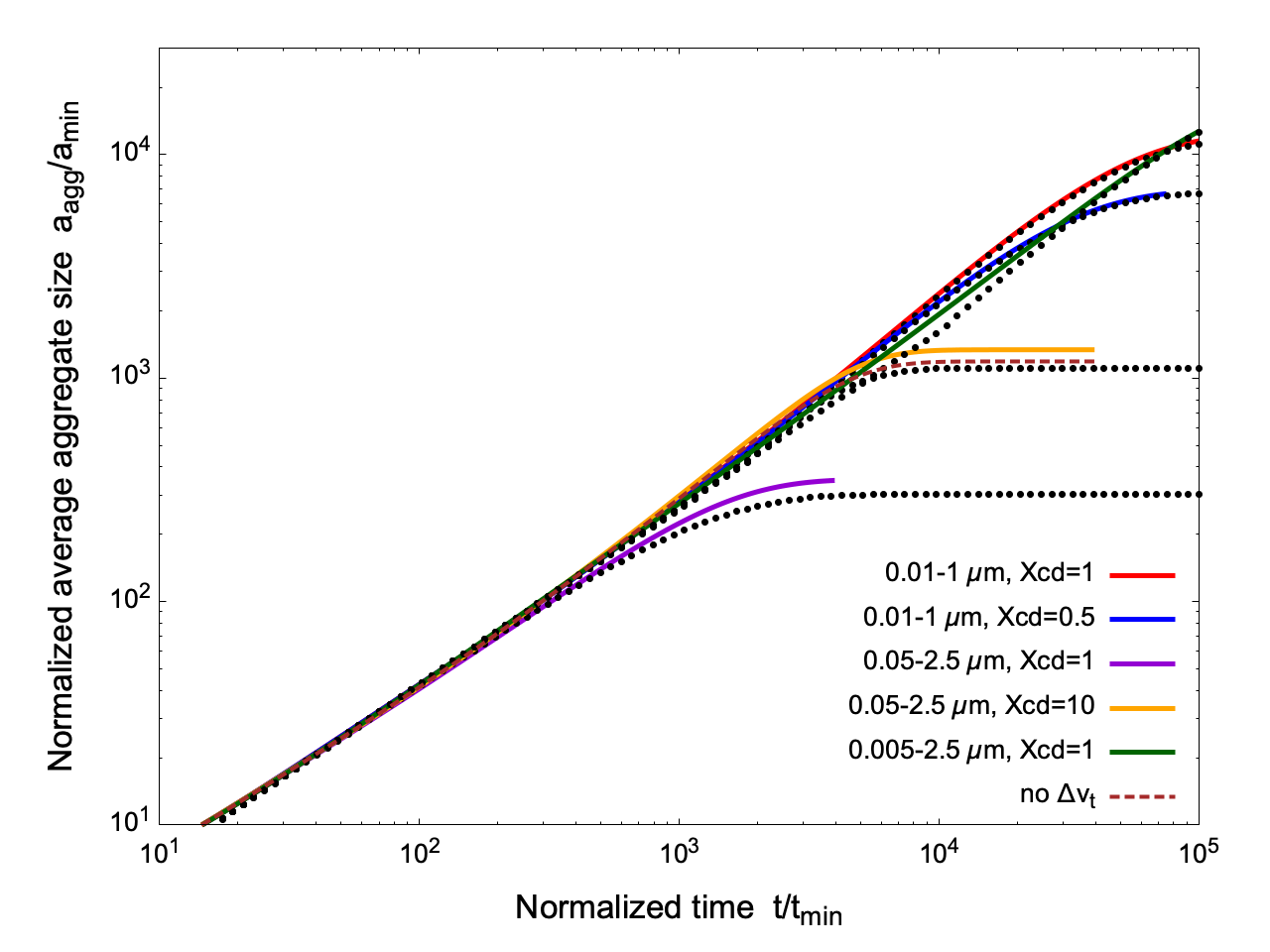}
  \subcaption{Average size of aggregates $a_{\mathrm{agg}}$}
  \label{Including size}
\end{minipage}
\begin{minipage}[t]{0.5\hsize}
  \centering
  \includegraphics[width=8.5cm,pagebox=cropbox,clip]{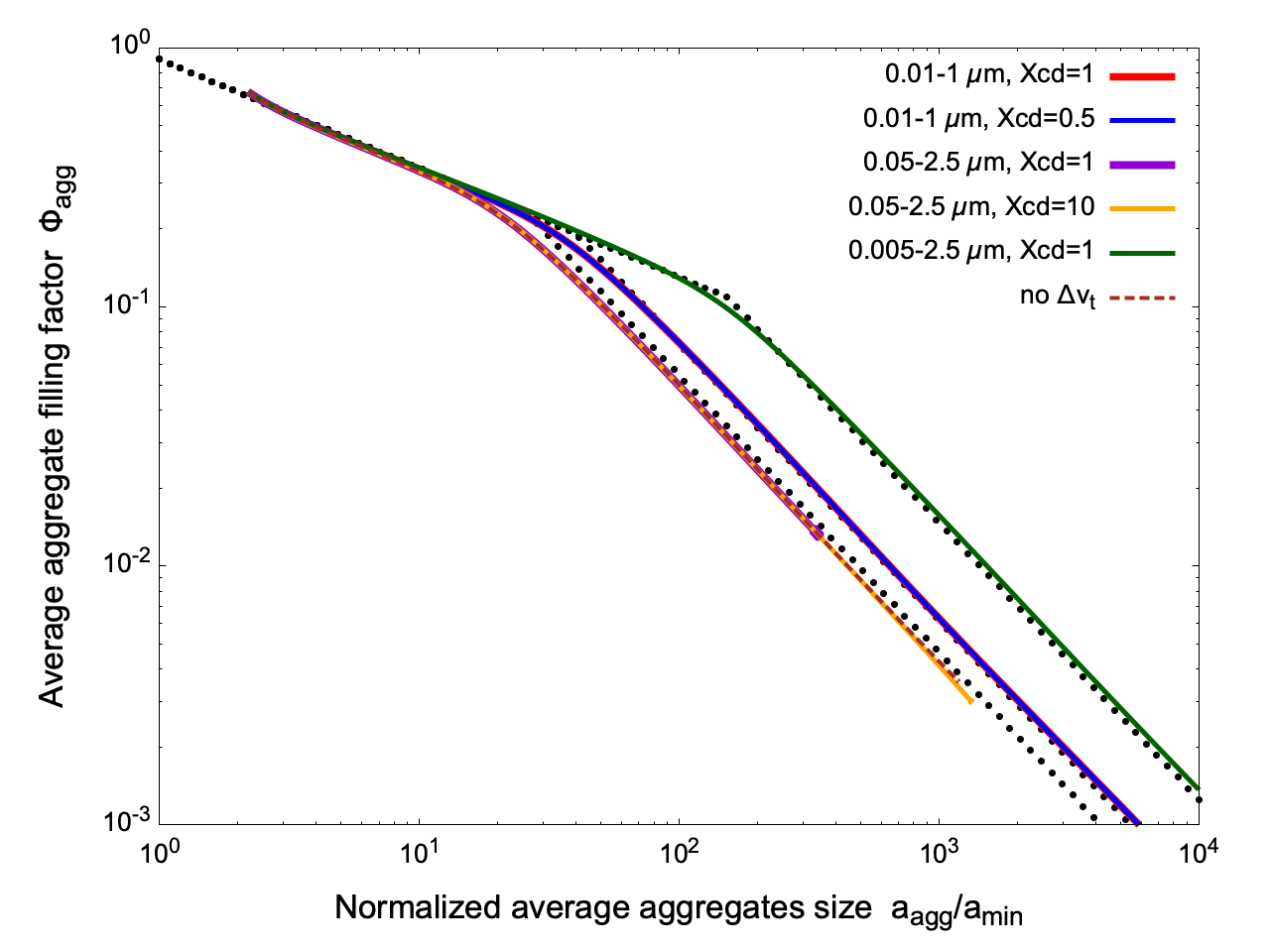}
  \subcaption{Average filling factor of aggregates $\phi_{\mathrm{agg}}$}
  \label{Including filling factor}
\end{minipage}
\end{tabular}
  \caption{Evolution of (a) aggregate size \color{red} and (b) filling factor with dust aggregation by turbulent motion and their accretion onto chondrules for $\beta=-3.5$ \color{black}. Different colors represent the following different initial conditions: $(a_{\mathrm{min}},a_{\mathrm{max}})=(0.01 \ \mathrm{\mu m}, 1 \ \mathrm{\mu m})$ and $\rho_{d,0}=\rho_{c,0}$ (red); $(a_{\mathrm{min}},a_{\mathrm{max}})=(0.01 \ \mathrm{\mu m}, 1 \ \mathrm{\mu m})$ and $\rho_{d,0}=0.5\rho_{c,0}$ (blue); $(a_{\mathrm{min}},a_{\mathrm{max}})=(0.05 \ \mathrm{\mu m}, 2.5 \ \mathrm{\mu m})$ and $\rho_{d,0}=\rho_{c,0}$ (violet); $(a_{\mathrm{min}},a_{\mathrm{max}})=(0.05 \ \mathrm{\mu m}, 2.5 \ \mathrm{\mu m})$ and $\rho_{d,0}=10\rho_{c,0}$ (yellow); and $(a_{\mathrm{min}},a_{\mathrm{max}})=(0.005 \ \mathrm{\mu m}, 2.5 \ \mathrm{\mu m})$ and $\rho_{d,0}=\rho_{c,0}$ (green). \color{red} The dashed brown lines represent the results for the initial condition of $(a_{\mathrm{min}},a_{\mathrm{max}})=(0.05 \ \mathrm{\mu m}, 2.5 \ \mathrm{\mu m})$ and $\rho_{d,0}=10\rho_{c,0}$ without dust aggregation by turbulent motion. We fixed the chondrule density, $\rho_{c}=0.005\rho_{g}$. \color{black} (a) The black dotted lines indicate Eq. \ref{fitting3}. (b) The black dotted lines indicate Eq. \ref{fitting2}.}
  \label{Including}
\end{figure}

\clearpage

\begin{figure}[H]
\begin{tabular}{cc}
\begin{minipage}[t]{0.5\hsize}
  \centering
  \includegraphics[width=8.5cm,pagebox=cropbox,clip]{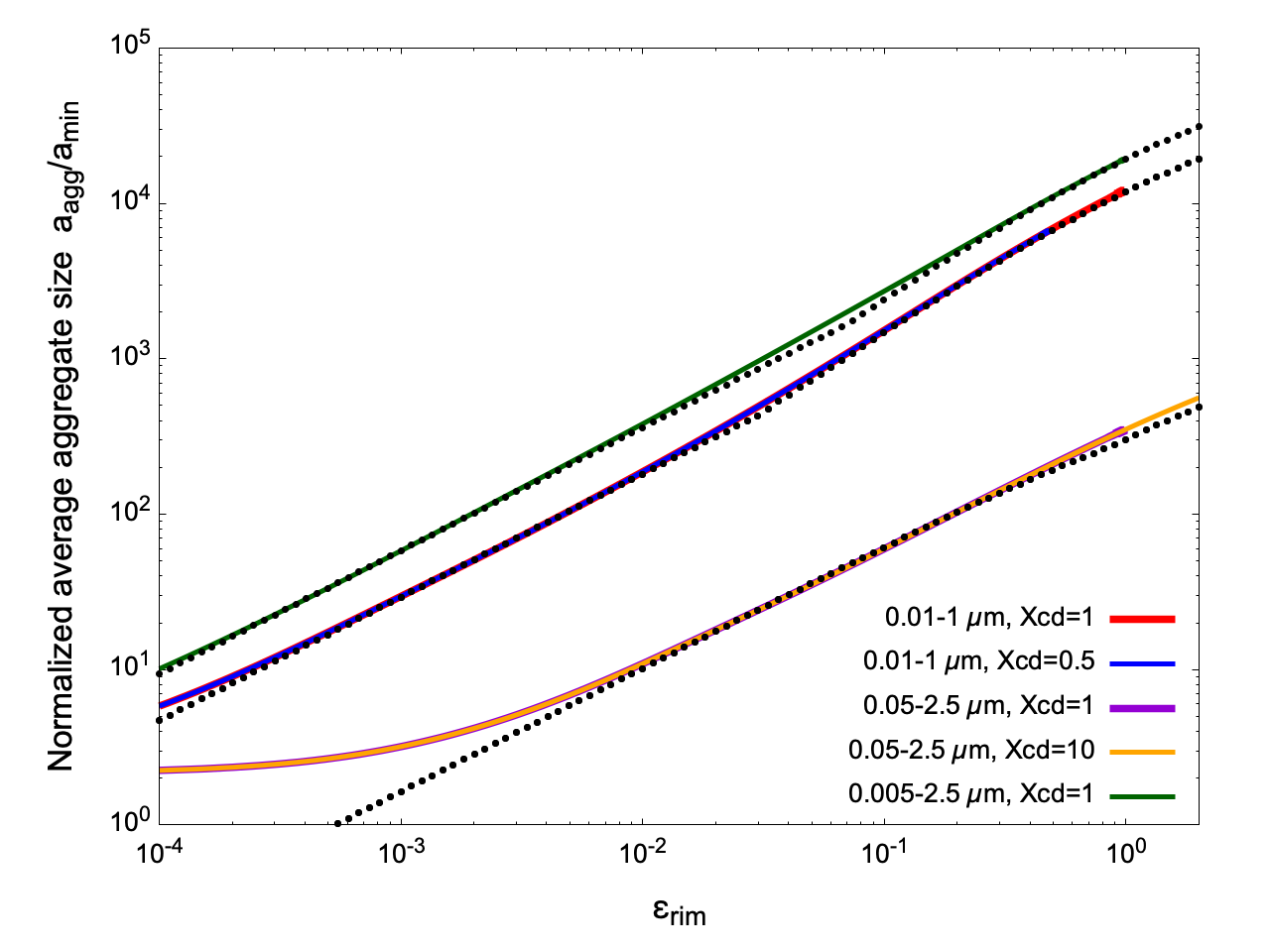}
  \subcaption{Average size of aggregates $a_{\mathrm{agg}}$}
  \label{e-rim size}
\end{minipage}
\begin{minipage}[t]{0.5\hsize}
  \centering
  \includegraphics[width=8.5cm,pagebox=cropbox,clip]{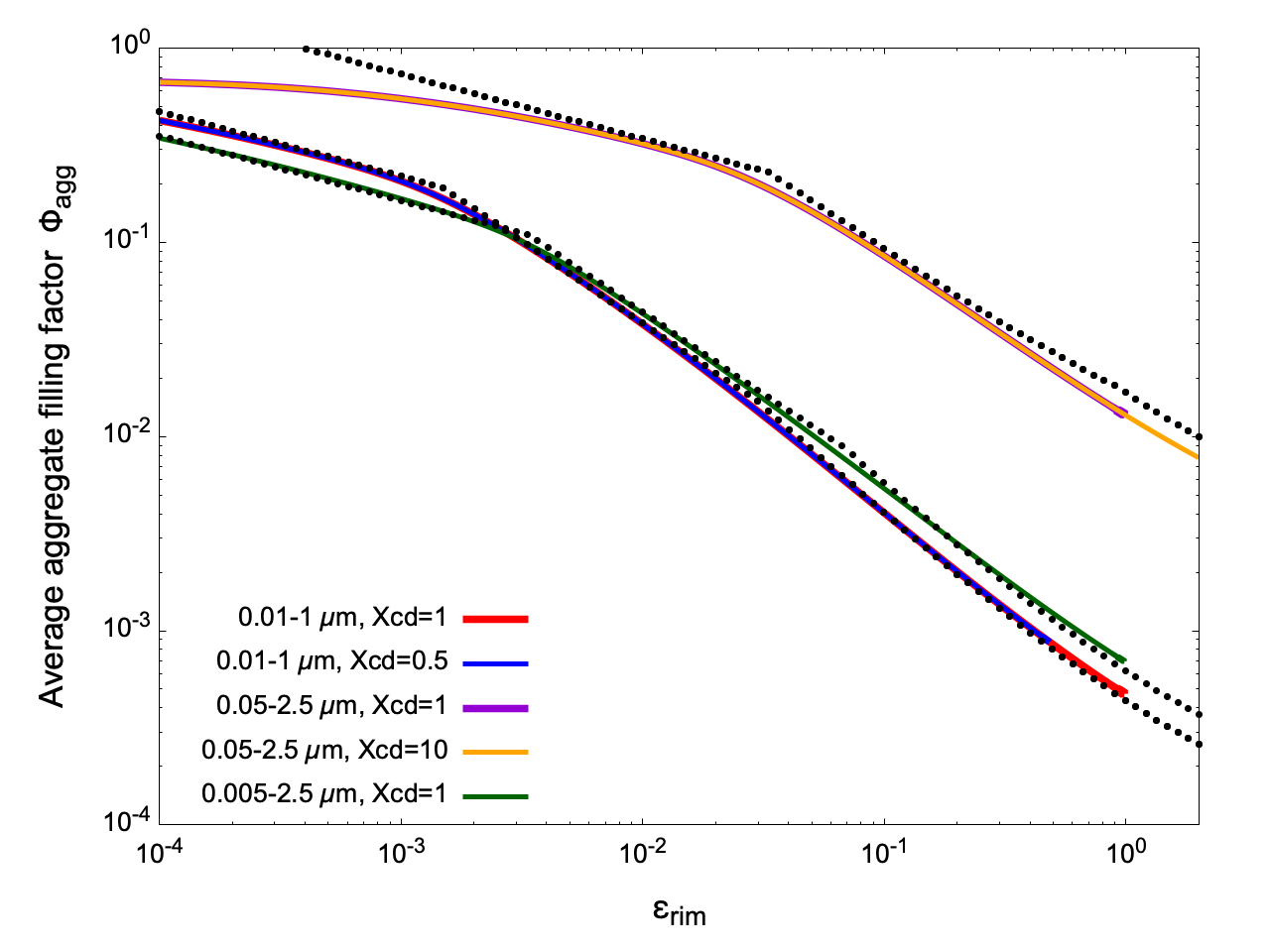}
  \subcaption{Average filling factor of aggregates $\phi_{\mathrm{agg}}$}
  \label{e-rim filling factor}
\end{minipage}
\end{tabular}
  \caption{(a) average size \color{red} and (b) average filling factor of aggregate against the rim to chondrule mass ratio $\epsilon_{\mathrm{rim}}$ for $\beta=-3.5$. The colors of the lines represent the same initial conditions as those in Figure \ref{Including}. The blue line for $(a_{\mathrm{min}},a_{\mathrm{max}})=(0.05 \ \mathrm{\mu m}, 2.5 \ \mathrm{\mu m})$ and $\rho_{d,0}=\rho_{c,0}$ completely overlaps the violet line for $(a_{\mathrm{min}},a_{\mathrm{max}})=(0.05 \ \mathrm{\mu m}, 2.5 \ \mathrm{\mu m})$ and $\rho_{d,0}=10\rho_{c,0}$. (a) The black dotted lines indicate Eq. \ref{fitting4}. (b) The black dotted lines are obtained by substituting Eq. \ref{fitting4} into Eq. \ref{fitting2}. \color{black} }
  \label{e-rim}
\end{figure}

\clearpage

\begin{figure}[H]
\begin{tabular}{cc}
\begin{minipage}[t]{0.5\hsize}
  \centering
  \includegraphics[width=8.5cm,pagebox=cropbox,clip]{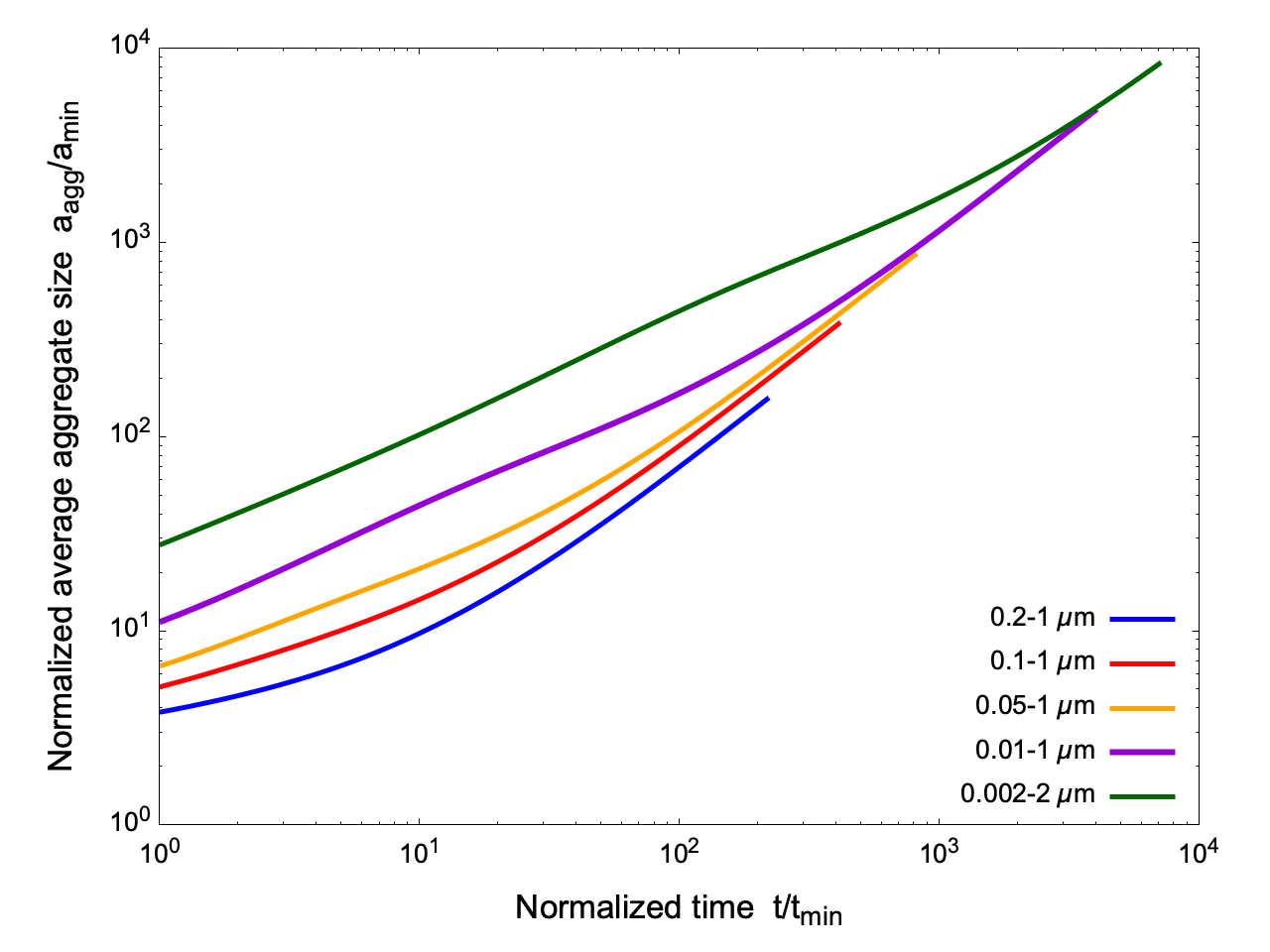}
  \subcaption{Evolution of the average aggregate size}
  \label{aggregate size 2.5a}
\end{minipage}
\begin{minipage}[t]{0.5\hsize}
  \centering
  \includegraphics[width=8.5cm,pagebox=cropbox,clip]{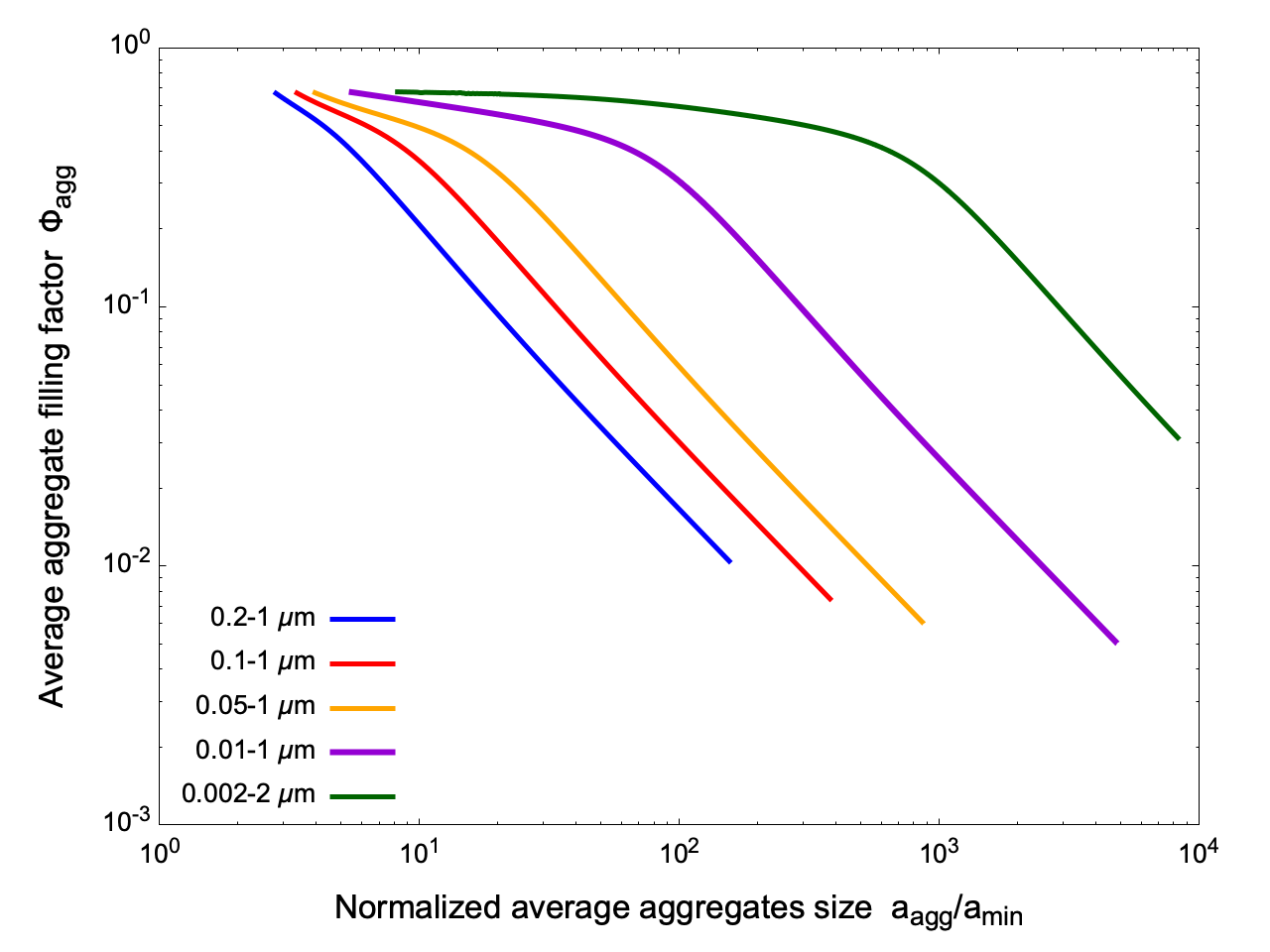}
  \subcaption{Evolution of the average filling factor}
  \label{filling factor 2.5a}
\end{minipage}
\end{tabular}
\caption{Evolution of aggregates for the case of $\beta=-2.5$ normalized by $a_{\mathrm{min}}$ and $t_{\mathrm{min}}$. Different colors represent different size distributions of the initial monomer grains; $(a_{\mathrm{min}},a_{\mathrm{max}})=(0.2 \ \mathrm{\mu m}, 1 \ \mathrm{\mu m})$ (blue), $(0.1 \ \mathrm{\mu m}, 1 \ \mathrm{\mu m})$ (red), $(0.05 \ \mathrm{\mu m}, 1 \ \mathrm{\mu m})$ (yellow), $(0.01 \ \mathrm{\mu m}, 1 \ \mathrm{\mu m})$ (violet), and $(0.002 \ \mathrm{\mu m}, 2 \ \mathrm{\mu m})$ (green).}
\label{evolution of aggregates 2.5a}
\end{figure}

\clearpage

\begin{figure}[H]
\begin{tabular}{cc}
\begin{minipage}[t]{0.5\hsize}
  \centering
  \includegraphics[width=8.5cm,pagebox=cropbox,clip]{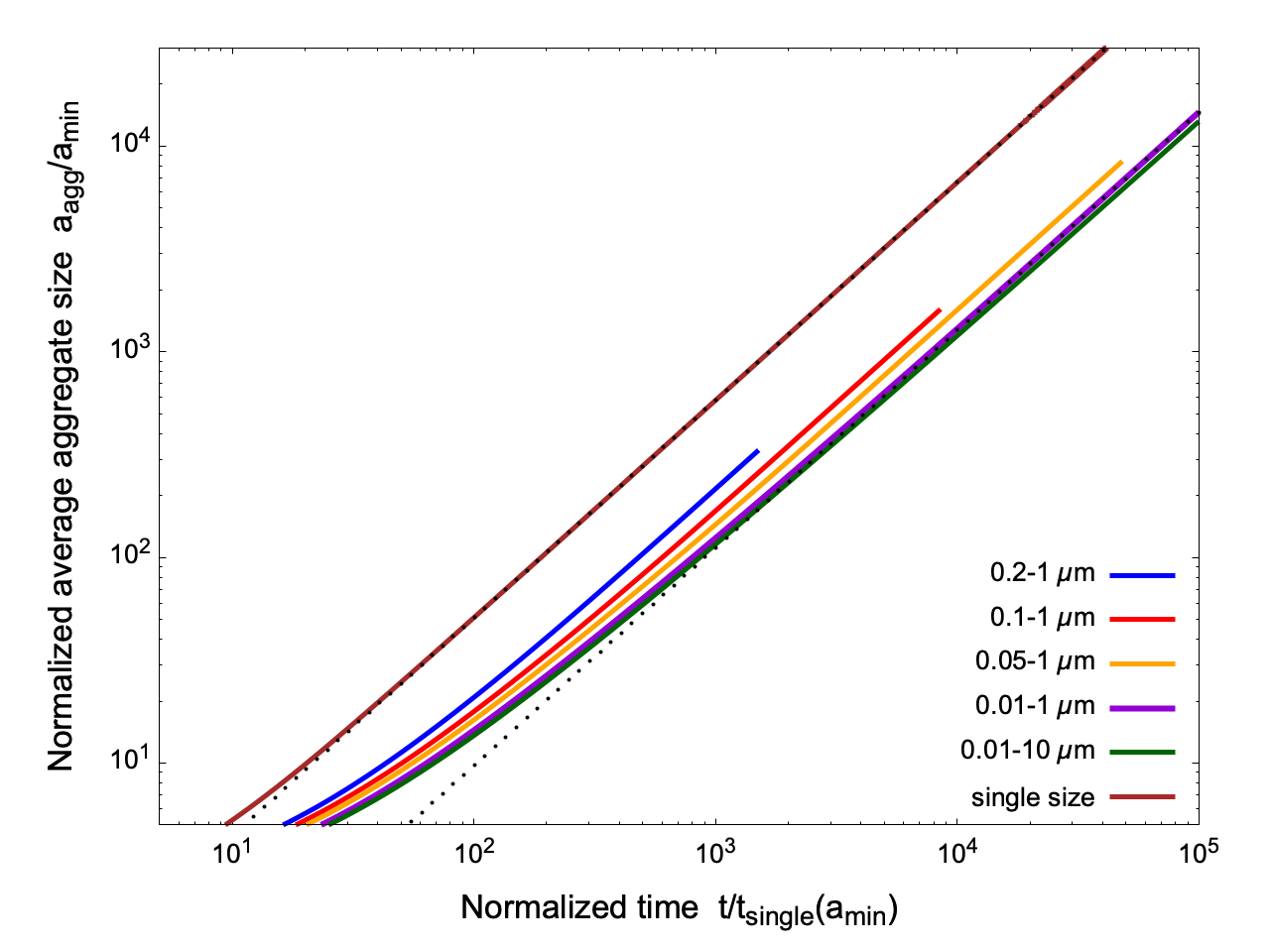}
  \subcaption{Evolution of the average aggregate size}
  \label{aggregate size 4.5b}
\end{minipage}
\begin{minipage}[t]{0.5\hsize}
  \centering
  \includegraphics[width=8.5cm,pagebox=cropbox,clip]{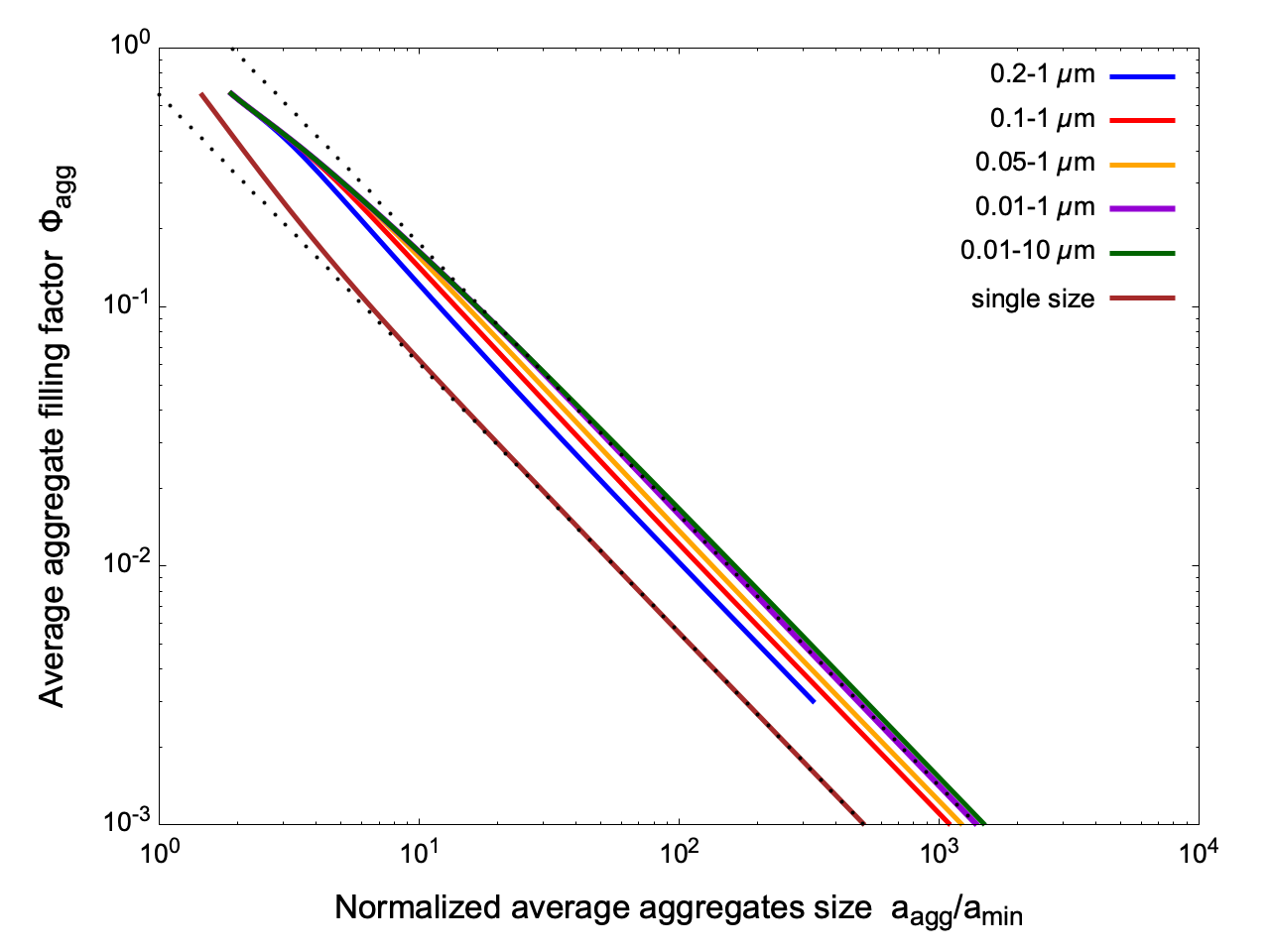}
  \subcaption{Evolution of the average filling factor}
  \label{filling factor 4.5b}
\end{minipage}
\end{tabular}
\caption{ \color{red} Evolution of aggregates for the case of $\beta=-4.5$ normalized by $a_{\mathrm{min}}$ and $t_{\mathrm{single}}(a_{\mathrm{min}})$. Different colors represent different size distributions of the initial monomer grains; $(a_{\mathrm{min}},a_{\mathrm{max}})=(0.2 \ \mathrm{\mu m}, 1 \ \mathrm{\mu m})$ (blue), $(0.1 \ \mathrm{\mu m}, 1 \ \mathrm{\mu m})$ (red), $(0.05 \ \mathrm{\mu m}, 1 \ \mathrm{\mu m})$ (yellow), $(0.01 \ \mathrm{\mu m}, 1 \ \mathrm{\mu m})$ (violet), and $(0.01 \ \mathrm{\mu m}, 10 \ \mathrm{\mu m})$ (green). The black dotted lines indicate the fitting formulae for single size distribution, Eqs. \ref{fit-a-single} and \ref{fit-phi-single}, and for $(a_{\mathrm{min}},a_{\mathrm{max}})=(0.01 \ \mathrm{\mu m},1 \ \mathrm{\mu m})$, Eqs. \ref{fit-a-4.5} and \ref{fit-phi-4.5}. Eqs. \ref{fit-a-single} and \ref{fit-phi-single} are obtained from the results during during $100a_{\mathrm{single}}<a_{\mathrm{agg}}<10000a_{\mathrm{single}}$. Eqs. \ref{fit-a-4.5} and \ref{fit-phi-4.5} are obtained from the results during $100a_{\mathrm{min}}<a_{\mathrm{agg}.}<10000a_{\mathrm{min}}$. \color{black} }
\label{evolution of aggregates 4.5}
\end{figure}

\clearpage

\begin{figure}[H]
\begin{tabular}{cc}
\begin{minipage}[t]{0.5\hsize}
  \centering
  \includegraphics[width=8.5cm,pagebox=cropbox,clip]{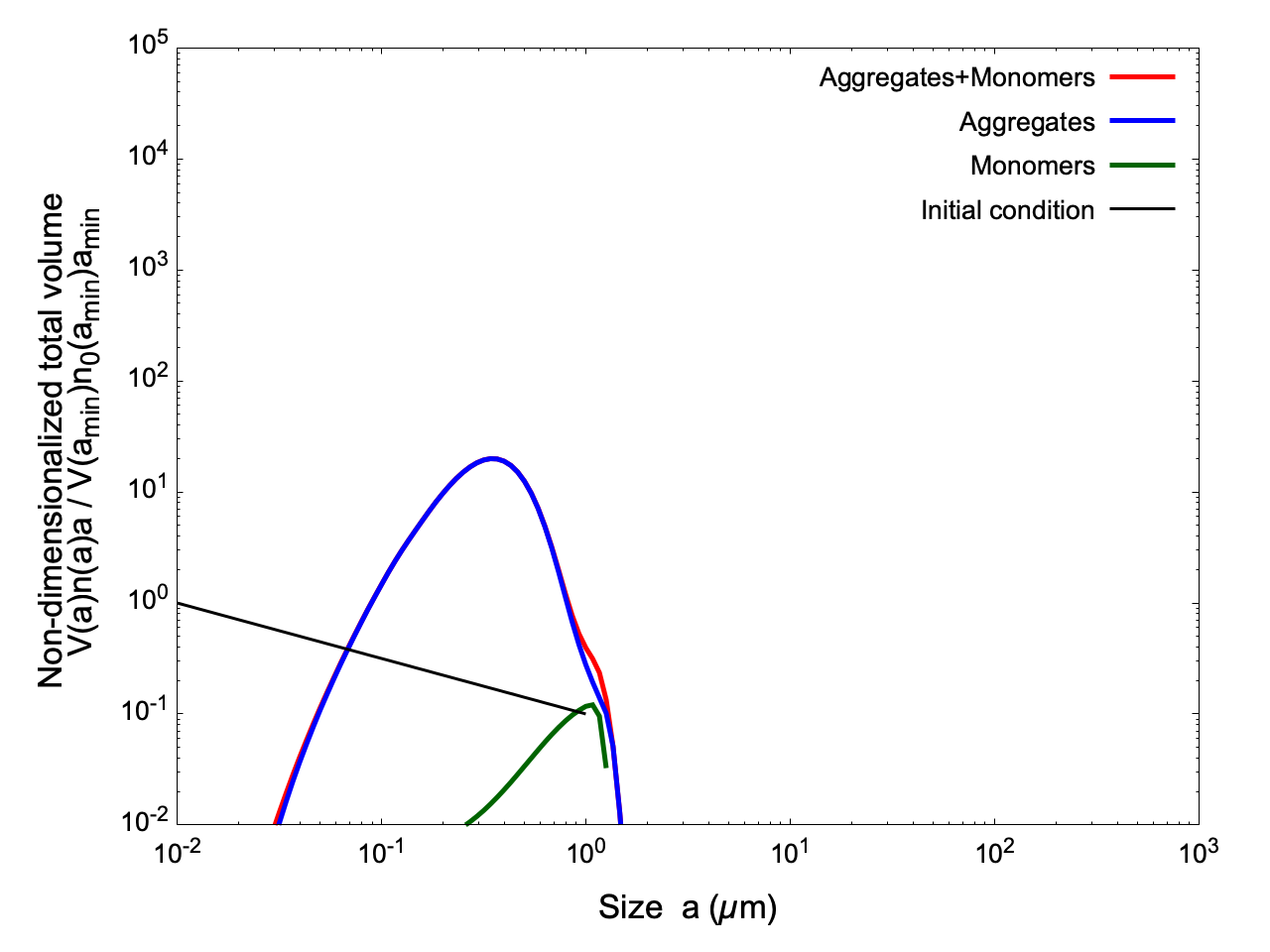}
  \subcaption{Average aggregate size is $0.2 \ \mathrm{\mu m}$}
  \label{volume distribution a 4.5}
\end{minipage}
\begin{minipage}[t]{0.5\hsize}
  \centering
  \includegraphics[width=8.5cm,pagebox=cropbox,clip]{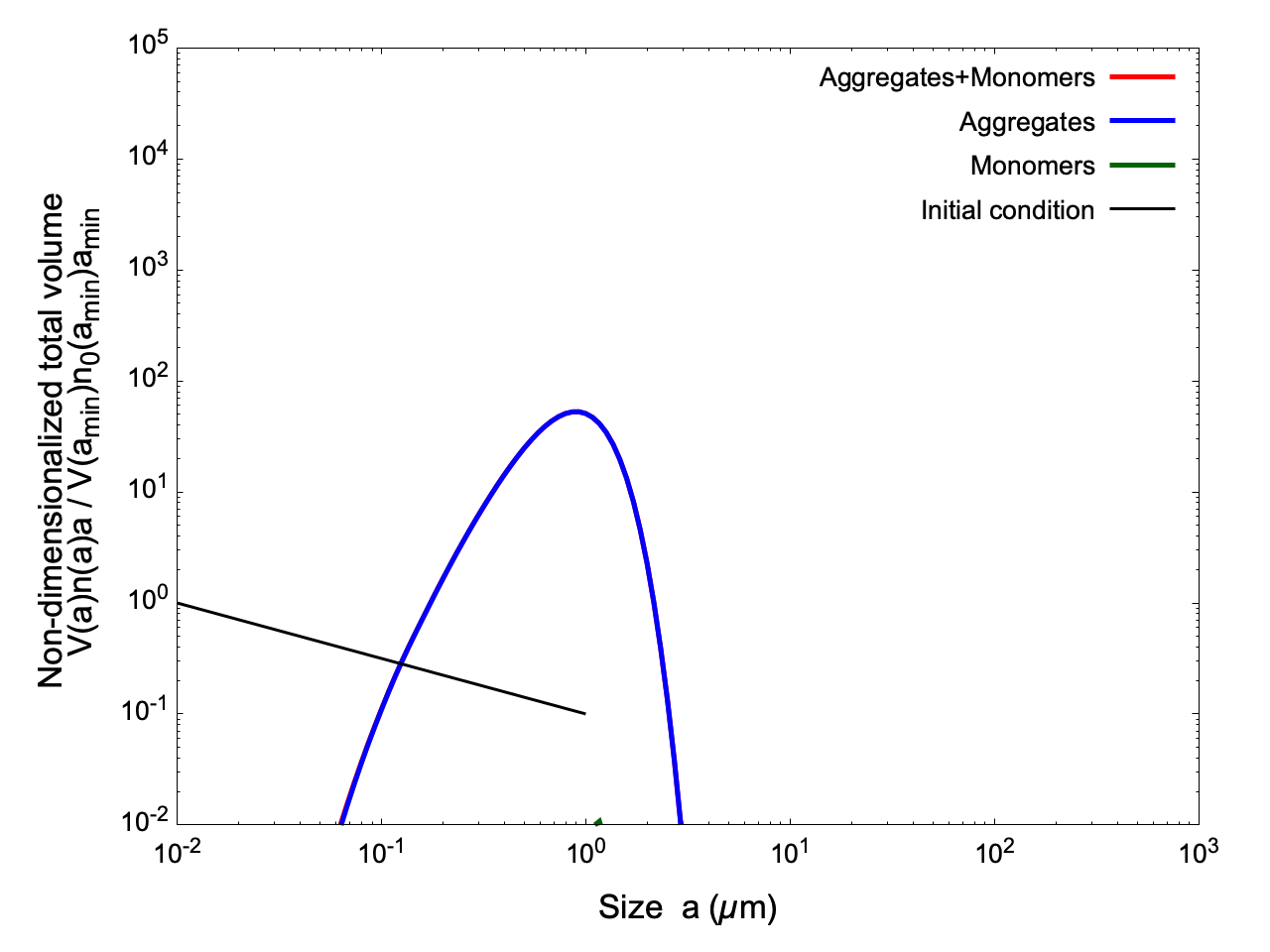}
  \subcaption{Average aggregate size is $0.5 \ \mathrm{\mu m}$}
  \label{volume distribution b 4.5}
\end{minipage}\\ \\
\begin{minipage}[t]{0.5\hsize}
  \centering
  \includegraphics[width=8.5cm,pagebox=cropbox,clip]{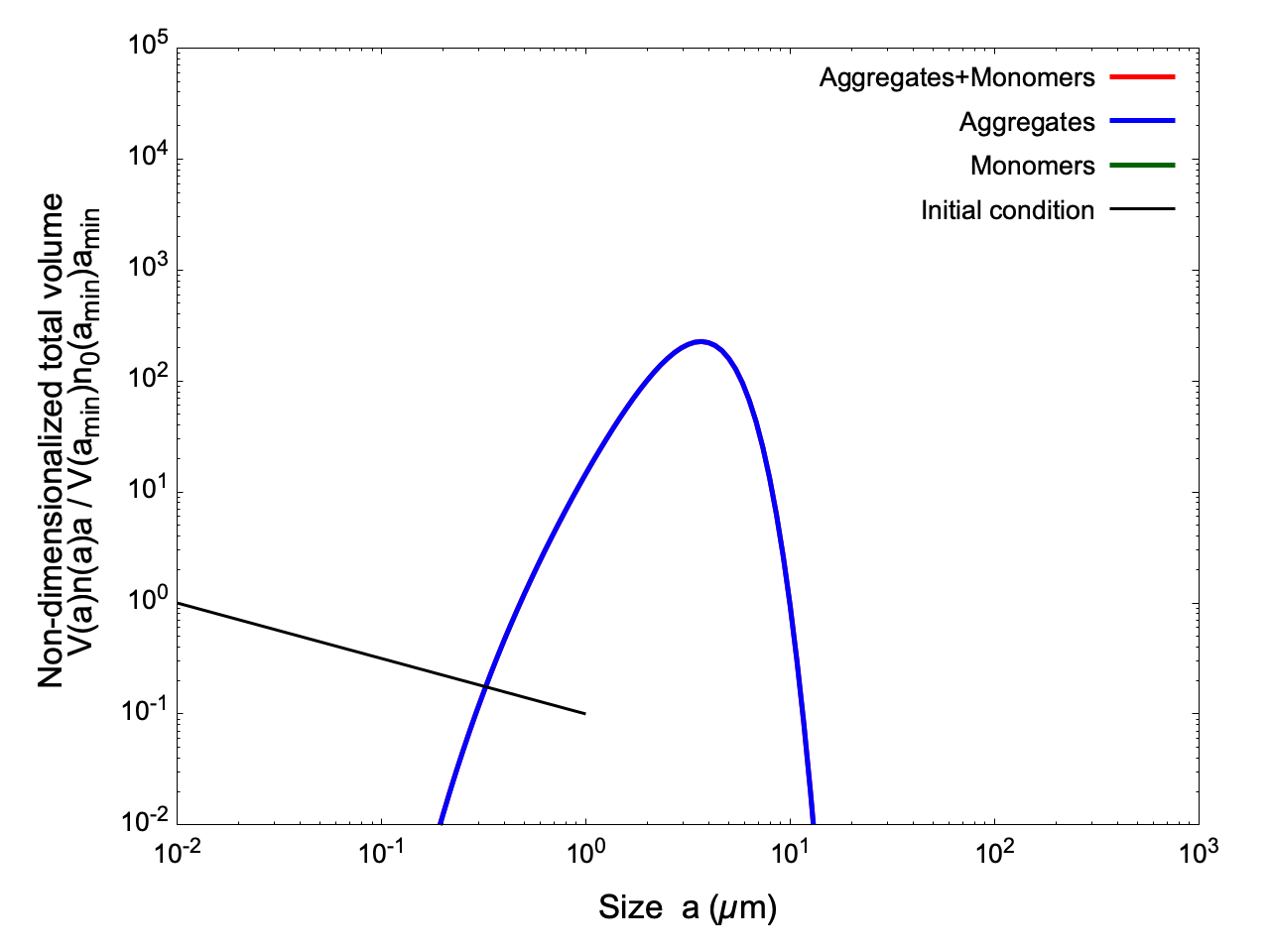}
  \subcaption{Average aggregate size is $2 \ \mathrm{\mu m}$}
  \label{volume distribution c 4.5}
\end{minipage}
\begin{minipage}[t]{0.5\hsize}
  \centering
  \includegraphics[width=8.5cm,pagebox=cropbox,clip]{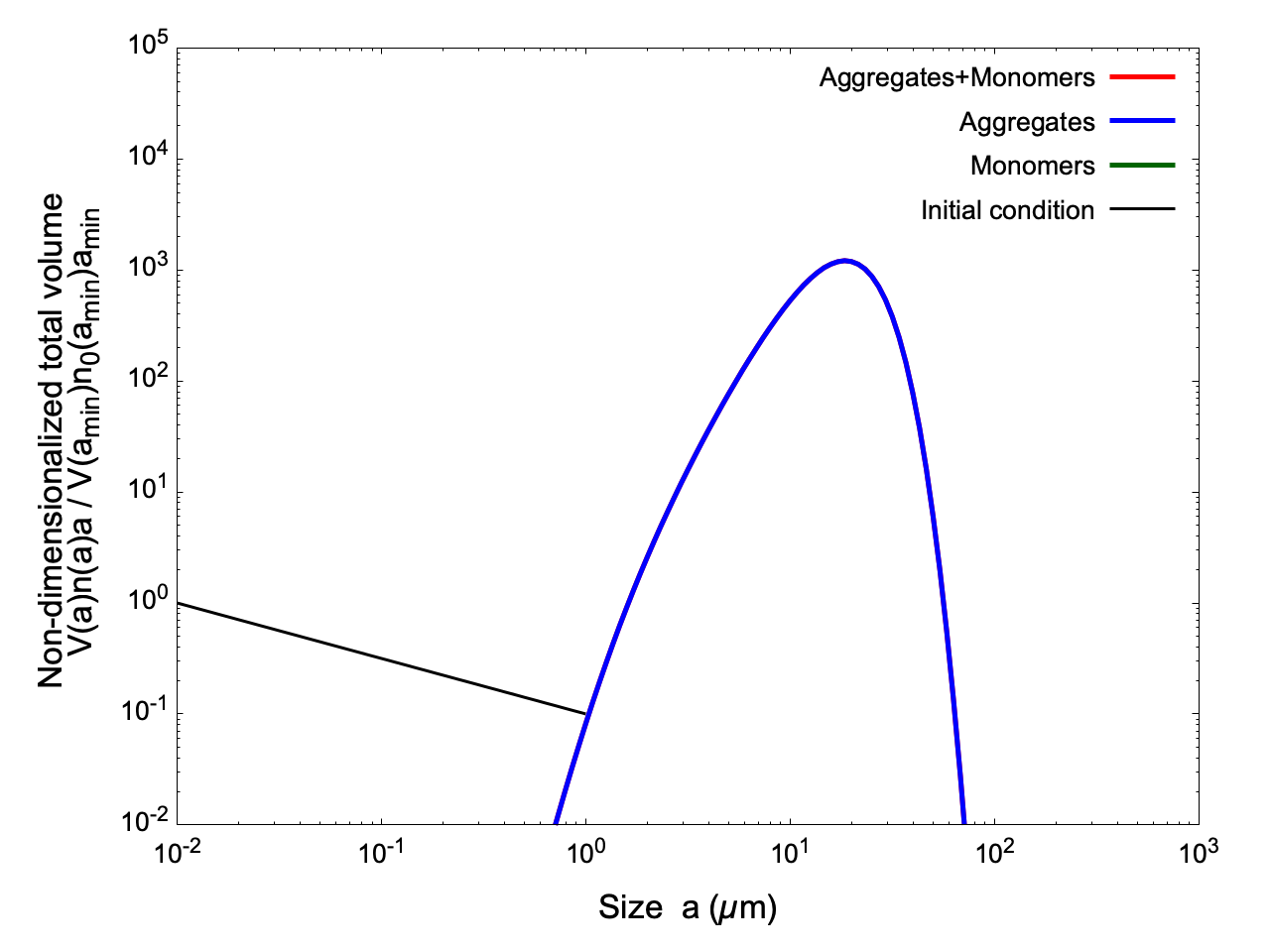}
  \subcaption{Average aggregate size is $10 \ \mathrm{\mu m}$}
  \label{volume distribution d 4.5} 
  \end{minipage}
\end{tabular}
\caption{Evolution of the volume distribution for the case of $\beta=-4.5$. The remaining details are the same as in Figure \ref{volume distribution 3.5}. }
\label{volume distribution 4.5}
\end{figure}

\clearpage

\begin{figure}[H]
\begin{tabular}{cc}
\begin{minipage}[t]{0.5\hsize}
  \centering
  \includegraphics[width=8.5cm,pagebox=cropbox,clip]{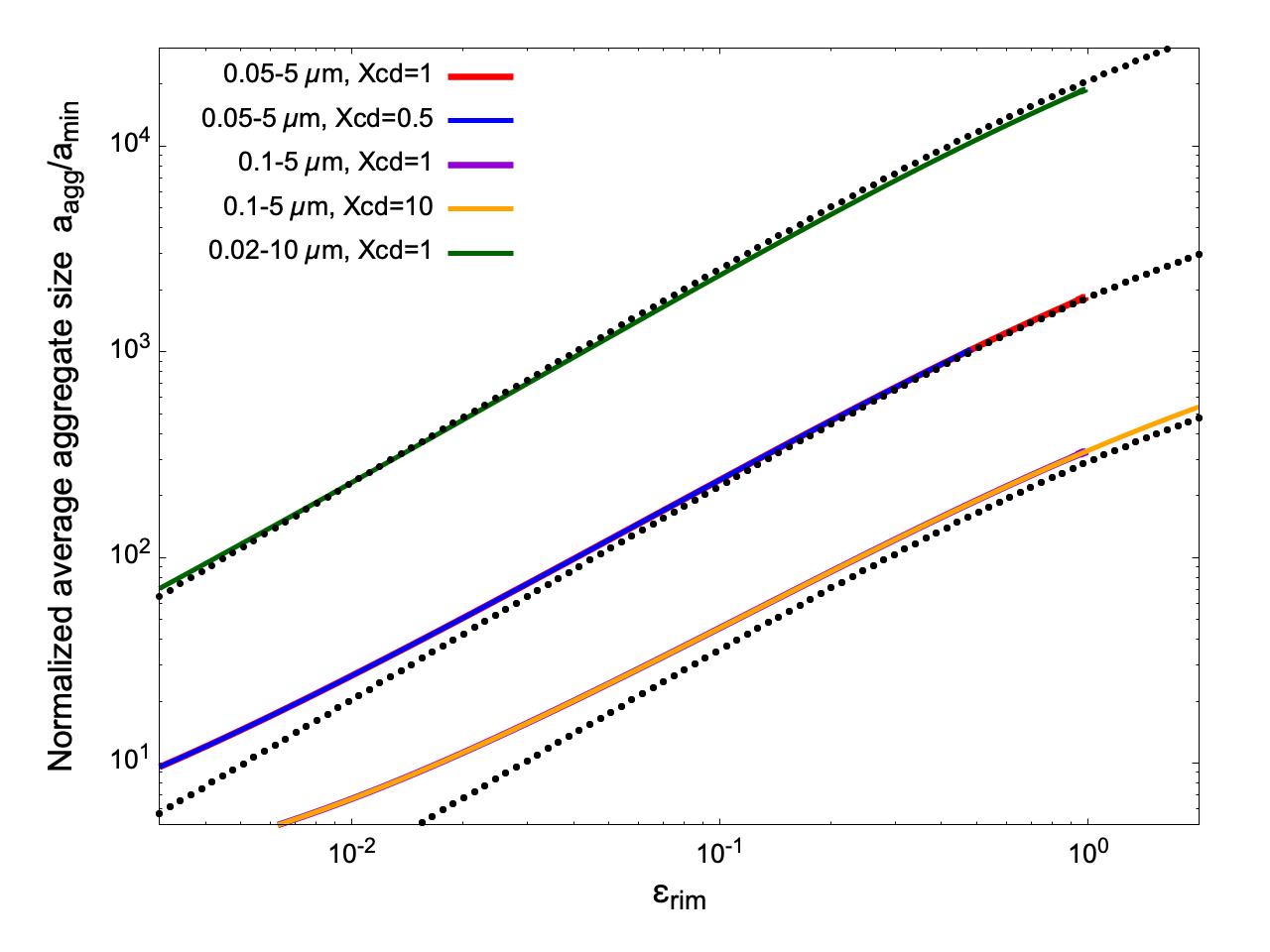}
  \subcaption{Average size of aggregates $a_{\mathrm{agg}}$}
  \label{e-rim size 4.5}
\end{minipage}
\begin{minipage}[t]{0.5\hsize}
  \centering
  \includegraphics[width=8.5cm,pagebox=cropbox,clip]{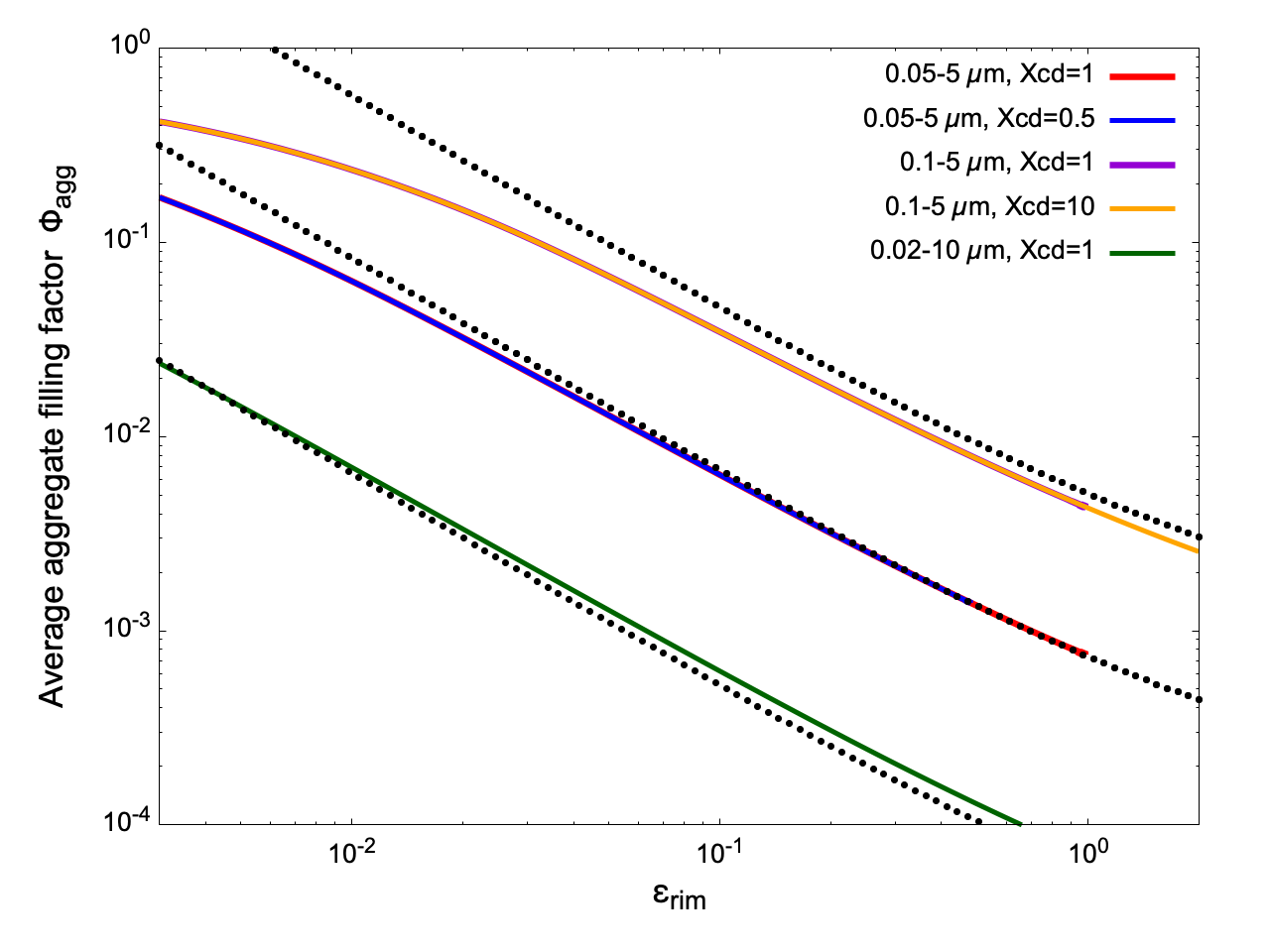}
  \subcaption{Average filling factor of aggregates $\phi_{\mathrm{agg}}$}
  \label{e-rim filling factor 4.5}
\end{minipage}
\end{tabular}
  \caption{ \color{red} (a) average size and (b) average filling factor of aggregate against the rim to chondrule mass ratio $\epsilon_{\mathrm{rim}}$ for $\beta=-4.5$. Different colors represent the following different initial conditions: $(a_{\mathrm{min}},a_{\mathrm{max}})=(0.05 \ \mathrm{\mu m}, 5 \ \mathrm{\mu m})$ and $\rho_{d,0}=\rho_{c,0}$ (red); $(a_{\mathrm{min}},a_{\mathrm{max}})=(0.05 \ \mathrm{\mu m}, 5 \ \mathrm{\mu m})$ and $\rho_{d,0}=0.5\rho_{c,0}$ (blue); $(a_{\mathrm{min}},a_{\mathrm{max}})=(0.1 \ \mathrm{\mu m}, 5 \ \mathrm{\mu m})$ and $\rho_{d,0}=\rho_{c,0}$ (violet); $(a_{\mathrm{min}},a_{\mathrm{max}})=(0.1 \ \mathrm{\mu m}, 5 \ \mathrm{\mu m})$ and $\rho_{d,0}=10\rho_{c,0}$ (yellow); and $(a_{\mathrm{min}},a_{\mathrm{max}})=(0.02 \ \mathrm{\mu m}, 10 \ \mathrm{\mu m})$ and $\rho_{d,0}=\rho_{c,0}$ (green). We fixed the chondrule density, $\rho_{c}=0.005\rho_{g}$. (a) The black dotted lines indicate Eq. \ref{fitting4.5acc}. (b) The black dotted lines are obtained by substituting Eq. \ref{fitting4.5acc} into Eq. \ref{fit-phi-4.5}. \color{black} }
  \label{e-rim 4.5}
\end{figure}

\clearpage

\begin{figure}[H]
\begin{tabular}{cc}
\begin{minipage}[t]{0.5\hsize}
  \centering
  \includegraphics[width=8.5cm,pagebox=cropbox,clip]{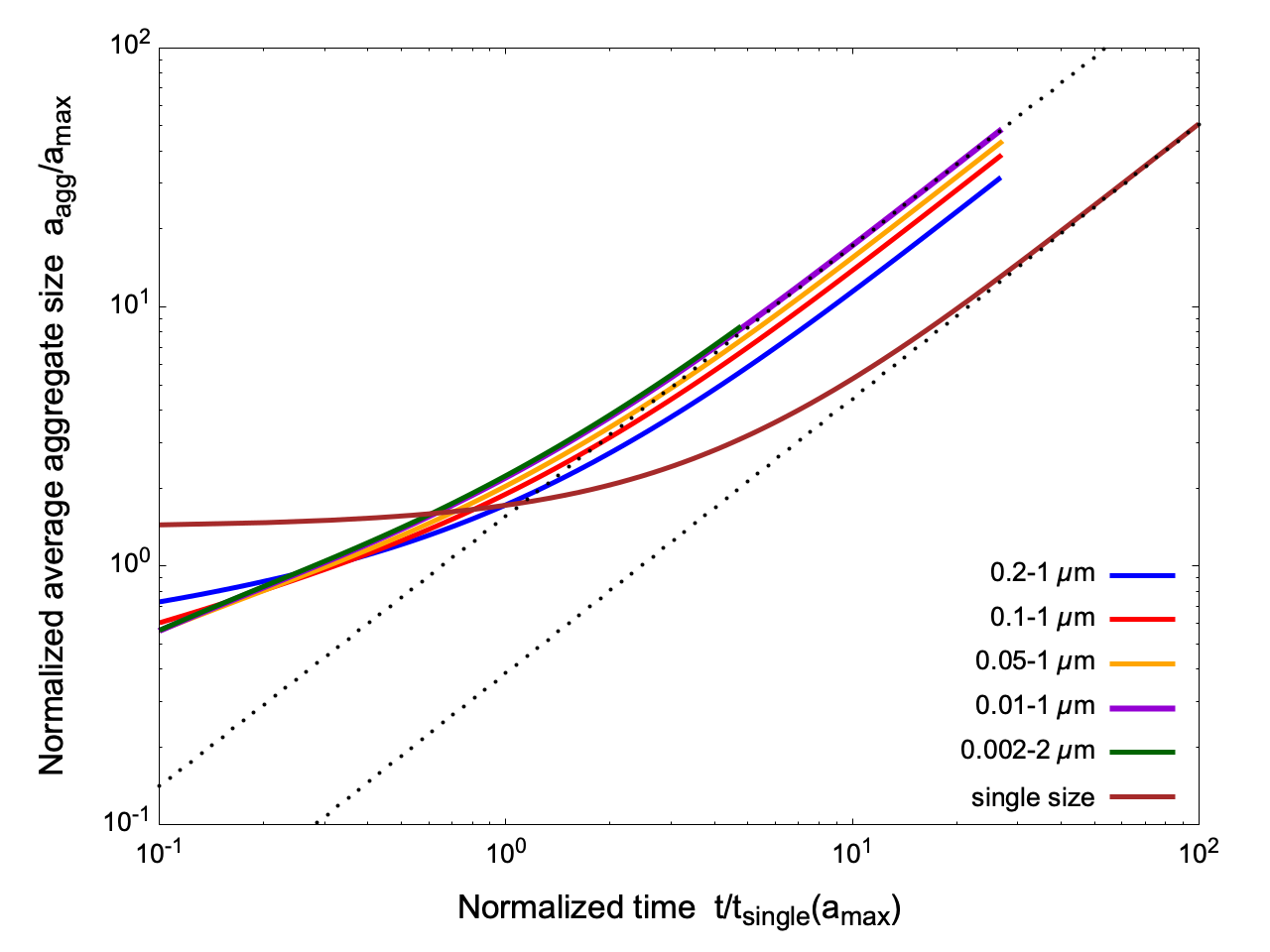}
  \subcaption{Evolution of the average aggregate size}
  \label{aggregate size 2.5b}
\end{minipage}
\begin{minipage}[t]{0.5\hsize}
  \centering
  \includegraphics[width=8.5cm,pagebox=cropbox,clip]{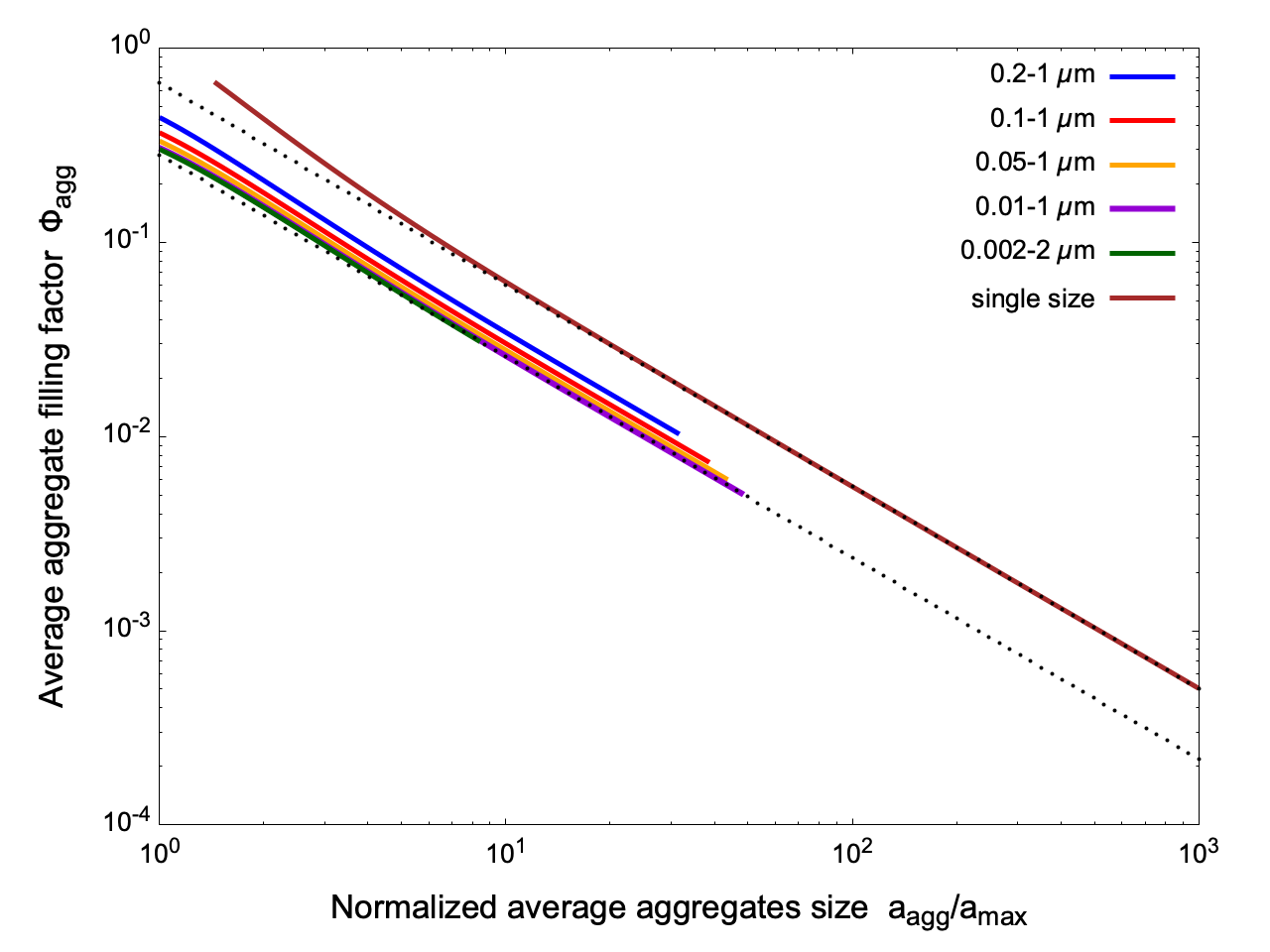}
  \subcaption{Evolution of the average filling factor}
  \label{filling factor 2.5b}
\end{minipage}
\end{tabular}
\caption{ \color{red} Evolution of aggregates for the case of $\beta=-2.5$ normalized by $a_{\mathrm{max}}$ and $t_{\mathrm{single}}(a_{\mathrm{max}})$. The colors of the lines represent the same initial conditions as those in Figure \ref{evolution of aggregates 2.5a}. The black dotted lines indicate the fitting formulae for single size distribution, Eqs. \ref{fit-a-single} and \ref{fit-phi-single}, and for $(a_{\mathrm{min}},a_{\mathrm{max}})=(0.01 \ \mathrm{\mu m},1 \ \mathrm{\mu m})$, Eqs. \ref{fit-a-2.5} and \ref{fit-phi-2.5}. Eqs. \ref{fit-a-2.5} and \ref{fit-phi-2.5} are obtained from the results during $10a_{\mathrm{max}}<a_{\mathrm{agg}}$. \color{black} }
\label{evolution of aggregates 2.5}
\end{figure}

\clearpage

\begin{figure}[H]
\begin{tabular}{cc}
\begin{minipage}[t]{0.5\hsize}
  \centering
  \includegraphics[width=8.5cm,pagebox=cropbox,clip]{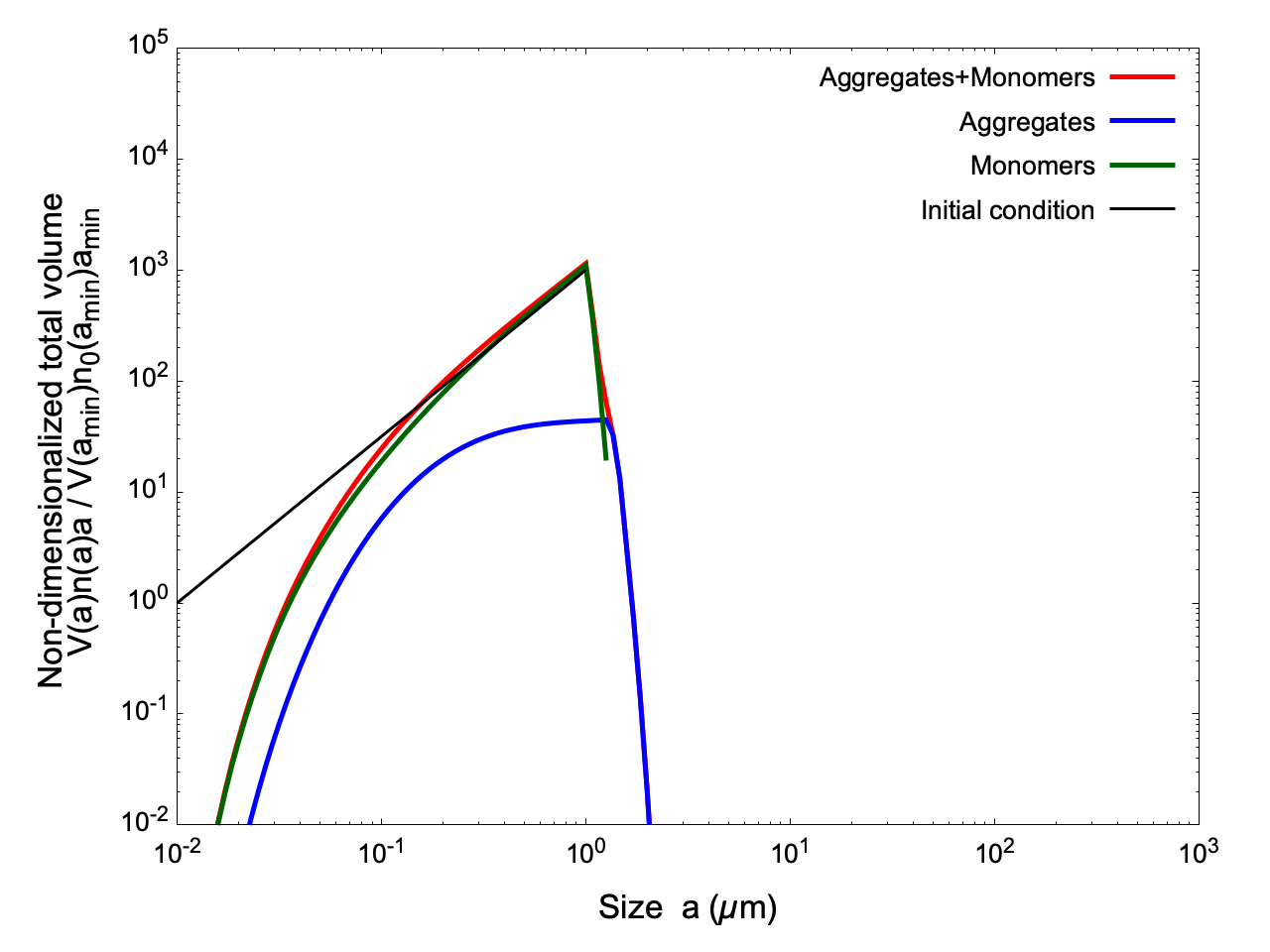}
  \subcaption{Average aggregate size is $0.2 \ \mathrm{\mu m}$}
  \label{volume distribution a 2.5}
\end{minipage}
\begin{minipage}[t]{0.5\hsize}
  \centering
  \includegraphics[width=8.5cm,pagebox=cropbox,clip]{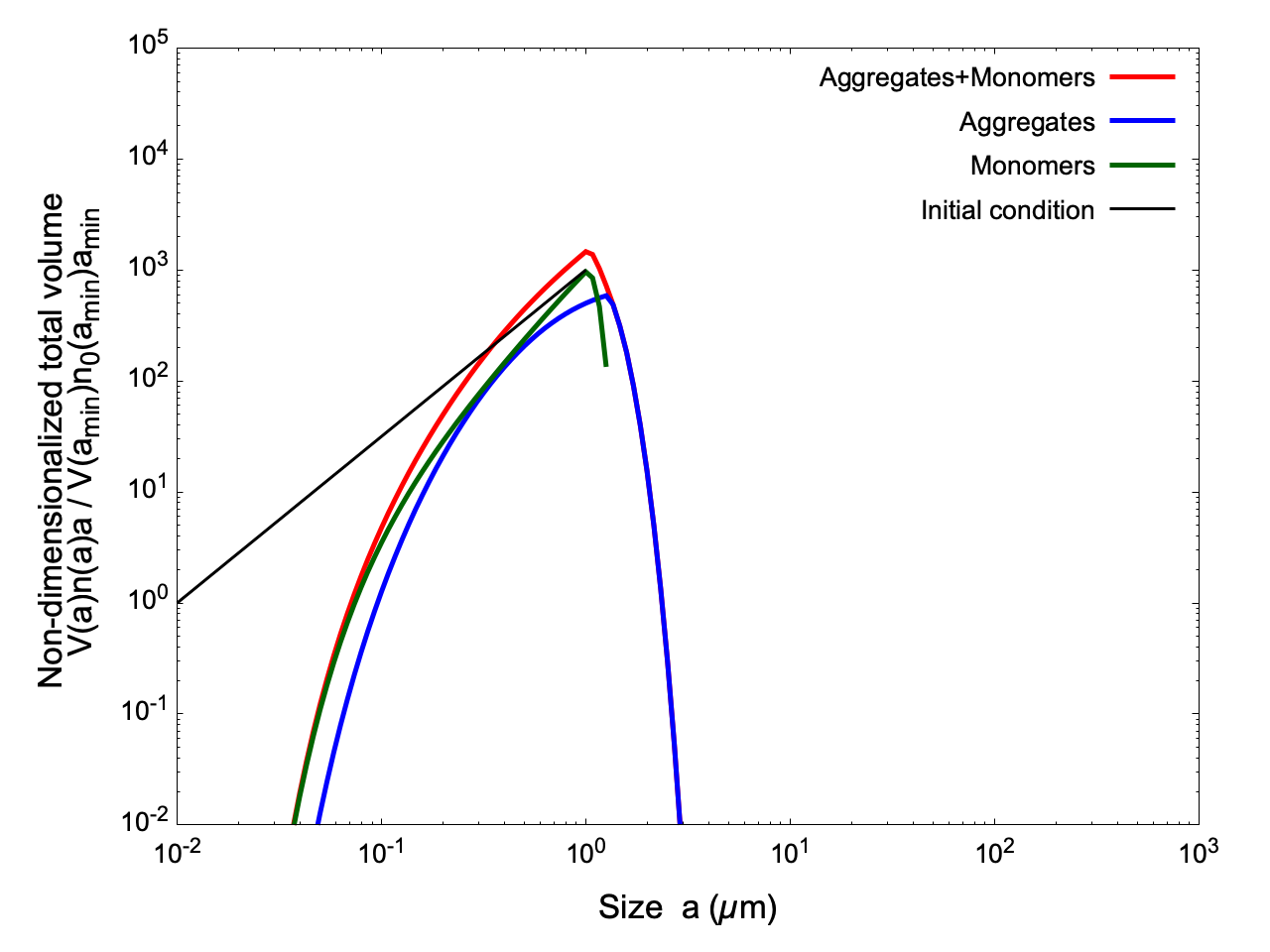}
  \subcaption{Average aggregate size is $0.5 \ \mathrm{\mu m}$}
  \label{volume distribution b 2.5}
\end{minipage}\\ \\
\begin{minipage}[t]{0.5\hsize}
  \centering
  \includegraphics[width=8.5cm,pagebox=cropbox,clip]{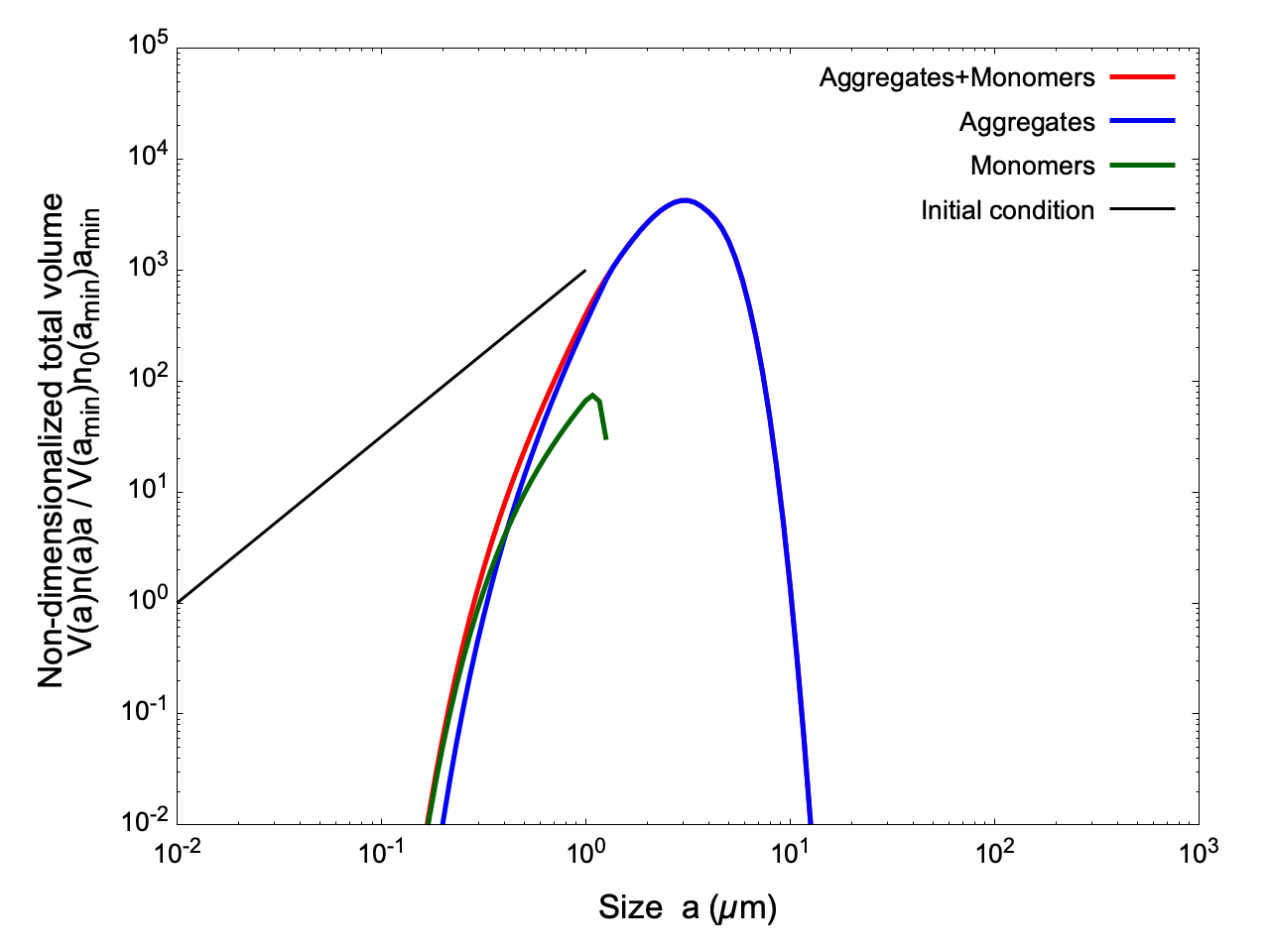}
  \subcaption{Average aggregate size is $2 \ \mathrm{\mu m}$}
  \label{volume distribution c 2.5}
\end{minipage}
\begin{minipage}[t]{0.5\hsize}
  \centering
  \includegraphics[width=8.5cm,pagebox=cropbox,clip]{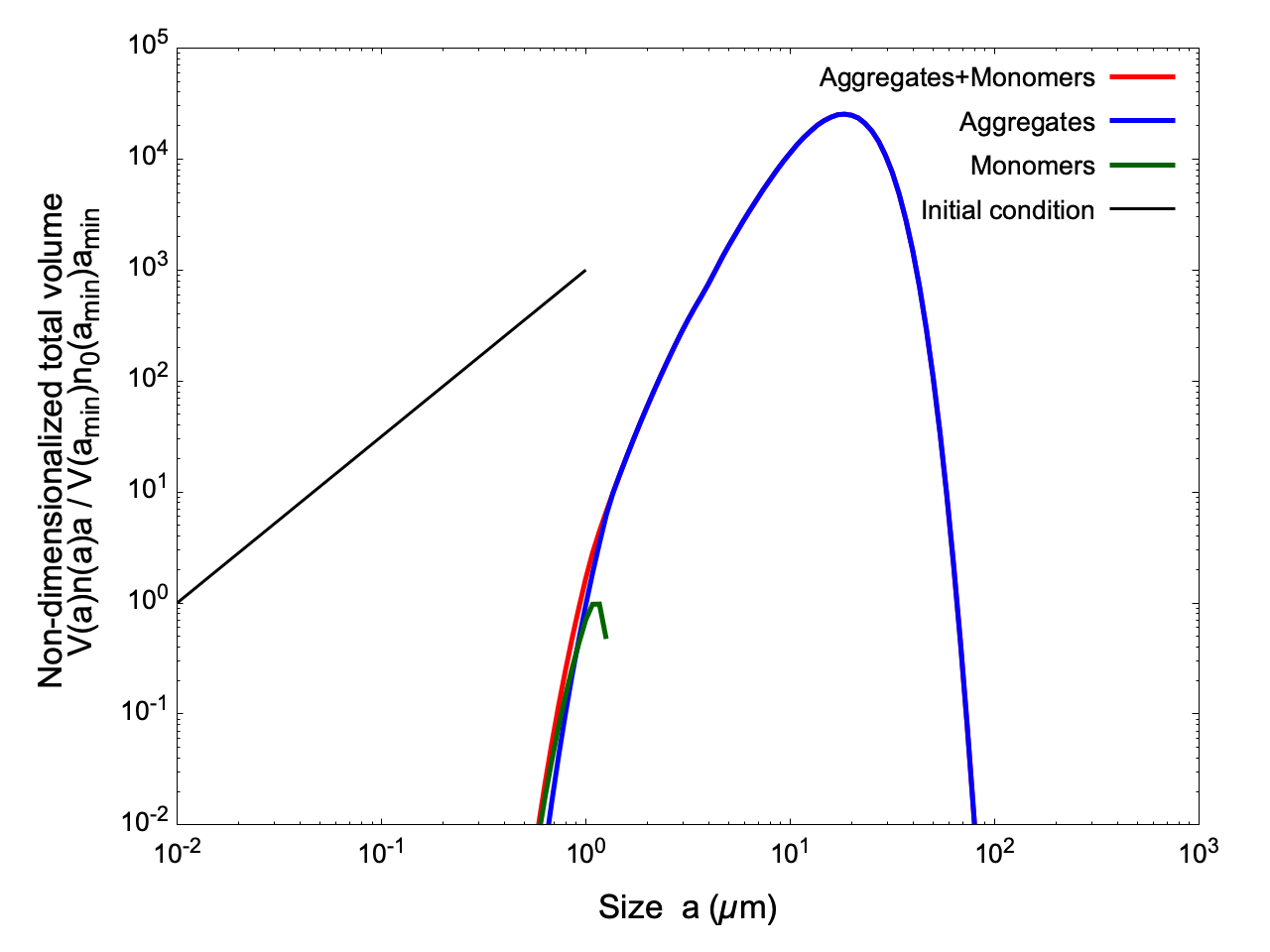}
  \subcaption{Average aggregate size is $10 \ \mathrm{\mu m}$}
  \label{volume distribution d 2.5} 
  \end{minipage}
\end{tabular}
\caption{Evolution of the volume distribution for the case of $\beta=-2.5$. The remaining details are the same as in Figure \ref{volume distribution 3.5}. }
\label{volume distribution 2.5}
\end{figure}

\clearpage

\begin{figure}[H]
\begin{tabular}{cc}
\begin{minipage}[t]{0.5\hsize}
  \centering
  \includegraphics[width=8.5cm,pagebox=cropbox,clip]{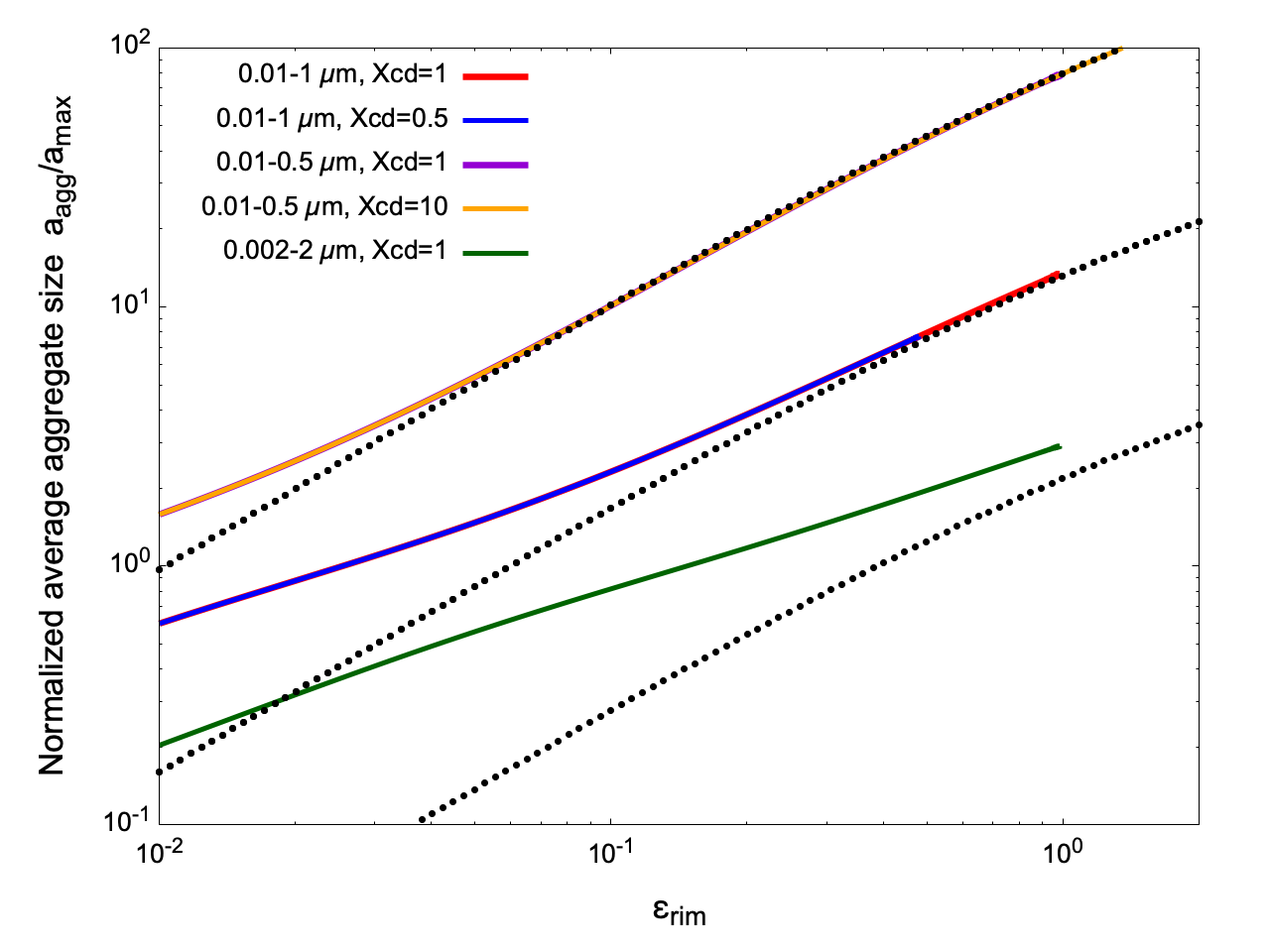}
  \subcaption{Average size of aggregates $a_{\mathrm{agg}}$}
  \label{e-rim size 2.5}
\end{minipage}
\begin{minipage}[t]{0.5\hsize}
  \centering
  \includegraphics[width=8.5cm,pagebox=cropbox,clip]{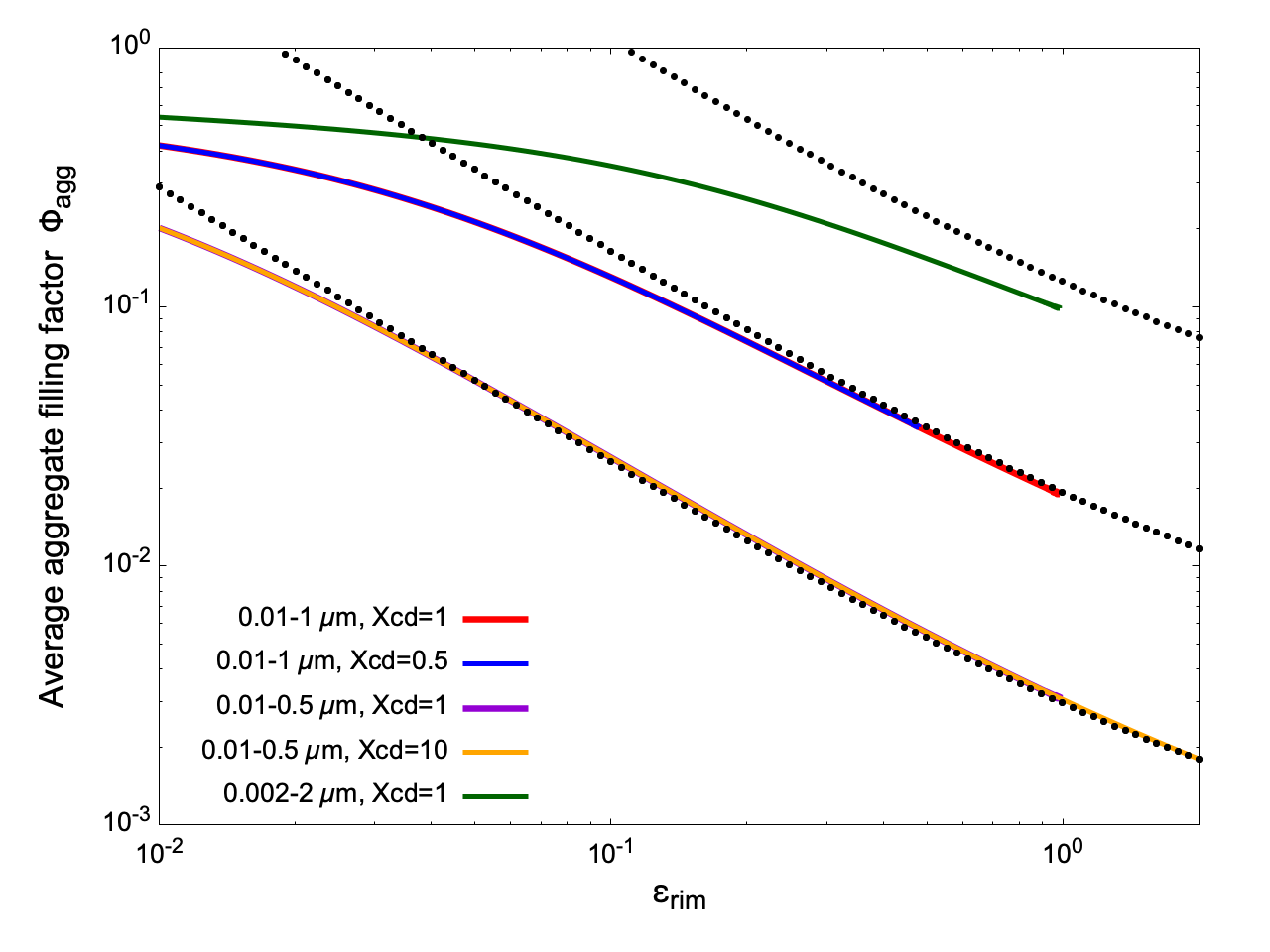}
  \subcaption{Average filling factor of aggregates $\phi_{\mathrm{agg}}$}
  \label{e-rim filling factor 2.5}
\end{minipage}
\end{tabular}
  \caption{ \color{red} (a) average size and (b) average filling factor of aggregate against the rim to chondrule mass ratio $\epsilon_{\mathrm{rim}}$ for $\beta=-2.5$. Different colors represent the following different initial conditions: $(a_{\mathrm{min}},a_{\mathrm{max}})=(0.01 \ \mathrm{\mu m}, 1 \ \mathrm{\mu m})$ and $\rho_{d,0}=\rho_{c,0}$ (red); $(a_{\mathrm{min}},a_{\mathrm{max}})=(0.01 \ \mathrm{\mu m}, 1 \ \mathrm{\mu m})$ and $\rho_{d,0}=0.5\rho_{c,0}$ (blue); $(a_{\mathrm{min}},a_{\mathrm{max}})=(0.01 \ \mathrm{\mu m}, 0.5 \ \mathrm{\mu m})$ and $\rho_{d,0}=\rho_{c,0}$ (violet); $(a_{\mathrm{min}},a_{\mathrm{max}})=(0.01 \ \mathrm{\mu m}, 0.5 \ \mathrm{\mu m})$ and $\rho_{d,0}=10\rho_{c,0}$ (yellow); and $(a_{\mathrm{min}},a_{\mathrm{max}})=(0.002 \ \mathrm{\mu m}, 2 \ \mathrm{\mu m})$ and $\rho_{d,0}=\rho_{c,0}$ (green). We fixed the chondrule density, $\rho_{c}=0.005\rho_{g}$. (a) The black dotted lines indicate Eq. \ref{fitting2.5acc}. (b) The black dotted lines are obtained by substituting Eq. \ref{fitting2.5acc} into Eq. \ref{fit-phi-2.5}. \color{black} }
  \label{e-rim 2.5}
\end{figure}

\clearpage

\begin{figure}[H]
  \centering
  \includegraphics[width=12cm,pagebox=cropbox,clip]{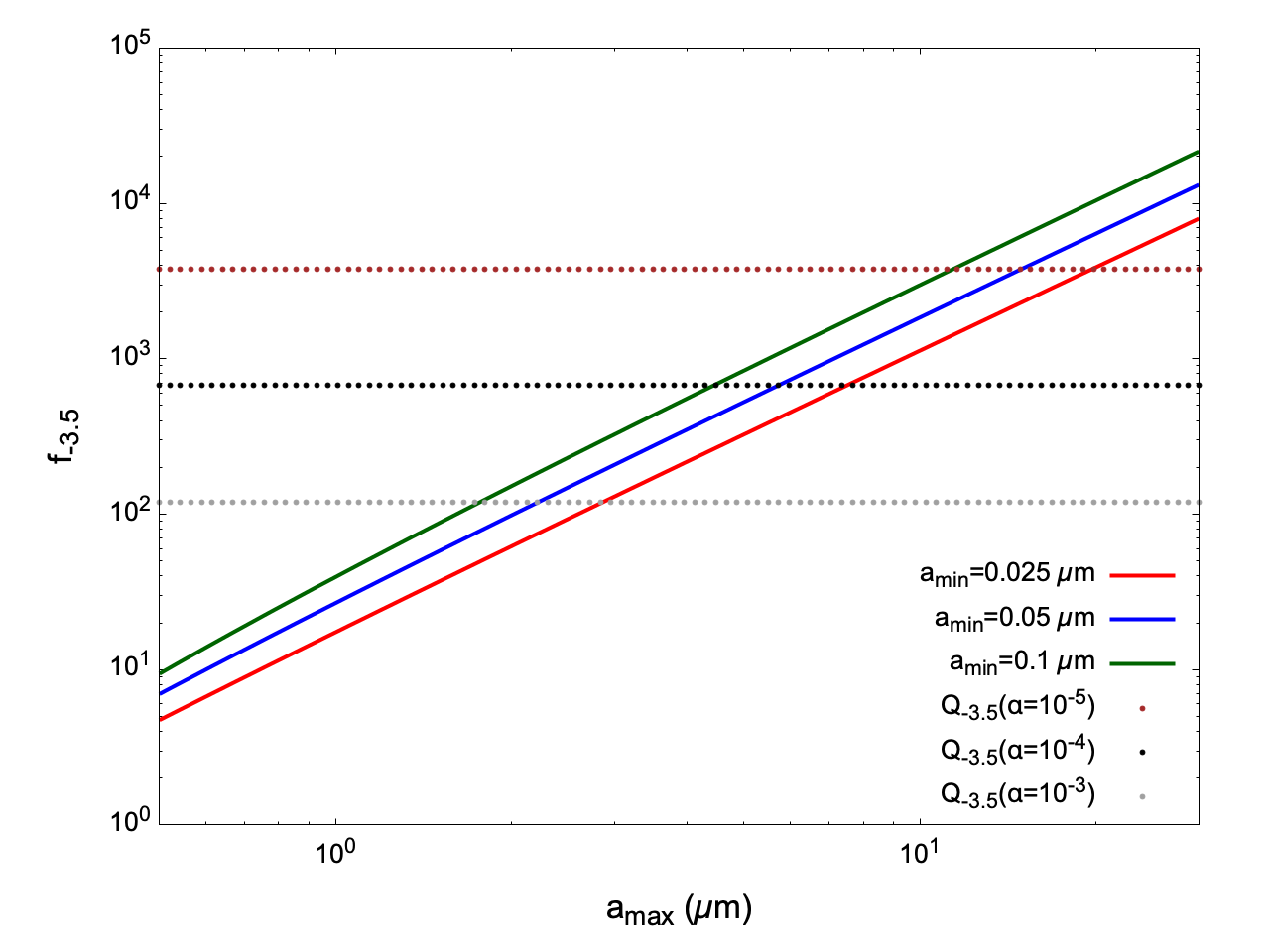}
  \caption{Value of $f_{-3.5}(a_{\mathrm{max}}, a_{\mathrm{min}})$ (Eq. \ref{f_3.5_eq}) with respect to the maximum size of monomer grains, $a_{\mathrm{max}}$. We fixed $a_{\mathrm{min}}$; $a_{\mathrm{min}}=0.025 \ \mathrm{\mu m}$ (red), $a_{\mathrm{min}}=0.05 \ \mathrm{\mu m}$ (blue), and $a_{\mathrm{min}}=0.1 \ \mathrm{\mu m}$ (green). \color{red} The brown ($3.78 \times 10^{3}$), black ($6.73 \times 10^{2}$), and gray ($1.20 \times 10^{2}$) dotted lines indicate the threshold values $Q_{-3.5}$ (Eq. \ref{Q_3.5_eq}) for $\alpha=10^{-5}$, $10^{-4}$, and $10^{-3}$, respectively (gas surface density and rime mass fraction are constant, $\Sigma_{g}=680 \ \mathrm{g/cm^{2}}$ and $\epsilon_{\mathrm{rim}}=0.4$). \color{black} The aggregate-accretion case lies below the dotted lines and the monomer-accretion case is above them. }
  \label{f-3.5}
\end{figure}

\clearpage

\begin{figure}[H]
  \begin{tabular}{cc}
\begin{minipage}[t]{0.5\hsize}
  \centering
  \includegraphics[width=8.5cm,pagebox=cropbox,clip]{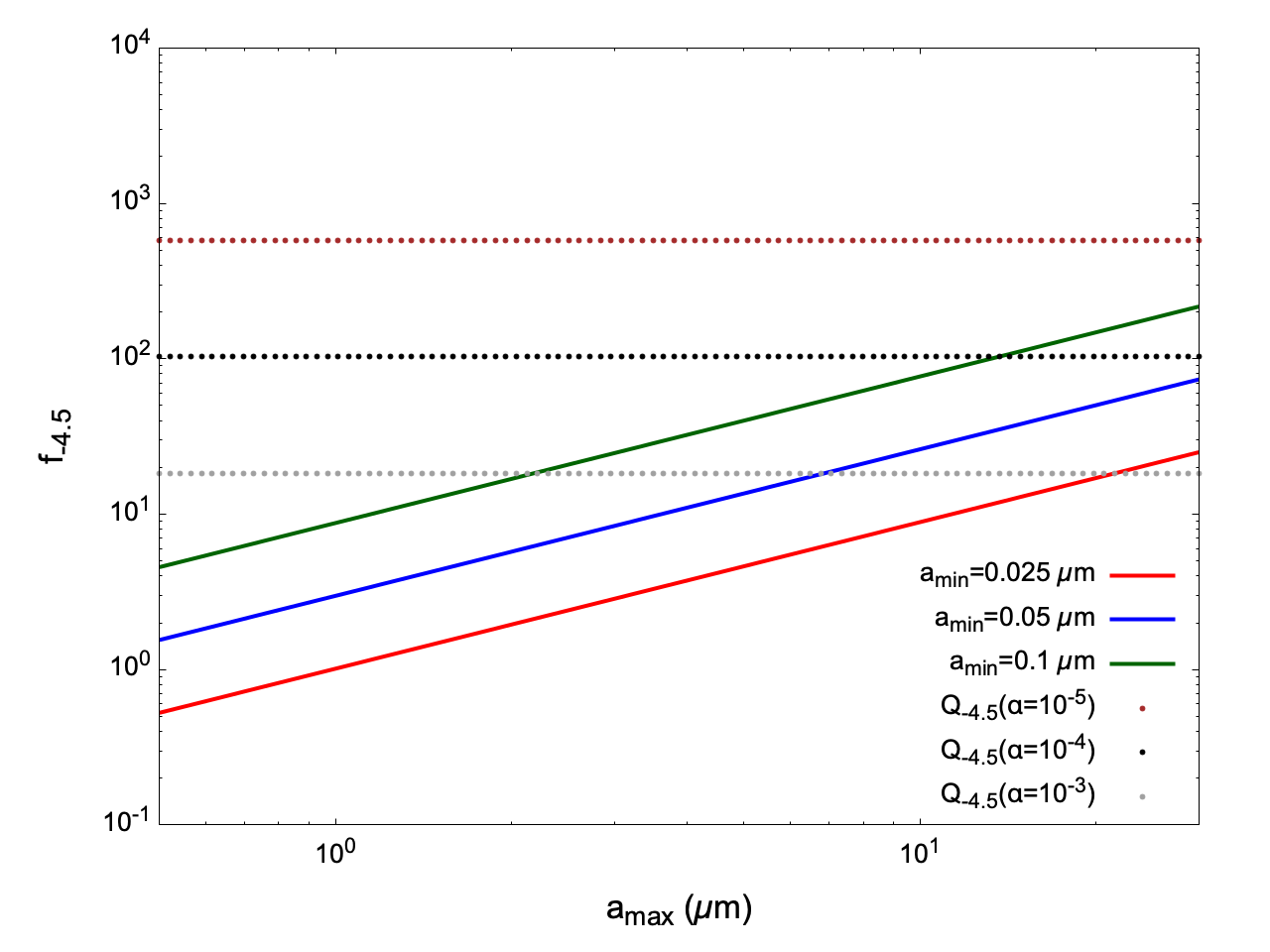}
  \subcaption{$\beta=-4.5$}
\end{minipage}
\begin{minipage}[t]{0.5\hsize}
  \centering
  \includegraphics[width=8.5cm,pagebox=cropbox,clip]{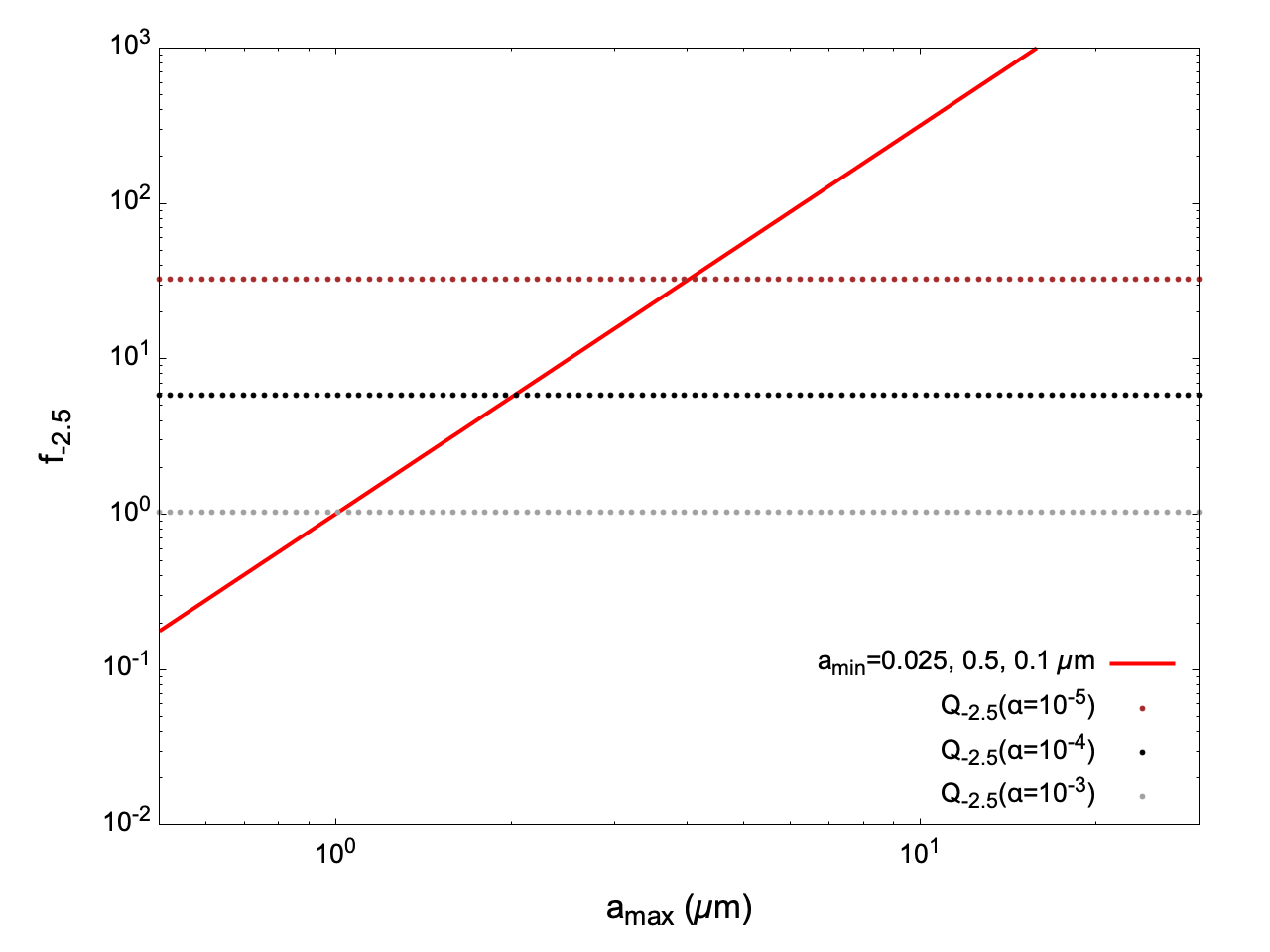}
  \subcaption{$\beta=-2.5$}
\end{minipage}
\end{tabular}
  \caption{Value of $f_{-4.5}(a_{\mathrm{max}}, a_{\mathrm{min}})$ (Eq. \ref{f_4.5_eq}) and $f_{-2.5}(a_{\mathrm{max}}, a_{\mathrm{min}})$ (Eq. \ref{f_2.5_eq}) with respect to the maximum size of monomer grains. We fixed $a_{\mathrm{min}}$; $a_{\mathrm{min}}=0.025 \ \mathrm{\mu m}$ (red), $a_{\mathrm{min}}=0.05 \ \mathrm{\mu m}$ (blue), and $a_{\mathrm{min}}=0.1 \ \mathrm{\mu m}$ (green) for $\beta=-4.5$. When $\beta=-2.5$, $f_{-2.5}(a_{\mathrm{max}}, a_{\mathrm{min}})$ does not depend on $a_{\mathrm{min}}$, so we only show the red line. \color{red}  The threshold dotted lines indicate $Q_{-4.5}$ (Eq. \ref{Q_4.5_eq}); $5.77 \times 10^{2}$ (brown for $\alpha=10^{-5}$), $1.03 \times 10^{2}$ (black for $\alpha=10^{-4}$), and $1.82 \times 10$ (gray for $\alpha=10^{-3}$) in the left figure, and $Q_{-2.5}$ (Eq. \ref{Q_2.5_eq}); $3.25 \times 10^{1}$ (brown for $\alpha=10^{-5}$), $5.79$ (black for $\alpha=10^{-4}$), and $1.03$ (gray for $\alpha=10^{-3}$) in the right figure, respectively (gas surface density and rime mass fraction are constant, $\Sigma_{g}=680 \ \mathrm{g/cm^{2}}$ and $\epsilon_{\mathrm{rim}}=0.4$). \color{black}  The aggregate-accretion case lies below the dotted lines and the monomer-accretion case is above them. }
  \label{f-4.5,2.5}
\end{figure}

\clearpage

\begin{figure}[H]
  \centering
  \includegraphics[width=12cm,pagebox=cropbox,clip]{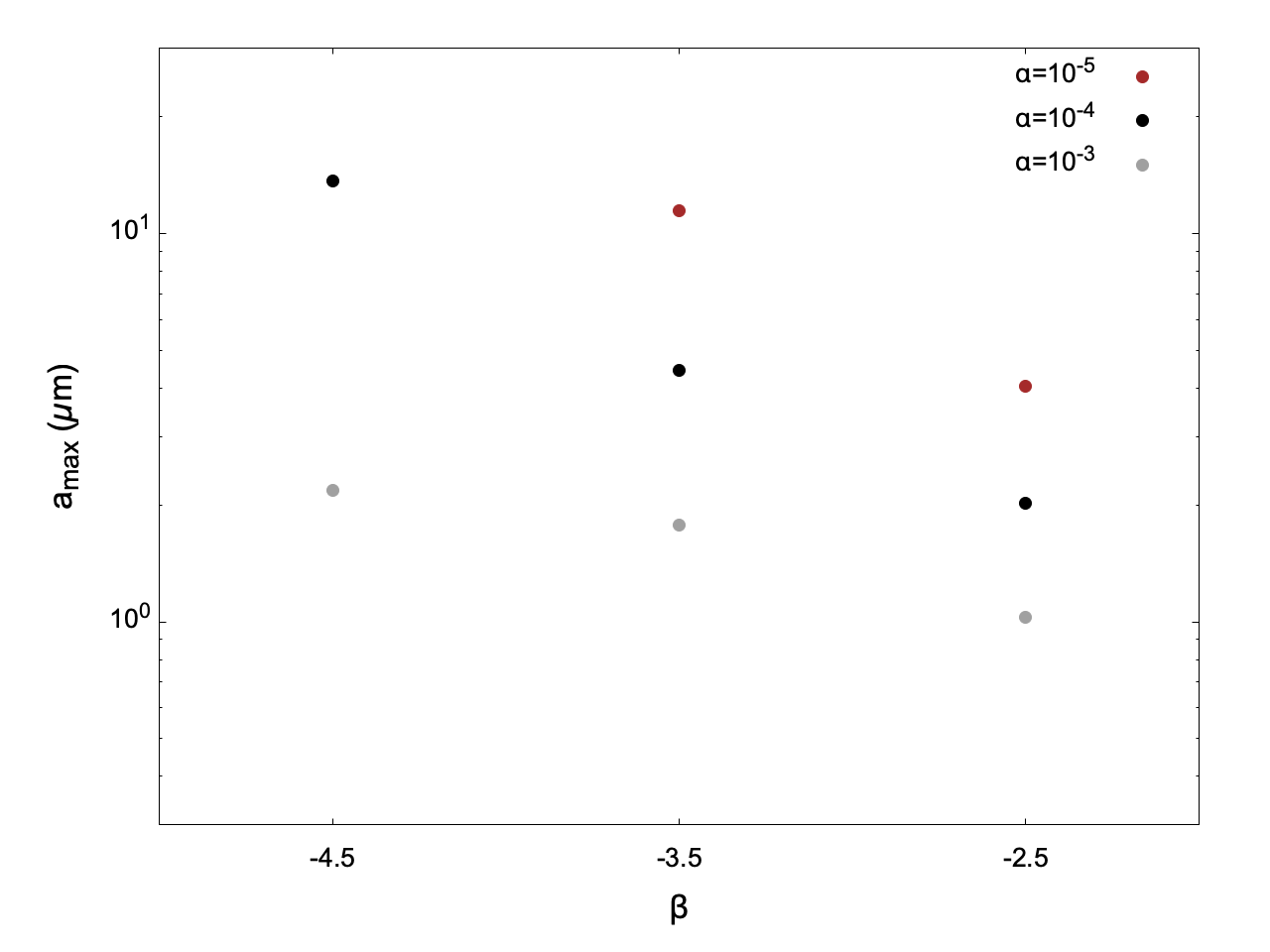}
  \caption{Threshold maximum monomer size for the monomer-accretion case when $a_{\mathrm{min}}=0.1 \ \mathrm{\mu m}$. The different colors of the points indicate different values of $\alpha$; brown ($\alpha=10^{-5}$), black ($\alpha=10^{-4}$), and gray ($\alpha=10^{-3}$).}
  \label{amax-beta}
\end{figure}

\clearpage

\begin{figure}[H]
  \centering
  \includegraphics[width=15cm,pagebox=cropbox,clip]{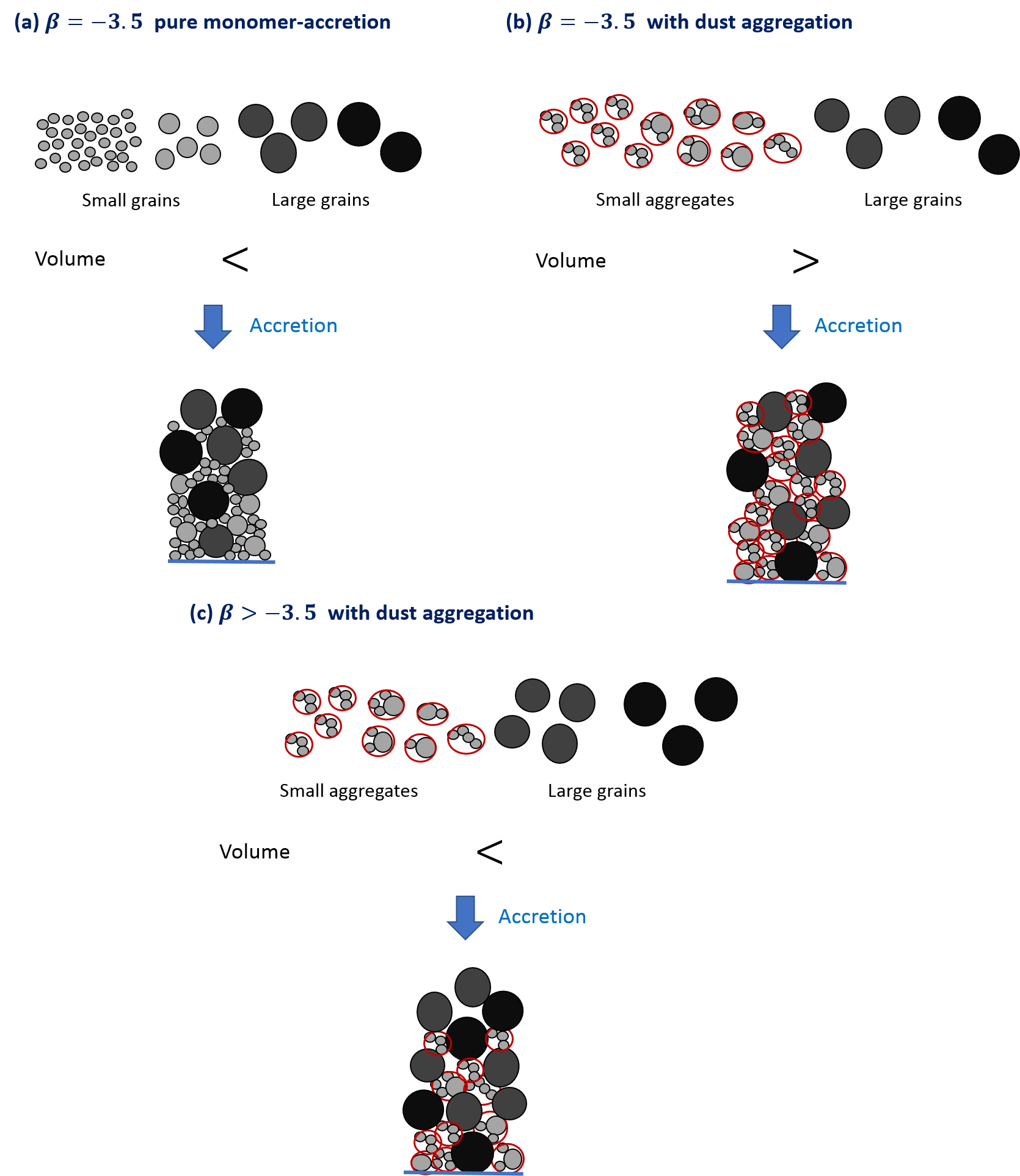}
  \caption{Structures of the rims that are predicted from the volume distributions (Figures \ref{volume distribution 3.5} and \ref{volume distribution 2.5}). (a) $\beta=-3.5$ and pure monomer accretion without dust aggregation. In this case, the larger grains control the overall structure of the rim. The smaller grains pass through the voids and preferentially accumulate near chondrule surface. Grain size coarsening is observed \citep{Xiang.etal2019} (b) $\beta=-3.5$ with dust aggregation. In this case, the small aggregates in the red circles that are composed of the smaller grains control the overall structure of the rim. Grain size coarsening might not be observed. (c) $\beta>-3.5$ with dust aggregation. In this case, the larger grains control the overall structure of the rim (see the volume distribution in Figure \ref{volume distribution 2.5}). Grain size coarsening might be observed.}
  \label{prediction}
\end{figure}

\clearpage

\bibliography{paper}

\begin{thebibliography}{40}
\expandafter\ifx\csname natexlab\endcsname\relax\def\natexlab#1{#1}\fi
\providecommand{\url}[1]{\texttt{#1}}
\providecommand{\href}[2]{#2}
\providecommand{\path}[1]{#1}
\providecommand{\DOIprefix}{doi:}
\providecommand{\ArXivprefix}{arXiv:}
\providecommand{\URLprefix}{URL: }
\providecommand{\Pubmedprefix}{pmid:}
\providecommand{\doi}[1]{\href{http://dx.doi.org/#1}{\path{#1}}}
\providecommand{\Pubmed}[1]{\href{pmid:#1}{\path{#1}}}
\providecommand{\bibinfo}[2]{#2}
\ifx\xfnm\relax \def\xfnm[#1]{\unskip,\space#1}\fi
%Type = Article
\bibitem[{Adachi et~al.(1976)Adachi, Hayashi and Nakazawa}]{Adachi.etal1976}
\bibinfo{author}{Adachi, I.}, \bibinfo{author}{Hayashi, C.},
  \bibinfo{author}{Nakazawa, K.}, \bibinfo{year}{1976}.
\newblock \bibinfo{title}{The gas drag effect on the elliptical motion of a
  solid body in the primordial solar nebula.}
\newblock \bibinfo{journal}{Progress of Theoretical Physics}
  \bibinfo{volume}{56}, \bibinfo{pages}{1756--1771}.
\newblock \DOIprefix\doi{10.1143/PTP.56.1756}.
%Type = Article
\bibitem[{Alexander et~al.(2008)Alexander, Grossman, Ebel and
  Ciesla}]{Alexander.etal2008}
\bibinfo{author}{Alexander, C.M.O.}, \bibinfo{author}{Grossman, J.N.},
  \bibinfo{author}{Ebel, D.S.}, \bibinfo{author}{Ciesla, F.J.},
  \bibinfo{year}{2008}.
\newblock \bibinfo{title}{The {{Formation Conditions}} of {{Chondrules}} and
  {{Chondrites}}}.
\newblock \bibinfo{journal}{Science} \bibinfo{volume}{320},
  \bibinfo{pages}{1617}.
\newblock \DOIprefix\doi{10.1126/science.1156561}.
%Type = Article
\bibitem[{Andrews(2020)}]{Andrews2020}
\bibinfo{author}{Andrews, S.M.}, \bibinfo{year}{2020}.
\newblock \bibinfo{title}{Observations of {{Protoplanetary Disk Structures}}}.
\newblock \bibinfo{journal}{Annual Review of Astronomy and Astrophysics}
  \bibinfo{volume}{58}, \bibinfo{pages}{483--528}.
\newblock \DOIprefix\doi{10.1146/annurev-astro-031220-010302}.
%Type = Article
\bibitem[{Arakawa(2017)}]{Arakawa2017}
\bibinfo{author}{Arakawa, S.}, \bibinfo{year}{2017}.
\newblock \bibinfo{title}{Ejection of {{Chondrules}} from {{Fluffy Matrices}}}.
\newblock \bibinfo{journal}{The Astrophysical Journal} \bibinfo{volume}{846},
  \bibinfo{pages}{118}.
\newblock \DOIprefix\doi{10.3847/1538-4357/aa8564}.
%Type = Article
\bibitem[{Arakawa and Nakamoto(2016)}]{Arakawa.Nakamoto2016}
\bibinfo{author}{Arakawa, S.}, \bibinfo{author}{Nakamoto, T.},
  \bibinfo{year}{2016}.
\newblock \bibinfo{title}{Rocky {{Planetesimal Formation}} via {{Fluffy
  Aggregates}} of {{Nanograins}}}.
\newblock \bibinfo{journal}{The Astrophysical Journal Letters}
  \bibinfo{volume}{832}, \bibinfo{pages}{L19}.
\newblock \DOIprefix\doi{10.3847/2041-8205/832/2/L19}.
%Type = Article
\bibitem[{Ashworth(1977)}]{Ashworth1977}
\bibinfo{author}{Ashworth, J.R.}, \bibinfo{year}{1977}.
\newblock \bibinfo{title}{Matrix textures in unequilibrated ordinary
  chondrites}.
\newblock \bibinfo{journal}{Earth and Planetary Science Letters}
  \bibinfo{volume}{35}, \bibinfo{pages}{25--34}.
\newblock \DOIprefix\doi{10.1016/0012-821X(77)90024-3}.
%Type = Article
\bibitem[{Beitz et~al.(2013)Beitz, Blum, Mathieu, Pack and
  Hezel}]{Beitz.etal2013}
\bibinfo{author}{Beitz, E.}, \bibinfo{author}{Blum, J.},
  \bibinfo{author}{Mathieu, R.}, \bibinfo{author}{Pack, A.},
  \bibinfo{author}{Hezel, D.C.}, \bibinfo{year}{2013}.
\newblock \bibinfo{title}{Experimental investigation of the nebular formation
  of chondrule rims and the formation of chondrite parent bodies}.
\newblock \bibinfo{journal}{Geochimica et Cosmochimica Acta}
  \bibinfo{volume}{116}, \bibinfo{pages}{41--51}.
\newblock \DOIprefix\doi{10.1016/j.gca.2012.04.059}.
%Type = Article
\bibitem[{Bland et~al.(2011)Bland, Howard, Prior, Wheeler, Hough and
  Dyl}]{Bland.etal2011}
\bibinfo{author}{Bland, P.A.}, \bibinfo{author}{Howard, L.E.},
  \bibinfo{author}{Prior, D.J.}, \bibinfo{author}{Wheeler, J.},
  \bibinfo{author}{Hough, R.M.}, \bibinfo{author}{Dyl, K.A.},
  \bibinfo{year}{2011}.
\newblock \bibinfo{title}{Earliest rock fabric formed in the {{Solar System}}
  preserved in a chondrule rim}.
\newblock \bibinfo{journal}{Nature Geoscience} \bibinfo{volume}{4},
  \bibinfo{pages}{244--247}.
\newblock \DOIprefix\doi{10.1038/ngeo1120}.
%Type = Article
\bibitem[{Brearley(1993)}]{Brearley1993}
\bibinfo{author}{Brearley, A.J.}, \bibinfo{year}{1993}.
\newblock \bibinfo{title}{Matrix and fine-grained rims in the unequilibrated
  {{CO3}} chondrite, {{ALHA77307}} - {{Origins}} and evidence for diverse,
  primitive nebular dust components}.
\newblock \bibinfo{journal}{Geochimica et Cosmochimica Acta}
  \bibinfo{volume}{57}, \bibinfo{pages}{1521--1550}.
\newblock \DOIprefix\doi{10.1016/0016-7037(93)90011-K}.
%Type = Article
\bibitem[{Brearley et~al.(1999)Brearley, Hanowski and
  Whalen}]{Brearley.etal1999a}
\bibinfo{author}{Brearley, A.J.}, \bibinfo{author}{Hanowski, N.P.},
  \bibinfo{author}{Whalen, J.F.}, \bibinfo{year}{1999}.
\newblock \bibinfo{title}{Fine-{{Grained Rims}} in {{CM Carbonaceous
  Chondrites}}: {{A Comparison}} of {{Rims}} in {{Murchison}} and {{ALH}}
  81002} , \bibinfo{pages}{1460}.
%Type = Article
\bibitem[{Carballido(2011)}]{Carballido2011}
\bibinfo{author}{Carballido, A.}, \bibinfo{year}{2011}.
\newblock \bibinfo{title}{Accretion of dust by chondrules in a
  {{MHD}}-turbulent solar nebula}.
\newblock \bibinfo{journal}{Icarus} \bibinfo{volume}{211},
  \bibinfo{pages}{876--884}.
\newblock \DOIprefix\doi{10.1016/j.icarus.2010.10.018}.
%Type = Article
\bibitem[{Cuzzi(2004)}]{Cuzzi2004}
\bibinfo{author}{Cuzzi, J.N.}, \bibinfo{year}{2004}.
\newblock \bibinfo{title}{Blowing in the wind: {{III}}. {{Accretion}} of dust
  rims by chondrule-sized particles in a turbulent protoplanetary nebula}.
\newblock \bibinfo{journal}{Icarus} \bibinfo{volume}{168},
  \bibinfo{pages}{484--497}.
\newblock \DOIprefix\doi{10.1016/j.icarus.2003.12.008}.
%Type = Article
\bibitem[{Dominik and Tielens(1995)}]{Dominik.Tielens1995}
\bibinfo{author}{Dominik, C.}, \bibinfo{author}{Tielens, A.G.G.M.},
  \bibinfo{year}{1995}.
\newblock \bibinfo{title}{Resistance to rolling in the adhesive contact of two
  elastic spheres}.
\newblock \bibinfo{journal}{Philosophical Magazine, Part A}
  \bibinfo{volume}{72}, \bibinfo{pages}{783--803}.
\newblock \DOIprefix\doi{10.1080/01418619508243800}.
%Type = Article
\bibitem[{Dominik and Tielens(1997)}]{Dominik.Tielens1997}
\bibinfo{author}{Dominik, C.}, \bibinfo{author}{Tielens, A.G.G.M.},
  \bibinfo{year}{1997}.
\newblock \bibinfo{title}{The {{Physics}} of {{Dust Coagulation}} and the
  {{Structure}} of {{Dust Aggregates}} in {{Space}}}.
\newblock \bibinfo{journal}{The Astrophysical Journal} \bibinfo{volume}{480},
  \bibinfo{pages}{647--673}.
\newblock \DOIprefix\doi{10.1086/303996}.
%Type = Article
\bibitem[{Gunkelmann et~al.(2017)Gunkelmann, Kataoka, Dullemond and
  Urbassek}]{Gunkelmann.etal2017}
\bibinfo{author}{Gunkelmann, N.}, \bibinfo{author}{Kataoka, A.},
  \bibinfo{author}{Dullemond, C.P.}, \bibinfo{author}{Urbassek, H.M.},
  \bibinfo{year}{2017}.
\newblock \bibinfo{title}{Low-velocity collisions of chondrules: {{How}} a thin
  dust cover helps enhance the sticking probability}.
\newblock \bibinfo{journal}{Astronomy and Astrophysics} \bibinfo{volume}{599},
  \bibinfo{pages}{L4}.
\newblock \DOIprefix\doi{10.1051/0004-6361/201630155}.
%Type = Article
\bibitem[{Haenecour et~al.(2018)Haenecour, Floss, Zega, Croat, Wang, Jolliff
  and Carpenter}]{Haenecour.etal2018}
\bibinfo{author}{Haenecour, P.}, \bibinfo{author}{Floss, C.},
  \bibinfo{author}{Zega, T.J.}, \bibinfo{author}{Croat, T.K.},
  \bibinfo{author}{Wang, A.}, \bibinfo{author}{Jolliff, B.L.},
  \bibinfo{author}{Carpenter, P.}, \bibinfo{year}{2018}.
\newblock \bibinfo{title}{Presolar silicates in the matrix and fine-grained
  rims around chondrules in primitive {{CO3}}.0 chondrites: {{Evidence}} for
  pre-accretionary aqueous alteration of the rims in the solar nebula}.
\newblock \bibinfo{journal}{Geochimica et Cosmochimica Acta}
  \bibinfo{volume}{221}, \bibinfo{pages}{379--405}.
\newblock \DOIprefix\doi{10.1016/j.gca.2017.06.004}.
%Type = Article
\bibitem[{Hanna and Ketcham(2018)}]{Hanna.Ketcham2018}
\bibinfo{author}{Hanna, R.D.}, \bibinfo{author}{Ketcham, R.A.},
  \bibinfo{year}{2018}.
\newblock \bibinfo{title}{Evidence for accretion of fine-grained rims in a
  turbulent nebula for {{CM Murchison}}}.
\newblock \bibinfo{journal}{Earth and Planetary Science Letters}
  \bibinfo{volume}{481}, \bibinfo{pages}{201--211}.
\newblock \DOIprefix\doi{10.1016/j.epsl.2017.10.029}.
%Type = Article
\bibitem[{Hayashi(1981)}]{Hayashi1981}
\bibinfo{author}{Hayashi, C.}, \bibinfo{year}{1981}.
\newblock \bibinfo{title}{Structure of the {{Solar Nebula}}, {{Growth}} and
  {{Decay}} of {{Magnetic Fields}} and {{Effects}} of {{Magnetic}} and
  {{Turbulent Viscosities}} on the {{Nebula}}}.
\newblock \bibinfo{journal}{Progress of Theoretical Physics Supplement}
  \bibinfo{volume}{70}, \bibinfo{pages}{35--53}.
\newblock \DOIprefix\doi{10.1143/PTPS.70.35}.
%Type = Article
\bibitem[{Lee(2000)}]{Lee2000}
\bibinfo{author}{Lee, M.H.}, \bibinfo{year}{2000}.
\newblock \bibinfo{title}{On the {{Validity}} of the {{Coagulation Equation}}
  and the {{Nature}} of {{Runaway Growth}}}.
\newblock \bibinfo{journal}{Icarus} \bibinfo{volume}{143},
  \bibinfo{pages}{74--86}.
\newblock \DOIprefix\doi{10.1006/icar.1999.6239}.
%Type = Article
\bibitem[{Leitner et~al.(2016)Leitner, Vollmer, Floss, Zipfel and
  Hoppe}]{Leitner.etal2016}
\bibinfo{author}{Leitner, J.}, \bibinfo{author}{Vollmer, C.},
  \bibinfo{author}{Floss, C.}, \bibinfo{author}{Zipfel, J.},
  \bibinfo{author}{Hoppe, P.}, \bibinfo{year}{2016}.
\newblock \bibinfo{title}{Ancient stardust in fine-grained chondrule dust rims
  from carbonaceous chondrites}.
\newblock \bibinfo{journal}{Earth and Planetary Science Letters}
  \bibinfo{volume}{434}, \bibinfo{pages}{117--128}.
\newblock \DOIprefix\doi{10.1016/j.epsl.2015.11.028}.
%Type = Article
\bibitem[{Liffman(2019)}]{Liffman2019}
\bibinfo{author}{Liffman, K.}, \bibinfo{year}{2019}.
\newblock \bibinfo{title}{Fine-grained rim formation - {{High}} speed, kinetic
  dust aggregation in the early {{Solar System}}}.
\newblock \bibinfo{journal}{Geochimica et Cosmochimica Acta}
  \bibinfo{volume}{264}, \bibinfo{pages}{118--129}.
\newblock \DOIprefix\doi{10.1016/j.gca.2019.08.009}.
%Type = Article
\bibitem[{Mathis et~al.(1977)Mathis, Rumpl and Nordsieck}]{Mathis.etal1977}
\bibinfo{author}{Mathis, J.S.}, \bibinfo{author}{Rumpl, W.},
  \bibinfo{author}{Nordsieck, K.H.}, \bibinfo{year}{1977}.
\newblock \bibinfo{title}{The size distribution of interstellar grains}.
\newblock \bibinfo{journal}{The Astrophysical Journal} \bibinfo{volume}{217},
  \bibinfo{pages}{425--433}.
\newblock \DOIprefix\doi{10.1086/155591}.
%Type = Article
\bibitem[{Matsumoto et~al.(2021)Matsumoto, Hasegawa, Matsuda and
  Liu}]{Matsumoto.etal2021}
\bibinfo{author}{Matsumoto, Y.}, \bibinfo{author}{Hasegawa, Y.},
  \bibinfo{author}{Matsuda, N.}, \bibinfo{author}{Liu, M.C.},
  \bibinfo{year}{2021}.
\newblock \bibinfo{title}{Formation of rims around chondrules via porous
  aggregate accretion}.
\newblock \bibinfo{journal}{Icarus} \bibinfo{volume}{367},
  \bibinfo{pages}{114538}.
\newblock \DOIprefix\doi{10.1016/j.icarus.2021.114538}.
%Type = Article
\bibitem[{Matsumoto et~al.(2019)Matsumoto, Wakita, Hasegawa and
  Oshino}]{Matsumoto.etal2019}
\bibinfo{author}{Matsumoto, Y.}, \bibinfo{author}{Wakita, S.},
  \bibinfo{author}{Hasegawa, Y.}, \bibinfo{author}{Oshino, S.},
  \bibinfo{year}{2019}.
\newblock \bibinfo{title}{Aggregate {{Growth}} and {{Internal Structures}} of
  {{Chondrite Parent Bodies Forming}} from {{Dense Clumps}}}.
\newblock \bibinfo{journal}{Astrophys. J.} \bibinfo{volume}{887},
  \bibinfo{pages}{248}.
\newblock \DOIprefix\doi{10.3847/1538-4357/ab5b06}.
%Type = Article
\bibitem[{Metzler et~al.(1992)Metzler, Bischoff and
  Stoeffler}]{Metzler.etal1992}
\bibinfo{author}{Metzler, K.}, \bibinfo{author}{Bischoff, A.},
  \bibinfo{author}{Stoeffler, D.}, \bibinfo{year}{1992}.
\newblock \bibinfo{title}{Accretionary dust mantles in {{CM}} chondrites -
  {{Evidence}} for solar nebula processes}.
\newblock \bibinfo{journal}{Geochimica et Cosmochimica Acta}
  \bibinfo{volume}{56}, \bibinfo{pages}{2873--2897}.
\newblock \DOIprefix\doi{10.1016/0016-7037(92)90365-P}.
%Type = Article
\bibitem[{Miura et~al.(2010)Miura, Tanaka, Yamamoto, Nakamoto, Yamada,
  Tsukamoto and Nozawa}]{Miura.etal2010}
\bibinfo{author}{Miura, H.}, \bibinfo{author}{Tanaka, K.K.},
  \bibinfo{author}{Yamamoto, T.}, \bibinfo{author}{Nakamoto, T.},
  \bibinfo{author}{Yamada, J.}, \bibinfo{author}{Tsukamoto, K.},
  \bibinfo{author}{Nozawa, J.}, \bibinfo{year}{2010}.
\newblock \bibinfo{title}{Formation of {{Cosmic Crystals}} in {{Highly
  Supersaturated Silicate Vapor Produced}} by {{Planetesimal Bow Shocks}}}.
\newblock \bibinfo{journal}{The Astrophysical Journal} \bibinfo{volume}{719},
  \bibinfo{pages}{642--654}.
\newblock \DOIprefix\doi{10.1088/0004-637X/719/1/642}.
%Type = Article
\bibitem[{Morfill et~al.(1998)Morfill, Durisen and Turner}]{Morfill.etal1998}
\bibinfo{author}{Morfill, G.E.}, \bibinfo{author}{Durisen, R.H.},
  \bibinfo{author}{Turner, G.W.}, \bibinfo{year}{1998}.
\newblock \bibinfo{title}{{{NOTE}}: An {{Accretion Rim Constraint}} on
  {{Chondrule Formation Theories}}}.
\newblock \bibinfo{journal}{Icarus} \bibinfo{volume}{134},
  \bibinfo{pages}{180--184}.
\newblock \DOIprefix\doi{10.1006/icar.1998.5948}.
%Type = Article
\bibitem[{Okuzumi et~al.(2009)Okuzumi, Tanaka and Sakagami}]{Okuzumi.etal2009}
\bibinfo{author}{Okuzumi, S.}, \bibinfo{author}{Tanaka, H.},
  \bibinfo{author}{Sakagami, M.a.}, \bibinfo{year}{2009}.
\newblock \bibinfo{title}{Numerical {{Modeling}} of the {{Coagulation}} and
  {{Porosity Evolution}} of {{Dust Aggregates}}}.
\newblock \bibinfo{journal}{The Astrophysical Journal} \bibinfo{volume}{707},
  \bibinfo{pages}{1247--1263}.
\newblock \DOIprefix\doi{10.1088/0004-637X/707/2/1247}.
%Type = Article
\bibitem[{Ormel and Cuzzi(2007)}]{Ormel.Cuzzi2007}
\bibinfo{author}{Ormel, C.W.}, \bibinfo{author}{Cuzzi, J.N.},
  \bibinfo{year}{2007}.
\newblock \bibinfo{title}{Closed-form expressions for particle relative
  velocities induced by turbulence}.
\newblock \bibinfo{journal}{Astronomy and Astrophysics} \bibinfo{volume}{466},
  \bibinfo{pages}{413--420}.
\newblock \DOIprefix\doi{10.1051/0004-6361:20066899}.
%Type = Article
\bibitem[{Ormel et~al.(2008)Ormel, Cuzzi and Tielens}]{Ormel.etal2008}
\bibinfo{author}{Ormel, C.W.}, \bibinfo{author}{Cuzzi, J.N.},
  \bibinfo{author}{Tielens, A.G.G.M.}, \bibinfo{year}{2008}.
\newblock \bibinfo{title}{Co-{{Accretion}} of {{Chondrules}} and {{Dust}} in
  the {{Solar Nebula}}}.
\newblock \bibinfo{journal}{ApJ} \bibinfo{volume}{679},
  \bibinfo{pages}{1588--1610}.
\newblock \DOIprefix\doi{10.1086/587836}.
%Type = Article
\bibitem[{Sears et~al.(1993)Sears, Benoit and Jie}]{Sears.etal1993}
\bibinfo{author}{Sears, D.W.G.}, \bibinfo{author}{Benoit, P.H.},
  \bibinfo{author}{Jie, L.}, \bibinfo{year}{1993}.
\newblock \bibinfo{title}{Two chondrule groups each with distinctive rims in
  {{Murchison}} recognized by cathodoluminescence}.
\newblock \bibinfo{journal}{Meteoritics} \bibinfo{volume}{28},
  \bibinfo{pages}{669--675}.
\newblock \DOIprefix\doi{10.1111/j.1945-5100.1993.tb00638.x}.
%Type = Article
\bibitem[{Shakura and Sunyaev(1973)}]{Shakura.Sunyaev1973}
\bibinfo{author}{Shakura, N.I.}, \bibinfo{author}{Sunyaev, R.A.},
  \bibinfo{year}{1973}.
\newblock \bibinfo{title}{Black holes in binary systems. {{Observational}}
  appearance.}
\newblock \bibinfo{journal}{Astronomy and Astrophysics} \bibinfo{volume}{24},
  \bibinfo{pages}{337--355}.
%Type = Article
\bibitem[{Simon et~al.(2018)Simon, Cuzzi, McCain, Cato, Christoffersen, Fisher,
  Srinivasan, Tait, Olson and Scargle}]{Simon.etal2018}
\bibinfo{author}{Simon, J.I.}, \bibinfo{author}{Cuzzi, J.N.},
  \bibinfo{author}{McCain, K.A.}, \bibinfo{author}{Cato, M.J.},
  \bibinfo{author}{Christoffersen, P.A.}, \bibinfo{author}{Fisher, K.R.},
  \bibinfo{author}{Srinivasan, P.}, \bibinfo{author}{Tait, A.W.},
  \bibinfo{author}{Olson, D.M.}, \bibinfo{author}{Scargle, J.D.},
  \bibinfo{year}{2018}.
\newblock \bibinfo{title}{Particle size distributions in chondritic meteorites:
  {{Evidence}} for pre-planetesimal histories}.
\newblock \bibinfo{journal}{Earth and Planetary Science Letters}
  \bibinfo{volume}{494}, \bibinfo{pages}{69--82}.
\newblock \DOIprefix\doi{10.1016/j.epsl.2018.04.021}.
%Type = Article
\bibitem[{Takayama and Tomeoka(2012)}]{Takayama.Tomeoka2012}
\bibinfo{author}{Takayama, A.}, \bibinfo{author}{Tomeoka, K.},
  \bibinfo{year}{2012}.
\newblock \bibinfo{title}{Fine-grained rims surrounding chondrules in the
  {{Tagish Lake}} carbonaceous chondrite: {{Verification}} of their formation
  through parent-body processes}.
\newblock \bibinfo{journal}{Geochimica et Cosmochimica Acta}
  \bibinfo{volume}{98}, \bibinfo{pages}{1--18}.
\newblock \DOIprefix\doi{10.1016/j.gca.2012.08.015}.
%Type = Article
\bibitem[{Toriumi(1989)}]{Toriumi1989}
\bibinfo{author}{Toriumi, M.}, \bibinfo{year}{1989}.
\newblock \bibinfo{title}{Grain size distribution of the matrix in the
  {{Allende}} chondrite}.
\newblock \bibinfo{journal}{Earth and Planetary Science Letters}
  \bibinfo{volume}{92}, \bibinfo{pages}{265--273}.
\newblock \DOIprefix\doi{10.1016/0012-821X(89)90051-4}.
%Type = Article
\bibitem[{{Trigo-Rodriguez} et~al.(2006){Trigo-Rodriguez}, Rubin and
  Wasson}]{Trigo-Rodriguez.etal2006}
\bibinfo{author}{{Trigo-Rodriguez}, J.M.}, \bibinfo{author}{Rubin, A.E.},
  \bibinfo{author}{Wasson, J.T.}, \bibinfo{year}{2006}.
\newblock \bibinfo{title}{Non-nebular origin of dark mantles around chondrules
  and inclusions in {{CM}} chondrites}.
\newblock \bibinfo{journal}{Geochimica et Cosmochimica Acta}
  \bibinfo{volume}{70}, \bibinfo{pages}{1271--1290}.
\newblock \DOIprefix\doi{10.1016/j.gca.2005.11.009}.
%Type = Article
\bibitem[{Umst{\"a}tter and Urbassek(2021)}]{Umstatter.Urbassek2021a}
\bibinfo{author}{Umst{\"a}tter, P.}, \bibinfo{author}{Urbassek, H.M.},
  \bibinfo{year}{2021}.
\newblock \bibinfo{title}{Granular mechanics simulations of collisions between
  chondritic aggregates}.
\newblock \bibinfo{journal}{Astron. Amp Astrophys. Vol. 652 IdA40
  NUMPAGES7NUMPAGES Pp} \bibinfo{volume}{652}, \bibinfo{pages}{A40}.
\newblock \DOIprefix\doi{10.1051/0004-6361/202141581}.
%Type = Article
\bibitem[{Xiang et~al.(2019)Xiang, Carballido, Hanna, Matthews and
  Hyde}]{Xiang.etal2019}
\bibinfo{author}{Xiang, C.}, \bibinfo{author}{Carballido, A.},
  \bibinfo{author}{Hanna, R.D.}, \bibinfo{author}{Matthews, L.S.},
  \bibinfo{author}{Hyde, T.W.}, \bibinfo{year}{2019}.
\newblock \bibinfo{title}{The initial structure of chondrule dust rims {{I}}:
  {{Electrically}} neutral grains}.
\newblock \bibinfo{journal}{Icarus} \bibinfo{volume}{321},
  \bibinfo{pages}{99--111}.
\newblock \DOIprefix\doi{10.1016/j.icarus.2018.10.014}.
%Type = Article
\bibitem[{Zanetta et~al.(2021)Zanetta, Leroux, Le~Guillou, Zanda and
  Hewins}]{Zanetta.etal2021}
\bibinfo{author}{Zanetta, P.M.}, \bibinfo{author}{Leroux, H.},
  \bibinfo{author}{Le~Guillou, C.}, \bibinfo{author}{Zanda, B.},
  \bibinfo{author}{Hewins, R.H.}, \bibinfo{year}{2021}.
\newblock \bibinfo{title}{Nebular thermal processing of accretionary
  fine-grained rims in the {{Paris CM}} chondrite}.
\newblock \bibinfo{journal}{Geochimica et Cosmochimica Acta}
  \bibinfo{volume}{295}, \bibinfo{pages}{135--154}.
\newblock \DOIprefix\doi{10.1016/j.gca.2020.12.015}.
%Type = Article
\bibitem[{Zega and Buseck(2003)}]{Zega.Buseck2003a}
\bibinfo{author}{Zega, T.J.}, \bibinfo{author}{Buseck, P.R.},
  \bibinfo{year}{2003}.
\newblock \bibinfo{title}{Fine-grained-rim mineralogy of the {{Cold Bokkeveld
  CM}} chondrite}.
\newblock \bibinfo{journal}{Geochim. Cosmochim. Acta} \bibinfo{volume}{67},
  \bibinfo{pages}{1711--1721}.
\newblock \DOIprefix\doi{10.1016/S0016-7037(02)01172-9}.

\end{thebibliography}

\appendix
\renewcommand{\thetable}{\Alph{section}.\arabic{table}}
\renewcommand{\thefigure}{\Alph{section}.\arabic{figure}}
\setcounter{figure}{0}
\addappheadtotoc

\clearpage

\color{red}

\section{Resolution test}
\label{resolution test}
Here we show an example of the tests on the accuracy of our numerical simulations for the aggregation of monodisperse monomer grains by the Brownian motion, which can be compared with the previous work by \citet{Okuzumi.etal2009}. We conduct more accurate simulations with the ratio of $10^{1/20}$ between the two adjacent bins for both mass and filling factor. 

Before the comparisons, we need to explain two schemes of the assignments of the new aggregates formed through the collisions to the bins. The total mass of the new aggregates with the mass of $m_{12}=m_{1}+m_{2}$ and filling factor of $\phi_{12}=m_{12}/(V_{1}+V_{2}+V_{\mathrm{void}})\rho_{\mathrm{mat}}$ formed by the collisions between two dust particles must be assigned to the four adjacent bins of $(m[i_{12}], \phi[j_{12}])$, $(m[i_{12}+1], \phi[j_{12}])$, $(m[i_{12}], \phi[j_{12}+1])$ and $(m[i_{12}+1], \phi[j_{12}+1])$, where $m[i_{12}]\leq m_{12} < m[i_{12}+1]$ and $\phi[j_{12}+1] < \phi_{12} \leq \phi[j_{12}]$. We define two coefficients, $h_{m,12}$ and $h_{\phi,12}$, as,
\begin{align}
h_{m,12}=&(m[i_{12}+1]-m_{12})/(m[i_{12}+1]-m[i_{12}]), \notag \\
h_{\phi,12}=&(1/\phi[j_{12}+1]-1/\phi_{12})/(1/\phi[j_{12}+1]-1/\phi[j_{12}]).
\end{align}
The first scheme, which we used to obtain the main results, distribute the mass $m_{12}$ of new aggregates into the four bins as shown in Table. \ref{table_scheme1}. The second scheme distributes the mass $m_{12}$ of new aggregates the four bins as shown in Table. \ref{table_scheme2}. Both schemes preserve the total mass and volume;
\begin{align}
h_{m,12}m[i_{12}]+(1-h_{m,12})m[i_{12}+1] = & m[i_{12}+1] - h_{m,12}(m[i_{12}+1]-m[i_{12}]) \notag \\
= & m[i_{12}+1] - \left( \frac{m[i_{12}+1]-m_{12}}{m[i_{12}+1]-m[i_{12}]} \right) (m[i_{12}+1]-m[i_{12}]) \notag \\
= & m_{12}, \\
\frac{h_{\phi,12}m_{12}}{\phi[j_{12}]}+\frac{(1-h_{\phi,12})m_{12}}{\phi[j_{12}+1]} = & \frac{m_{12}}{\phi[j_{12}+1]} -  h_{\phi,12}m_{12} \left( \frac{1}{\phi[j_{12}+1]} - \frac{1}{\phi[j_{12}]} \right) \notag \\
= & \frac{m_{12}}{\phi[j_{12}+1]} -  m_{12} \left(  \frac{1/\phi[j_{12}+1]-1/\phi_{12}}{1/\phi[j_{12}+1]-1/\phi[j_{12}]}  \right) \left( \frac{1}{\phi[j_{12}+1]} - \frac{1}{\phi[j_{12}]} \right) \notag \\
= & \frac{m_{12}}{\phi_{12}}.
\end{align}

\begin{figure}[H]
\begin{tabular}{cc}
\begin{minipage}[t]{0.5\hsize}
  \centering
  \includegraphics[width=8.5cm,pagebox=cropbox,clip]{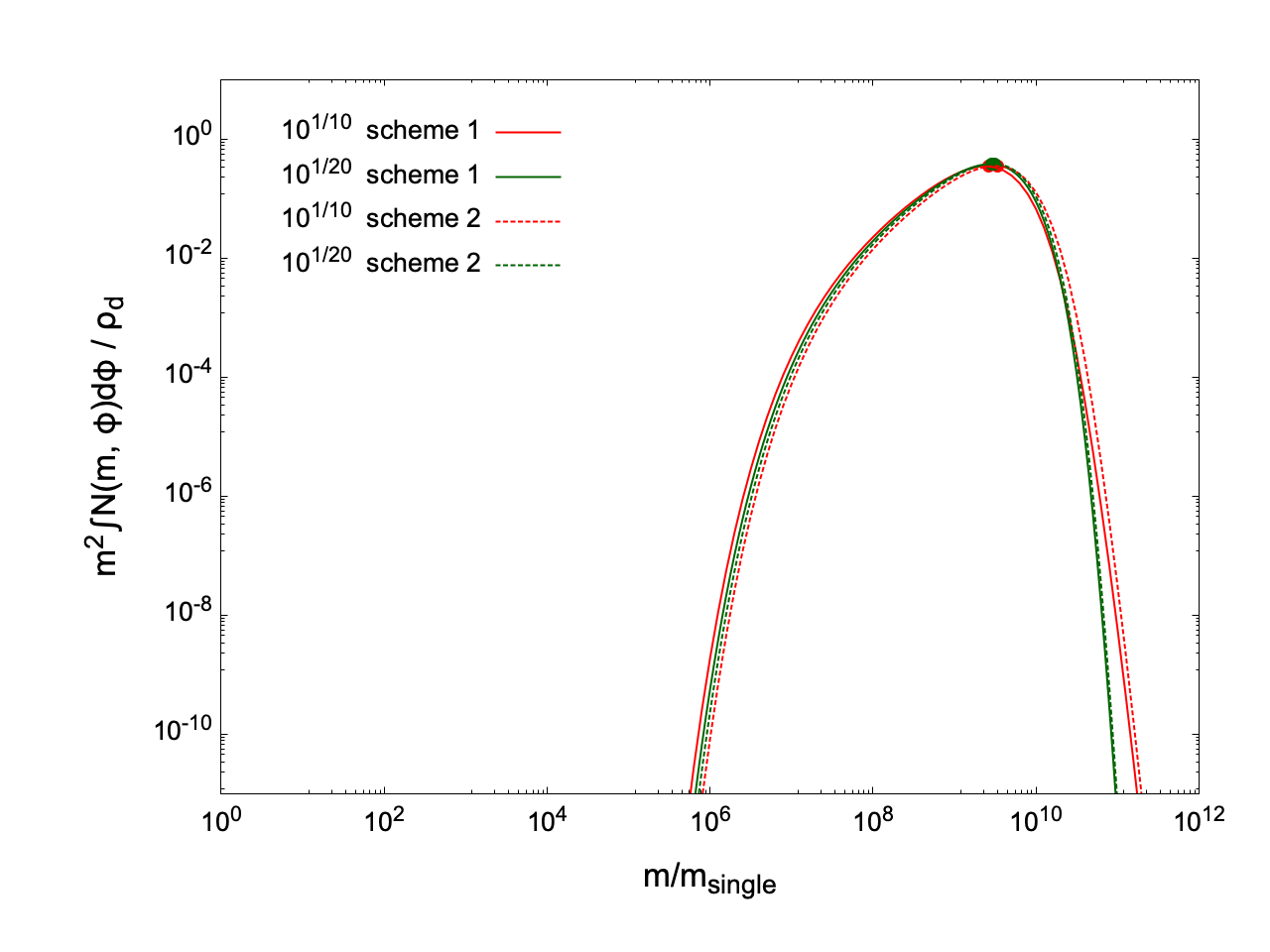}
  \subcaption{mass distribution when $t=10^{4}t_{\mathrm{B0}}=4\sqrt{2}\times 10^{4}t_{\mathrm{single}}$}
  \label{mass distribution single app}
\end{minipage}
\begin{minipage}[t]{0.5\hsize}
  \centering
  \includegraphics[width=8.5cm,pagebox=cropbox,clip]{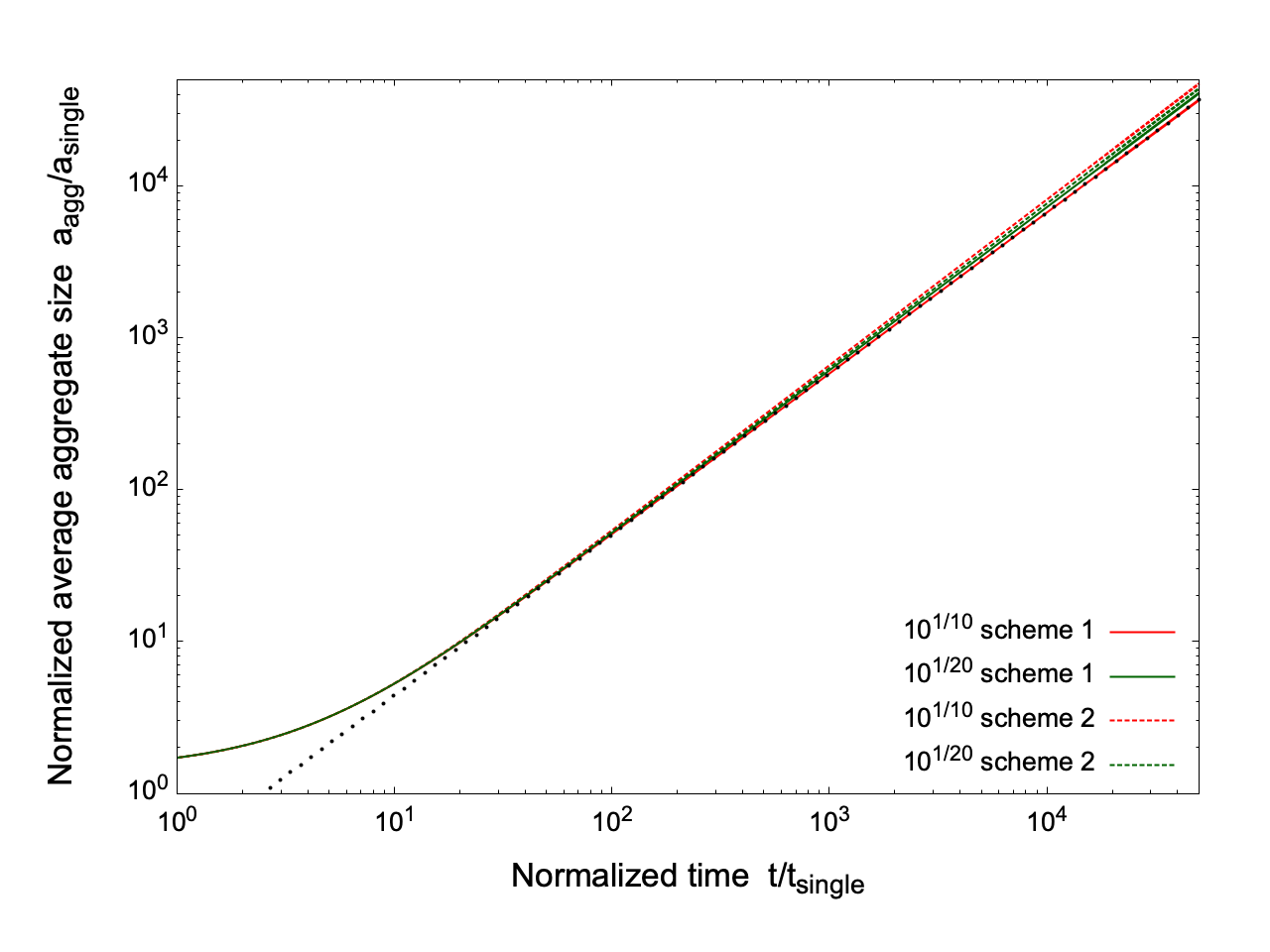}
  \subcaption{Average size of aggregates}
  \label{size evolution single app}
\end{minipage}
\end{tabular}
\caption{Comparison of the results for monodisperse monomer grains. (a) Mass distribution when $t=10^{4}t_{\mathrm{B0}}=4\sqrt{2}\times 10^{4}t_{\mathrm{single}}$, which can be compared with Figure 9 in \citet{Okuzumi.etal2009}. The circles indicate the mass-weighted average mass, $\iint m^{2}N(m,\phi)dmd\phi / \rho_{d}$. (b) Evolution of the average size of aggregates. The black dotted lines show the fitting formula in Eq. \ref{fit-a-single}.}
\label{test-single}
\end{figure}

Then, we compare the results for the case of monodisperse monomer grains. These results can be compared with the results from \citet{Okuzumi.etal2009}, especially their Figure 9. Figure \ref{mass distribution single app} shows the mass distribution, $m^{2} \int N(m,\phi)d\phi$, of dust particles when $t=10^{4}t_{\mathrm{B0}}=4\sqrt{2}\times 10^{4}t_{\mathrm{single}}$. \citet{Okuzumi.etal2009} used the different unit time from here, $t_{\mathrm{B0}}=(\pi a^{2} \times \rho_{d}/m_{\mathrm{gr}}(a) \times \sqrt{8k_{b}T / \pi m_{\mathrm{gr}}(a)})^{-1}$. Although the less accurate simulations show the increase in mass at the high mass tail, they almost match with the results from the more accurate calculations and \citet{Okuzumi.etal2009}. Figure \ref{size evolution single app} shows the evolution of the average size of aggregates. The numerical differences increase with time, but the differences of $a_{\mathrm{agg}}$ between the simulations with the higher and lower resolutions are at most around $10$ to $20\%$ within the simulation time of our study and do not affect our conclusions so much. Moreover, the uncertainties in the gas surface density and the turbulent intensity in the solar nebula affect much more than such numerical differences. 

\clearpage

\begin{table}[H]
\caption{The assignment of the mass $m_{12}$ of the new aggregates to the four adjacent bins by the first scheme.}
 \label{table_scheme1}
 \centering
 {\footnotesize
  \begin{tabular}{|c|c|r|}
   \hline
   \multicolumn{1}{|c|}{} & \multicolumn{1}{|c|}{$\phi[j_{12}]$} & \multicolumn{1}{|c|}{$\phi[j_{12}+1]$} \\ \hline 
   $m[i_{12}]$ & $h_{m,12}h_{\phi,12}m[i_{12}]$ & $h_{m,12}(1-h_{\phi,12})m[i_{12}]$ \ \ \ \ \  \\ \hline
   $m[i_{12}+1]$ & \ \ \ $(1-h_{m,12})h_{\phi,12}m[i_{12}+1]$ \ \ \ & $(1-h_{m,12})(1-h_{\phi,12})m[i_{12}+1]$  \\ \hline
  \end{tabular}
 }
\end{table}

\begin{table}[H]
\caption{The assignment of the mass $m_{12}$ of the new aggregates to the four adjacent bins by the second scheme.}
 \label{table_scheme2}
 \centering
 {\footnotesize
  \begin{tabular}{|c|c|r|}
   \hline
   \multicolumn{1}{|c|}{} & \multicolumn{1}{|c|}{$\phi[j_{12}]$} & \multicolumn{1}{|c|}{$\phi[j_{12}+1]$} \\ \hline 
   $m[i_{12}]$ & $h_{m,12}h_{\phi,12}m_{12}$ & $h_{m,12}(1-h_{\phi,12})m_{12}$ \ \ \ \ \  \\ \hline
   $m[i_{12}+1]$ & \ \ \ \ \ \ $(1-h_{m,12})h_{\phi,12}m_{12}$ \ \ \ \ \ \ & \ \ \ $(1-h_{m,12})(1-h_{\phi,12})m_{12}$ \ \ \   \\ \hline
  \end{tabular}
 }
\end{table}

\clearpage

\color{black}

\clearpage

\setcounter{figure}{0}
\setcounter{table}{0}

\section{Fitting formulae}
\label{appendixA}
\begin{figure}[H]
\begin{tabular}{cc}
\begin{minipage}[t]{0.5\hsize}
  \centering
  \includegraphics[width=8.5cm,pagebox=cropbox,clip]{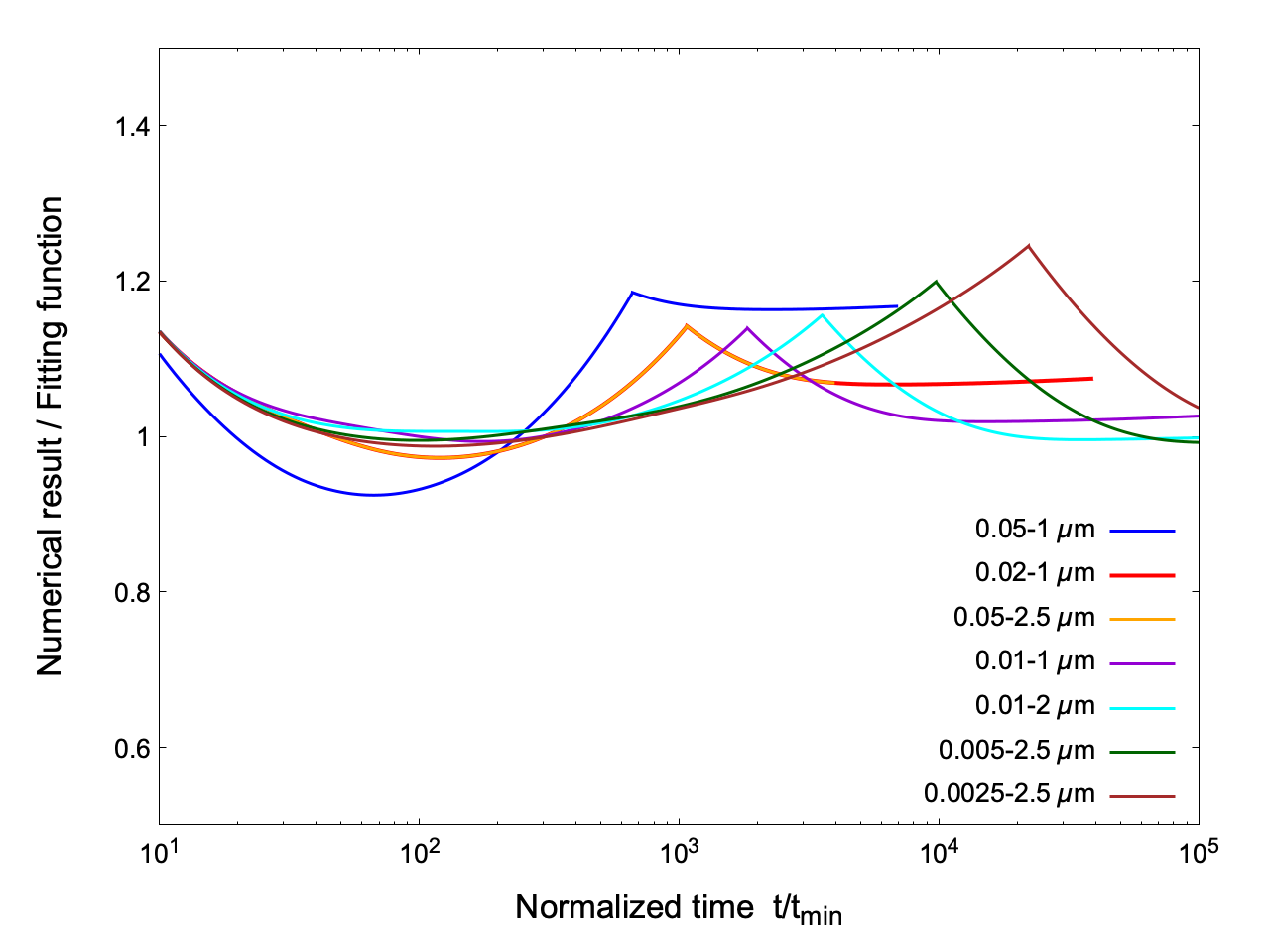}
  \subcaption{Average size of aggregates}
\end{minipage}
\begin{minipage}[t]{0.5\hsize}
  \centering
  \includegraphics[width=8.5cm,pagebox=cropbox,clip]{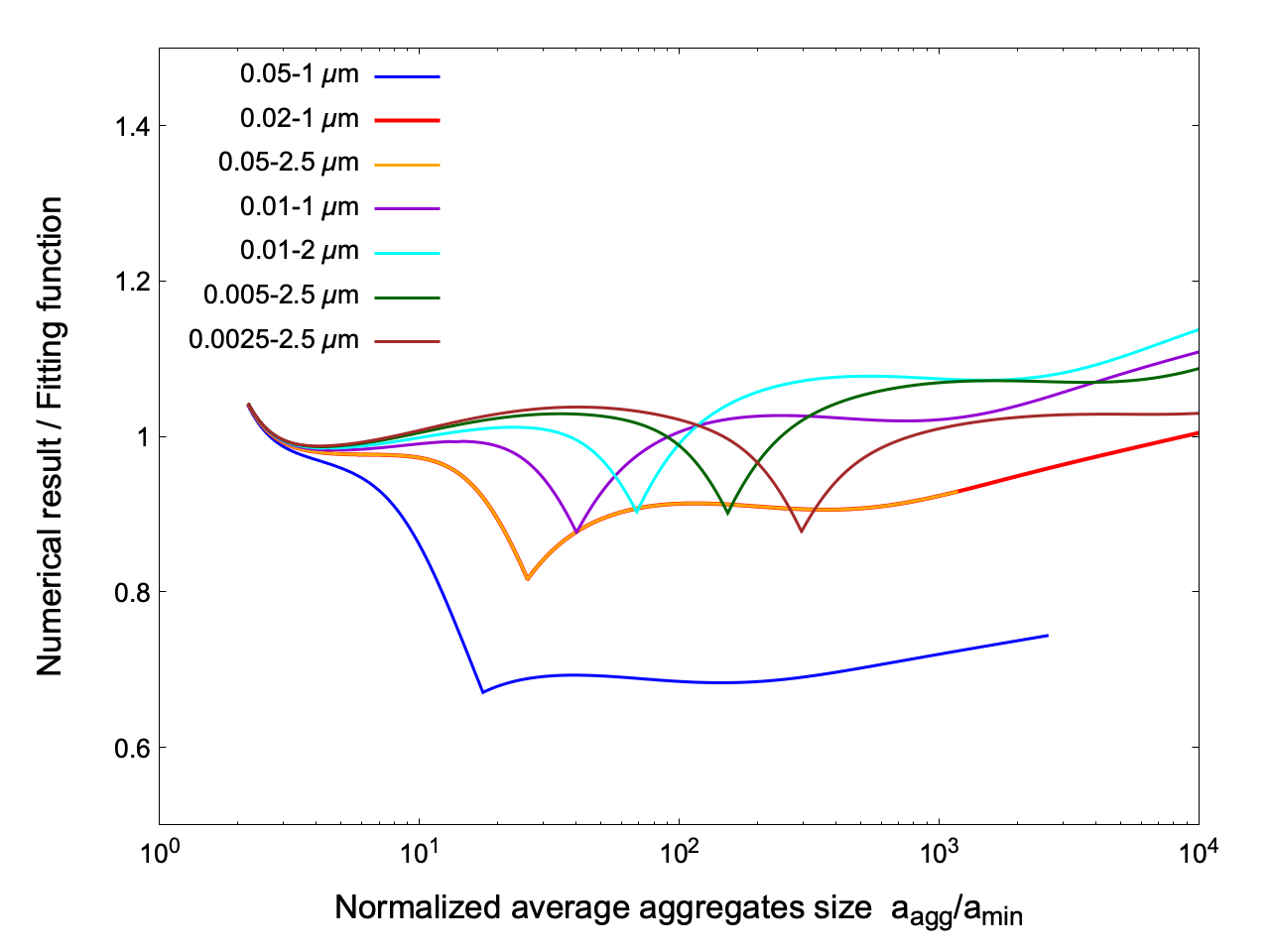}
  \subcaption{Average filling factor of aggregates}
\end{minipage}
\end{tabular}
\caption{Comparison of the numerical results with the fitting formulae. (a) Average size of aggregates. The colors of the lines represent the size distributions of the initial monomer grains, similar to that in Figure \ref{evolution of aggregates 3.5}. The vertical axis shows the ratio of the numerical results to the fitting formulae. (b)  Average fiiling factor of aggregates. }
\label{Comparison 3.5 App}
\end{figure}

\begin{figure}[H]
  \centering
  \includegraphics[width=12cm,pagebox=cropbox,clip]{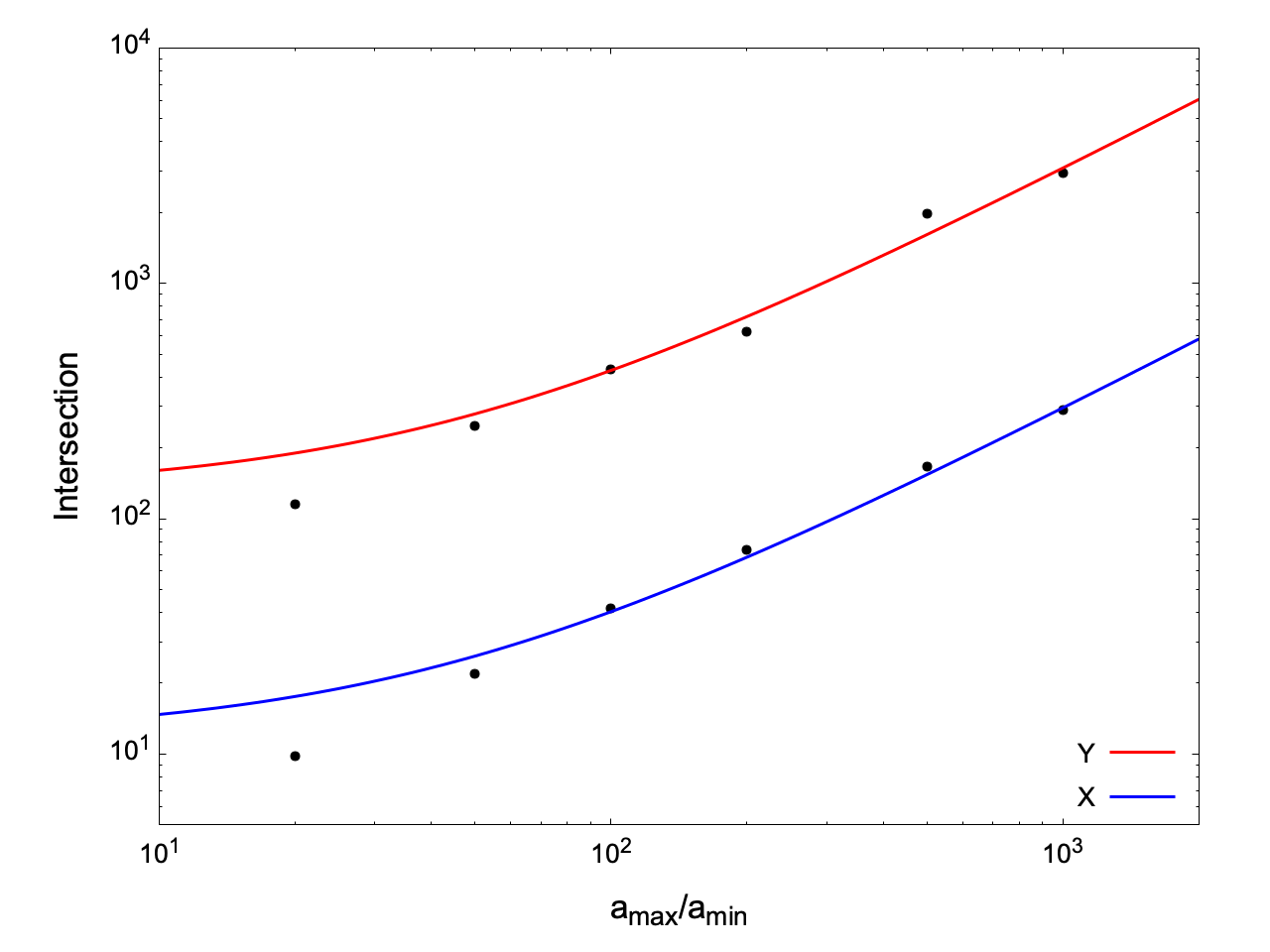}
  \caption{Comparison of the intersction between the monomer-aggregation stage and the BCCA-like stage with $Y$ (Eq. \ref{Y}) and $X$ (Eq. \ref{X}). }
  \label{YX}
\end{figure}

Using the nonlinear least-square method, we fitted the numerical results (Figure \ref{evolution of aggregates 3.5}) with exponential functions. These formulae have two parts; one represents the monomer-aggregation stage and the other represents the BCCA-like stage. We obtained the fitting formulae listed in Tables \ref{tableA1} and \ref{tableA2} from the results during $10a_{\mathrm{min}}<a_{\mathrm{agg}}<a_{\mathrm{max}}$ for the evolution of the average aggregate size, and $a_{\mathrm{min}}<a_{\mathrm{agg}}<0.4a_{\mathrm{max}}$ for the evolution of the average filling factor, which correspond to the monomer-aggregation stage, and from the results during $5a_{\mathrm{max}}<a_{\mathrm{agg}}$ for the evolution of the average aggregate size, and $3a_{\mathrm{max}}<a_{\mathrm{agg}}$ for the evolution of the average filling factor, which correspond to the BCCA-like stage. We note that the evolution of the average aggregate size lags behind that of the average filling factor. The formulae for the BCCA-like stage with various size ranges agreed well with each other. However, the formulae for the monomer-aggregation stage differed from each other. We have also provided another fitting formulae for monomer-aggregation stage obtained from the results during $10a_{\mathrm{min}}<a_{\mathrm{agg}}<100a_{\mathrm{min}}$ for the evolution of the average aggregate size, and $a_{\mathrm{min}}<a_{\mathrm{agg}}<20a_{\mathrm{min}}$ for the evolution of the average filling factor (Tables \ref{tableA3} and \ref{tableA4}). The later formulae of the monomer-aggregation stage agreed well among different size ranges, and they were also fitted well with the numerical results (Figure \ref{evolution of aggregates 3.5}). We used the later formulae in Sections 3 and 4 (Eqs. \ref{fitting1}, \ref{fitting2}). The average aggregate size ($a_{\mathrm{agg}}$) at the branch points (intersections) were linearly fitted (Eqs. \ref{Y} and \ref{X}). We compared the numerical results with the fitting formulae in Figures \ref{Comparison 3.5 App} and \ref{YX}. \color{red} The fitting formulae were fitted within around $30\%$ differences. \color{black}

\clearpage

\begin{table}[H]
\caption{Fitting formulae for the evolution of the average aggregate size. The formulae for the monomer-aggregation stage is obtained from the results during $10a_{\mathrm{min}}<a_{\mathrm{agg}}<a_{\mathrm{max}}$, and that for the BCCA-like stage is obtained from the results during $5a_{\mathrm{max}}<a_{\mathrm{agg}}$.}
 \label{tableA1}
 \centering
 {\footnotesize
  \begin{tabular}{cccc}
   \hline
   $a_{\mathrm{min}}$ &  $a_{\mathrm{max}}$ & Monomer-aggregation & BCCA-like \\
   ($\mathrm{\mu m}$) & ($\mathrm{\mu m}$) & $a_{\mathrm{agg}}/a_{\mathrm{min}}=A_{1}\left(t/t_{\mathrm{min}}\right)^{A_{2}}$ & $a_{\mathrm{agg}}/a_{\mathrm{min}}=A_{3}\left(t/t_{\mathrm{min}}\right)^{A_{4}}$ \\
   \hline \hline 
   $0.05$ & $1$ & $A_{1}=1.44, A_{2}=0.703$ & $A_{3}=0.246, A_{4}=1.05$ \\ 
   $0.02$ & $1$ & $A_{1}=1.26, A_{2}=0.758$ & $A_{3}=0.192, A_{4}=1.05$ \\ 
   $0.01$ & $1$ & $A_{1}=1.14, A_{2}=0.787$ & $A_{3}=0.157, A_{4}=1.05$ \\ 
   $0.01$ & $2$ & $A_{1}=1.04, A_{2}=0.806$ & $A_{3}=0.143, A_{4}=1.05$ \\ 
   $0.005$ & $2.5$ & $A_{1}=0.935, A_{2}=0.824$ & $A_{3}=0.112, A_{4}=1.04$ \\ 
   $0.0025$ & $2.5$ & $A_{1}=0.865, A_{2}=0.834$ & $A_{3}=0.0878, A_{4}=1.05$ \\ 
   \hline
  \end{tabular}
 }
\end{table}

\begin{table}[H]
\caption{Fitting formulae for the evolution of the average fillig factor. The formulae for the monomer-aggregation stage is obtained from the results during $a_{\mathrm{min}}<a_{\mathrm{agg}}<0.4a_{\mathrm{max}}$, and that for the  BCCA-like stage is obtained from the results during $3a_{\mathrm{max}}<a_{\mathrm{agg}}$. }
 \label{tableA2}
 \centering
 {\footnotesize
  \begin{tabular}{cccc}
   \hline
   $a_{\mathrm{min}}$ &  $a_{\mathrm{max}}$ & Monomer-aggregation & BCCA-like \\
   ($\mathrm{\mu m}$) & ($\mathrm{\mu m}$) & $\phi_{\mathrm{agg}}=B_{1}\left(a_{\mathrm{agg}}/a_{\mathrm{min}}\right)^{B_{2}}$ & $\phi_{\mathrm{agg}}=B_{3}\left(a_{\mathrm{agg}}/a_{\mathrm{min}}\right)^{B_{4}}$ \\
   \hline \hline 
   $0.05$ & $1$ & $B_{1}=0.984, B_{2}=-0.502$ & $B_{3}=3.97, B_{4}=-1.07$ \\ 
   $0.02$ & $1$ & $B_{1}=0.945, B_{2}=-0.459$ & $B_{3}=6.73, B_{4}=-1.07$ \\ 
   $0.01$ & $1$ & $B_{1}=0.934, B_{2}=-0.444$ & $B_{3}=9.97, B_{4}=-1.07$ \\ 
   $0.01$ & $2$ & $B_{1}=0.935, B_{2}=-0.438$ & $B_{3}=14.7, B_{4}=-1.07$ \\ 
   $0.005$ & $2.5$ & $B_{1}=0.931, B_{2}=-0.430$ & $B_{3}=24.4, B_{4}=-1.06$ \\ 
   $0.0025$ & $2.5$ & $B_{1}=0.931, B_{2}=-0.427$ & $B_{3}=35.1, B_{4}=-1.06$ \\ 
   \hline
  \end{tabular}
 }
\end{table}

\clearpage

\begin{table}[H]
\caption{Fitting formulae for the evolution of the average aggregate size. Here, the formulae for the monomer-aggregation stage is obtained from the results during $10a_{\mathrm{min}}<a_{\mathrm{agg}}<100a_{\mathrm{min}}$, and that for the BCCA-like stage does not change from that of Table \ref{tableA1}. For the monomer-aggregation stage of $(a_{\mathrm{min}}$, $a_{\mathrm{max}})=(0.05 \mathrm{\mu m}, 1 \mathrm{\mu m})$ and $(0.02 \mathrm{\mu m}, 1 \mathrm{\mu m})$, we show the average values of $A_{1}, A_{2}$ from the results of $(0.01 \mathrm{\mu m}, 1 \mathrm{\mu m})$ to $(0.0025 \mathrm{\mu m}, 2.5 \mathrm{\mu m})$, which were also used in Eq. \ref{fitting1}. }
 \label{tableA3}
 \centering
 {\footnotesize
  \begin{tabular}{ccccc}
   \hline
   $a_{\mathrm{min}}$ &  $a_{\mathrm{max}}$ & Monomer-aggregation & BCCA-like & Intersection \\
   ($\mathrm{\mu m}$) & ($\mathrm{\mu m}$) & $a_{\mathrm{agg}}/a_{\mathrm{min}}=A_{1}\left(t/t_{\mathrm{min}}\right)^{A_{2}}$ & $a_{\mathrm{agg}}/a_{\mathrm{min}}=A_{3}\left(t/t_{\mathrm{min}}\right)^{A_{4}}$ & $a_{\mathrm{agg}}/a_{\mathrm{min}}$ \\
   \hline \hline 
   $0.05$ & $1$ & $A_{1}=1.10, A_{2}=0.794$ & $A_{3}=0.246, A_{4}=1.05$ & 115 \\ 
   $0.02$ & $1$ & $A_{1}=1.10, A_{2}=0.794$ & $A_{3}=0.192, A_{4}=1.05$ & 247 \\ 
   $0.01$ & $1$ & $A_{1}=1.14, A_{2}=0.787$ & $A_{3}=0.157, A_{4}=1.05$ & 430 \\ 
   $0.01$ & $2$ & $A_{1}=1.12, A_{2}=0.792$ & $A_{3}=0.143, A_{4}=1.05$ & 621 \\  
   $0.005$ & $2.5$ & $A_{1}=1.07, A_{2}=0.800$ & $A_{3}=0.112, A_{4}=1.04$ & 1980 \\  
   $0.0025$ & $2.5$ & $A_{1}=1.08, A_{2}=0.797$ & $A_{3}=0.0878, A_{4}=1.05$ & 2930 \\ 
   \hline
  \end{tabular}
 }
\end{table}

\begin{table}[H]
\caption{Fitting formulae for the evolution of the average fillig factor. Here, the formulae for the monomer-aggregation stage is obtained from the results during $a_{\mathrm{min}}<a_{\mathrm{agg}}<20a_{\mathrm{min}}$, and that for the BCCA-like stage does not change from that of Table \ref{tableA2}. For the monomer-aggregation stage of $(a_{\mathrm{min}}$, $a_{\mathrm{max}})=(0.05 \mathrm{\mu m}, 1 \mathrm{\mu m})$ and $(0.02 \mathrm{\mu m}, 1 \mathrm{\mu m})$, we show the average values of $B_{1}, B_{2}$ from $(0.01 \mathrm{\mu m}, 1 \mathrm{\mu m})$ to $(0.0025 \mathrm{\mu m}, 2.5 \mathrm{\mu m})$, which were also used in Eq. \ref{fitting2}. }
 \label{tableA4}
 \centering
 {\footnotesize
  \begin{tabular}{ccccc}
   \hline
   $a_{\mathrm{min}}$ &  $a_{\mathrm{max}}$ & Monomer-aggregation & BCCA-like & Intersection \\
   ($\mathrm{\mu m}$) & ($\mathrm{\mu m}$) & $\phi_{\mathrm{agg}}=B_{1}\left(a_{\mathrm{agg}}/a_{\mathrm{min}}\right)^{B_{2}}$ & $\phi_{\mathrm{agg}}=B_{3}\left(a_{\mathrm{agg}}/a_{\mathrm{min}}\right)^{B_{4}}$ & $a_{\mathrm{agg}}/a_{\mathrm{min}}$ \\
   \hline \hline 
   $0.05$ & $1$ & $B_{1}=0.903, B_{2}=-0.420$ & $B_{3}=3.97, B_{4}=-1.07$ & 9.76 \\ 
   $0.02$ & $1$ & $B_{1}=0.903, B_{2}=-0.420$ & $B_{3}=6.73, B_{4}=-1.07$ & 22.0 \\ 
   $0.01$ & $1$ & $B_{1}=0.912, B_{2}=-0.429$ & $B_{3}=9.97, B_{4}=-1.07$ & 41.7 \\ 
   $0.01$ & $2$ & $B_{1}=0.904, B_{2}=-0.421$ & $B_{3}=14.7, B_{4}=-1.07$ & 73.5 \\ 
   $0.005$ & $2.5$ & $B_{1}=0.899, B_{2}=-0.415$ & $B_{3}=24.4, B_{4}=-1.06$ & 167 \\ 
   $0.0025$ & $2.5$ & $B_{1}=0.897, B_{2}=-0.413$ & $B_{3}=35.1, B_{4}=-1.06$ & 289 \\ 
   \hline
  \end{tabular}
 }
\end{table}

\clearpage

\clearpage

\clearpage

\section{Dust aggregation with accretion onto chondrules}
\label{appendixB}
Here, we derive Eq. \ref{fitting3}. We assumed that the average aggregate size, $a_{\mathrm{agg}}$, is related to the elapsed time as 
\begin{equation}
a_{\mathrm{agg}}/a_{0}=C\left\{\tau(t)\right\}^{\zeta},
\end{equation}
where $C$ and $\zeta$ are constants, and $\tau$ is defined as
\begin{equation}
\tau(t) = \int_0^t \frac{dt^{'}}{t_{\mathrm{coll}}(t^{'})} \ .
\label{tau}
\end{equation}
For example, during the monomer-aggregation stage for the case of $\beta=-3.5$ and without accretion onto chondrules, the following values were obtained: $a_{0}=a_{\mathrm{min}}$, $t_{\mathrm{coll}}=t_{\mathrm{min}}$, and $\zeta=0.794$, respectively (see Eq. \ref{fitting1}). In this case, $t_{\mathrm{coll}}(t)=\mathrm{const}$. However, if we consider accretion onto chondrules, $t_{\mathrm{coll}}(t)$ becomes a function of time,
\begin{align}
t_{\mathrm{coll}}(t) &= \left(\frac{\rho_{d,0}}{\rho_{d}(t)}\right)t_{\mathrm{coll}}(0) \notag \\
 &= \left[ \frac{X_{cd}+\exp \left\{ A_{\mathrm{coll}}\left(\frac{t}{ t_{\mathrm{coll}}(0) }\right) \right\}}{X_{cd}+1} \right]t_{\mathrm{coll}}(0) \ ,
\end{align}
where we used Eq. \ref{accretion} and
\begin{equation}
A_{\mathrm{coll}} = \left(\frac{3\pi \sqrt{\alpha} \mathrm{Re}^{1/4} \rho_{\mathrm{tot}}c_{s}}{8\Sigma_{g}}\right) \times t_{\mathrm{coll}}(0) \ .
\end{equation}
Integrating Eq. \ref{tau}, we obtained 
\begin{equation}
a_{\mathrm{agg}}/a_{0} = C\left[\left(\frac{1+X_{cd}}{X_{cd}}\right)\left(\frac{t}{t_{\mathrm{coll}}}\right) - \left(\frac{1+X_{cd}}{X_{cd}A_{\mathrm{coll}}}\right)\ln \left[ \frac{ X_{cd}+\exp \left\{ A_{\mathrm{coll}}\left(\frac{t}{t_{\mathrm{coll}}}\right) \right\}}{X_{cd}+1}\right] \ \right]^{\zeta},
\end{equation}
which corresponds to Eq. \ref{fitting3}.

\clearpage

\color{red}
\section{Why the lines overlap when $\beta=-2.5$ during the BCCA-like stage}
\label{beta2.5 reason}
The average aggregate size $a_{\mathrm{agg}}$ evolves approximately proportionally to $t$ during the BCCA-like stage, and when  $-3.0 \lesssim \beta$ (i.e., when monomer grain population can be approximated by the single size of $a_{\mathrm{max}}$), 
\begin{equation}
\frac{a_{\mathrm{agg}}}{a_{\mathrm{max}}} \approx C_{2} \left(\frac{t}{t_{\mathrm{single}}(a_{\mathrm{max}})}\right).
\end{equation}
According to Eqs. \ref{fit-a-single}, \ref{fit-a-2.5}, and Figure \ref{aggregate size 2.5b}, $C_{2}$ is a function of $a_{\mathrm{max}}/a_{\mathrm{min}}$ but almost unity, so we ignore this dependence here. Then, 
 \begin{equation}
\frac{a_{\mathrm{agg}}}{a_{\mathrm{min}}} \approx C_{2} \left(\frac{a_{\mathrm{max}}t_{\mathrm{min}}}{a_{\mathrm{min}}t_{\mathrm{single}}(a_{\mathrm{max}})}\right) \left(\frac{t}{t_{\mathrm{min}}}\right).
\label{min_app2.5}
\end{equation}
If we assume $a_{\mathrm{max}}^{4-\beta} \gg a_{\mathrm{min}}^{4-\beta}$ in Eq. \ref{def_t_min}, $t_{\mathrm{min}}$ can be approximated as 
\begin{equation}
t_{\mathrm{min}} \approx  \frac{1}{4\pi a_{\mathrm{min}}^{2} \times \frac{(4+\beta)\rho_{d}a_{\mathrm{min}}^{4+\beta}}{m_{\mathrm{gr}}(a_{\mathrm{min}}) a_{\mathrm{max}}^{4+\beta} } \times \sqrt{\frac{16k_{\mathrm{b}}T}{\pi m_{\mathrm{gr}}(a_{\mathrm{min}})}}} \propto a_{\mathrm{min}}^{-1.5-\beta}a_{\mathrm{max}}^{4+\beta}
\end{equation}
and
\begin{equation}
\frac{a_{\mathrm{max}}t_{\mathrm{min}}}{a_{\mathrm{min}}t_{\mathrm{single}}(a_{\mathrm{max}})} \propto \frac{a_{\mathrm{max}}^{2.5+\beta}}{a_{\mathrm{min}}^{2.5+\beta}}. \label{coefficient_app2.5}
\end{equation}
Only when $\beta=-2.5$, the dependence on both $a_{\mathrm{min}}$ and $a_{\mathrm{max}}$ vanishes and $a_{\mathrm{max}}t_{\mathrm{min}}/a_{\mathrm{min}}t_{\mathrm{single}}(a_{\mathrm{max}})$ is constant. In this case, the growth tracks during BCCA-like stage in $t/t_{\mathrm{min}}$-$a_{\mathrm{agg}}/a_{\mathrm{min}}$ plane (Figure \ref{aggregate size 2.5a}), which is approximately equivalent to  Eq. \ref{min_app2.5}, collapse into one line.
\color{black}

\end{document}